\documentclass[journal,10pt,twoside]{IEEEtran}
\usepackage{cite}
\usepackage{amsmath,amssymb,amsfonts,amsthm}
\usepackage{algorithmic}
\usepackage{graphicx}
\usepackage{textcomp}
\usepackage{xcolor}
\usepackage{placeins}

\usepackage{booktabs}
\usepackage{multirow}
\usepackage{array}
\usepackage{tabularx}
\usepackage{colortbl}
\usepackage{tikz}
\usepackage{pgfplots}
\usepgfplotslibrary{groupplots}
\usepackage{pgfplotstable}
\usepackage{hyperref}
\usepackage{url}
\usepackage{balance}
\usepackage{float}
\usepackage{subcaption}
\usepackage{rotating}
\usepackage{longtable}

\usetikzlibrary{shapes,arrows,positioning,matrix,fit,
                backgrounds,decorations.pathreplacing,
                calc,arrows.meta,shapes.geometric,shapes.misc,shadows,patterns}
\pgfplotsset{compat=1.18}

\definecolor{wellstudied}{RGB}{0,128,128}
\definecolor{emerging}{RGB}{218,165,32}
\definecolor{unexplored}{RGB}{120,120,120}
\definecolor{ieeblue}{RGB}{0,84,166}
\definecolor{lightgray}{RGB}{240,240,240}
\definecolor{lightyellow}{RGB}{255,250,220}
\definecolor{lightgreen}{RGB}{220,255,220}
\definecolor{lightblue}{RGB}{220,235,255}
\definecolor{lightred}{RGB}{255,230,220}
\definecolor{cellgreen}{RGB}{200,240,200}
\definecolor{cellyellow}{RGB}{255,245,200}
\definecolor{cellgray}{RGB}{210,210,210}

\hypersetup{colorlinks=true,linkcolor=ieeblue,citecolor=ieeblue,urlcolor=ieeblue}

\usepackage{float}
\begin{document}

\title{Closing the Semantic-Edge Gap: \\
        Tiny Language Models for 6G Wireless Intelligence}

\author{Srikanth~Kamath,~\IEEEmembership{Member,~IEEE}
        Arnav~Mathur,~\IEEEmembership{Student Member,~IEEE,}
        Joslyn~Sajan~George,~\IEEEmembership{Student Member,~IEEE,}
        and~Rahul~Jashvantbhai~Pandya,~\IEEEmembership{Senior~Member,~IEEE,}%

\thanks{~S.~Kamath is a PhD Scholar with the Department of Electrical,
Electronics, and Communication Engineering, IIT Dharwad.}%
\thanks{A.~Mathur is with the Department of Electronics and Communication
Engineering, Central University of Jammu, J\&K~181143, India, and is a
visiting research intern at IIT Dharwad, Karnataka~580011, India
(e-mail: 24beece15.ece@cujammu.ac.in).}%
\thanks{J.~Sajan~George is with the School of Electrical Engineering, Manipal Institute of Technology, Manipal,
Karnataka~576104, India, and is a visiting research intern at IIT Dharwad.}%

\thanks{R.~Pandya is an Associate Professor with the Department of Electrical,
Electronics, and Communication Engineering, IIT Dharwad.}%
}

\maketitle

\begin{abstract}
Sixth-generation (6G) wireless networks are envisioned as AI-native
systems in which semantic communication -- transmitting task-relevant
meaning rather than raw bits -- moves beyond Shannon's classical
bit-pipe model. Large language models (LLMs) dominate semantic
encoding but are categorically unsuitable for 6G user equipment and
Internet of Things (IoT) devices, given prohibitive memory footprint,
energy consumption, and inference latency. Tiny language models
(TinyLMs) -- compressed via TinyML techniques into
kilobyte-to-megabyte memory and milliwatt power budgets -- are the
missing bridge between LLM-level semantic encoding and 6G edge
hardware, yet no prior work systematically maps TinyML techniques onto
semantic communication architectures for this purpose. This survey
closes that research gap through a two-axis taxonomy connecting six
semantic-encoding compression families (quantization, pruning,
knowledge distillation, low-rank adaptation, neural architecture
search, hybrid pipelines) to five semantic communication architectures
(end-to-end joint source-channel coding, split learning, federated
learning, knowledge-graph-assisted, and multi-task/cross-modal
communication),
synthesized with a qualitative and quantitative meta-analysis of the
model-size-versus-semantic-fidelity Pareto frontier. Building on this
evidence base, the survey distills the identified research gaps into
concrete problem statements for future work and introduces the
recurring themes -- semantic (rather than bit-level) encoding, edge
deployability, and cross-layer design -- that connect the manuscript's
sections into a single narrative arc. Representative results include a
CNN-Transformer semantic encoder achieving 22~dB Peak Signal-to-Noise
Ratio (PSNR) at a 33.33\% semantic-representation size reduction; a
symbolic protocol machine reducing a full neural MAC protocol from
4.55~MB to 1~KB (99.98\% smaller) with zero task-performance loss;
federated bidirectional knowledge distillation (FedBKD) converging
under joint model-and-data heterogeneity where FedAvg-style averaging
underperforms; and knowledge-graph-assisted probability graphs cutting
transmission energy by 65\%. The survey
identifies nine open research challenges for TinyLM-enabled 6G
semantic communication, including two not previously articulated in
the literature.
\end{abstract}


\begin{figure*}[t]
\centering
\includegraphics[width=0.95\textwidth,height=0.85\textheight,keepaspectratio]{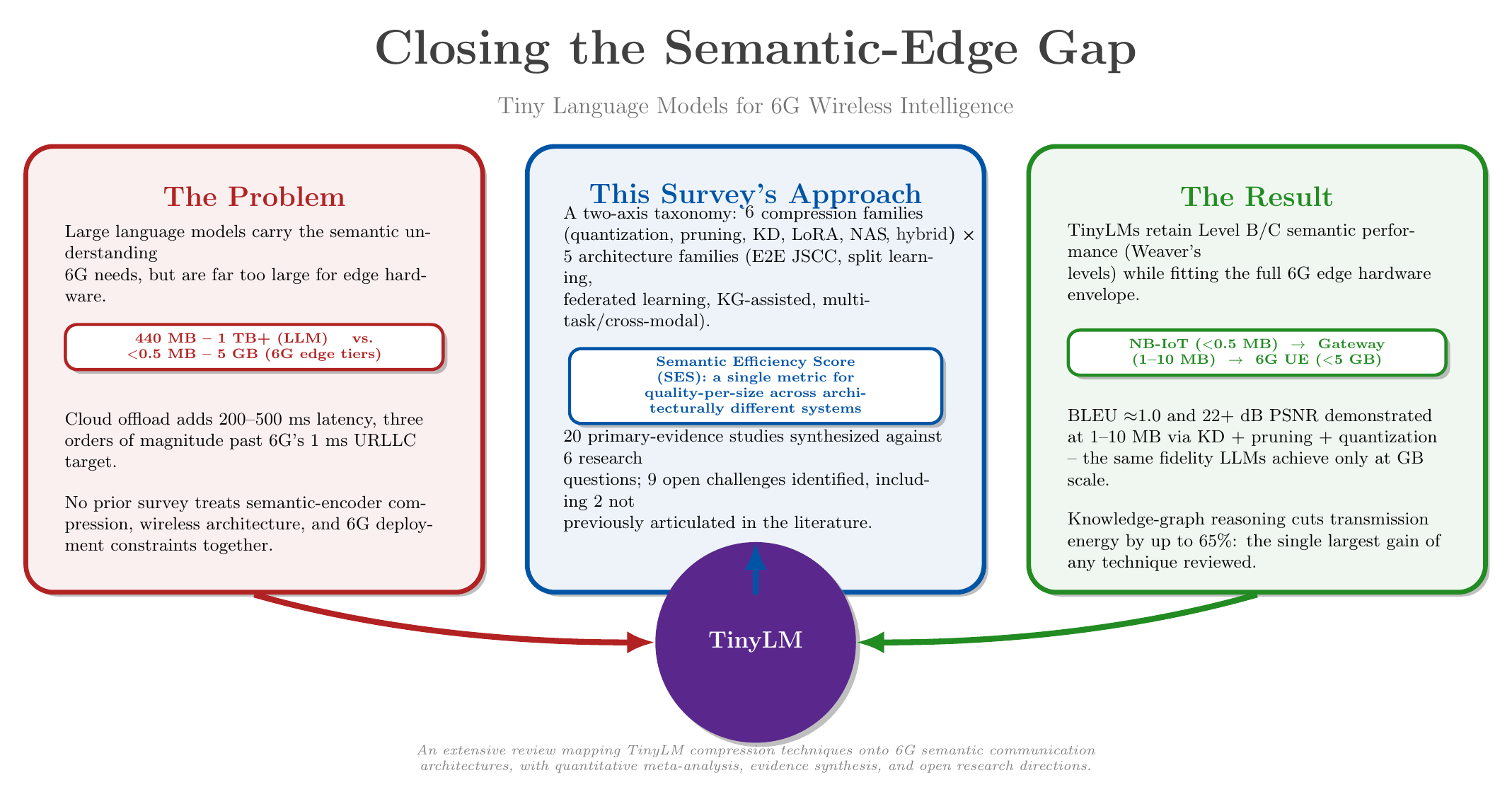}
\caption{Graphical abstract: the problem (LLM semantic power vs.\ 6G
edge hardware constraints), this survey's approach (two-axis
taxonomy, Semantic Efficiency Score, evidence synthesis), and the
key result (TinyLM deployment across the full 6G hardware envelope).}
\label{fig:graphical_abstract}
\end{figure*}

\section{Introduction}
\label{sec:intro}

\IEEEPARstart{T}{he} path from today's data-hungry, cloud-centric wireless networks to
an AI-native sixth generation (6G) runs through three stages: the raw
semantic power of large language models (LLMs), the deep-learning
architectures that adapt that power to communication tasks, and
finally the TinyLM semantic-encoding pipeline that makes the result
small enough to run on 6G edge hardware. This survey synthesizes
evidence from an extensive review of the TinyLM, semantic-encoding
compression, and semantic-communication literature, focusing on
studies that report both qualitative design insights and quantitative
deployment metrics relevant to 6G edge hardware. Twenty
primary-evidence studies meeting this scope are synthesized in detail
in Section~\ref{sec:evidence}, alongside a broader supporting corpus
referenced throughout the survey. The remainder of this section
motivates each stage of the LLM-to-TinyLM journey in turn and states
the Research Gap this survey fills.

\subsection{6G as an AI-Native Paradigm: Beyond the Bit-Pipe Model}

Cellular systems from the first through the fifth generation have
followed what this survey terms the \emph{bit-pipe paradigm}: the
network's job ends once a bit sequence is delivered with acceptably
low error, irrespective of what that sequence means or what task it
serves. The sixth generation of wireless communications, projected for
commercial deployment between 2030 and 2035, marks a fundamental
architectural move beyond this paradigm~\cite{Celik2024}: the IMT-2030
framework specifies peak data rates of 1~Tbps, sub-millisecond
end-to-end latency, a connectivity density of $10^7$~devices/km$^2$,
and -- unlike prior generations -- \emph{native AI intelligence}:
embedded machine learning, knowledge-based optimization, and on-device
learning integrated across the physical, access, and network
layers~\cite{Aloudat2025}, rather than the bolt-on analytics and
self-organizing-network (SON) features layered atop 4G/5G. These
targets cannot be satisfied by incrementally extending 5G
architectures; they necessitate a ground-up redesign in which
intelligence is a first-class citizen of the protocol
stack~\cite{Letaief2022_roadmap}.

Beyond enhanced mobile broadband (eMBB), ultra-reliable low-latency
communication (URLLC), and massive machine-type communication (mMTC)
-- the three service classes already defined by 5G -- IMT-2030
introduces immersive communications (holographic, haptic) and
integrated sensing and communication (ISAC), each placing demands on
semantic fidelity, latency, and energy that raw bit transmission
cannot meet~\cite{Yang2023_semcom}. The scale of this demand is
compounded by device growth: the IoT device population is projected to
grow from 14.4~billion in 2022 to over 39~billion by
2030~\cite{IoTAnalytics2025_StateOfIoT}, a growth rate that makes
brute-force bandwidth expansion economically and physically
unsustainable and instead requires transmitting only task-relevant
meaning~\cite{Zhu2023_PushingAI}. Realization of AI-native 6G therefore
demands semantic inference executing locally at the network edge --
though, as Section~\ref{sec:background_hw} details, the resulting hardware
envelope is tiered rather than uniform: milliwatt-power,
sub-megabyte-memory budgets bind at the NB-IoT tier,
while gateway and 6G UE tiers admit watt-level power and gigabyte-scale
memory~\cite{Shafique2021_TinyML}. The central engineering challenge
this survey addresses is precisely this tiered deployment gap: encoding
the semantic intelligence of modern LLMs into a form compatible with
6G edge hardware, at whichever tier a given device sits, without
catastrophic loss of semantic fidelity.

\subsection{Semantic Communication: From Theory to Practice}

Weaver's canonical three-level decomposition of
communication~\cite{ShannonWeaver1949} distinguishes three questions a
system can be designed to answer: Level~A (the technical problem) asks
how accurately symbols can be transmitted; Level~B (the semantic
problem) asks how precisely the transmitted symbols convey the
intended meaning; and Level~C (the effectiveness problem) asks how
successfully that conveyed meaning drives the desired behavior at the
receiver. Shannon's 1948 information theory~\cite{Shannon1948}
answered Level~A definitively but explicitly excluded Levels~B and~C,
noting that ``the semantic aspects of communication are irrelevant to
the engineering problem.'' For most of the seven decades that
followed, Levels~B and~C remained largely outside mainstream
communication-system design, since no tractable model existed for
representing ``meaning'' as an engineering quantity that a transmitter
and receiver could jointly optimize. This is precisely what changed
with the convergence of deep learning, transformer architectures, and
large-scale pre-trained language models in 2017--2022: neural networks
can now learn distributed representations that capture task-relevant
meaning directly from data, giving Level~B and~C communication a
concrete, trainable, and measurable form~\cite{Shi2023_ML6G}. This
survey is concerned with making that Level~B/C capability deployable
at the 6G edge.

Deep learning-based semantic communication systems build on this
capability by learning a \emph{semantic encoder} at the transmitter and
a matching \emph{semantic decoder} at the receiver: the encoder maps
raw source data to a compact representation that preserves
task-relevant meaning, and the decoder reconstructs that meaning --
rather than the original bits -- at the other end. Where the
channel-facing stage is learned jointly with this semantic
encoder/decoder pair, an approach known as joint source-channel coding
(JSCC), the resulting end-to-end architecture can outperform classical
designs that separate source coding, channel coding, and semantic
processing into independent stages, particularly at low
signal-to-noise ratio (SNR) where separate coding's asymptotic
optimality guarantees no longer hold~\cite{Gunduz2022_beyond};
Section~\ref{sec:architectures} discusses this JSCC role in detail. For
now, DeepSC~\cite{Xie2021_DeepSC,Sana2022_DeepSC} illustrates the
general pattern: it trains a Transformer-based semantic encoder and
decoder end-to-end over a simulated wireless channel, with a loss
function that directly penalizes semantic (meaning-level)
reconstruction error rather than bit error rate. At deployment, the
encoder transmits a compact semantic representation instead of the raw
sentence, and the decoder reconstructs the intended meaning even under
channel noise -- on the European Parliament corpus, DeepSC achieves Bilingual Evaluation Understudy (BLEU)
scores exceeding 0.9 at SNR as
low as 0~dB while transmitting 46\% fewer tokens than sending the
original, uncompressed
text~\cite{Wang2021_GLOBECOM}, a regime in which classical bit-exact
transmission degrades sharply. The same semantic-encoding principle
generalizes beyond text: S.~Jiang~\emph{et al.}'s knowledge-graph-based
encoder, which converts source sentences into entity-relation triples
and allocates transmission resources by triple importance, achieves
70\% data reduction relative to conventional sentence-level (non-KG)
transmission of the same content while maintaining semantic
accuracy~\cite{Jiang2022_KG}; and C.~Chaccour and W.~Saad's
contrastive-learning-based encoder, evaluated on simulated content of
varying complexity, separates semantically
redundant from semantically essential data before transmission,
reducing representation length by 57.22\% and improving semantic
impact by 71.9\%, in both cases relative to the ``vanilla'' (undisentangled)
semantic communication baseline the authors define in the same
study~\cite{Chaccour2022_contrastive} -- not relative to classical
bit-level transmission. Together, these results point to a common conclusion: it is the encoder's
\emph{learned notion of what is semantically essential} -- not the raw
modality or network architecture -- that determines how much a
semantic encoder can compress.

\subsection{The Core Problem: LLMs Cannot Fit on 6G Edge Devices}

Transformer-based LLMs are unsuitable for the most constrained 6G
edge tiers on parameter count alone -- though, as
Section~\ref{sec:background_hw} details, gateway and 6G UE tiers have enough
flash/RAM to host BERT-scale models directly, so this mismatch is a
tier-specific problem, not a blanket one across all 6G user equipment
and IoT devices: BERT has
110~million parameters~\cite{Devlin2019_BERT}, GPT-3 has 175~billion~\cite{Brown2020_GPT3}, and GPT-4 is widely
reported (though not officially confirmed) to be on the order of
$10^{12}$ parameters. Even the smallest of these, BERT, still exceeds
the flash budget of an NB-IoT-class microcontroller such as the
Cortex-M33 by more than two orders of magnitude once quantized to INT8
precision -- and GPT-3/GPT-4-class models are three to four orders of
magnitude larger still. Offloading inference to the cloud sidesteps
the memory problem but introduces a different one: 200--500~ms
round-trip latency, violating 6G's 1~ms URLLC target by three orders
of magnitude. Energy compounds the problem at the device itself: a 6G IoT sensor
performing \emph{continuous} inference at 100~mW -- not duty-cycled,
and not counting the radio -- exhausts a nominal 1000~mAh cell at the
3.6~V nominal voltage typical of the lithium thionyl chloride
(Li-SOCl$_2$) primary cells standard in multi-year NB-IoT sensor
deployments (3.6~Wh capacity) in
roughly 36~hours, wholly unacceptable against ten-year smart-city
deployment targets~\cite{Abadade2023_TinyML} unless inference is
duty-cycled aggressively enough to bring the \emph{average} power
draw down by three to four orders of magnitude -- itself only
possible if the model is small enough to complete each inference
quickly rather than running continuously, which is precisely the
TinyLM design constraint this survey addresses. Energy-harvesting platforms (solar, RF), which deliver only
10--100~$\mu$W, restrict on-device inference to sub-1~mW neural
architectures~\cite{Shafique2021_TinyML}.

The model class required to close this three-way gap -- parameter
count, latency, and energy -- is the TinyLM: an LLM-derived semantic
encoder compressed to fit within the resource envelope of a specific
6G edge tier (Section~\ref{sec:background_hw}) while preserving as much of the
original transformer's semantic encoding capability as that tier's
budget allows.

\subsection{Research Gap and Motivation}

Positioning this survey requires walking through three bodies of work
in sequence -- 6G semantic communication, large language models, and
tiny language models -- since the gap this survey fills sits precisely
at a junction none of them individually occupies.

\textbf{6G semantic communication.} A substantial body of work already
establishes semantic communication as 6G's answer to the bit-pipe
paradigm, developing architectures for end-to-end joint source-channel
coding, split learning, federated learning, and knowledge-graph-assisted
encoding~\cite{Yang2023_semcom,Luo2022_semcom,Wheeler2023_semcom}. This
literature is architecture-focused: it asks how a semantic encoder
should be structured and trained, largely independent of what hardware
that encoder must ultimately run on.

\textbf{Large language models.} A separate and much larger body of work
establishes that large language models are the strongest available
semantic encoders, since their pre-trained representations capture
task-relevant meaning far more effectively than hand-designed or
narrowly-trained alternatives. As Section~\ref{sec:intro} established,
however, this semantic power is inseparable from a parameter count and
inference cost that categorically exceeds any 6G edge device's budget.

\textbf{Tiny language models.} A third, TinyML-rooted body of work
compresses neural models in general -- via quantization, pruning,
distillation, and neural architecture search (NAS) -- to fit
resource-constrained hardware~\cite{Shafique2021_TinyML,Abadade2023_TinyML}.
This literature is compression-focused: it asks how to shrink a model
without asking what that model is being asked to do downstream, and it
rarely targets language models specifically, let alone the semantic
communication task.

\textbf{The integration gap.} No existing work systematically connects
these three threads: applying TinyML-style compression specifically to
LLM-derived semantic encoders -- yielding TinyLMs -- and mapping the
resulting TinyLMs onto the semantic communication architectures
reviewed above, across the tiered 6G edge deployment space of
Section~\ref{sec:background_hw}. Concretely, no prior work (1)~systematically
maps TinyLM compression techniques onto semantic communication
architectures across the 6G design space, or (2)~provides a
quantitative meta-analysis of the resulting
model-size-versus-semantic-fidelity Pareto frontier. This survey fills
both gaps, and in doing so surfaces the open research problems --
discussed throughout and consolidated in
Section~\ref{sec:challenges} -- that this TinyLM/6G-semantic-communication
intersection raises for future work, particularly for IoT and
edge-network deployment.

\begin{figure*}[t]
\centering
\includegraphics[width=0.92\textwidth,height=0.85\textheight,keepaspectratio]{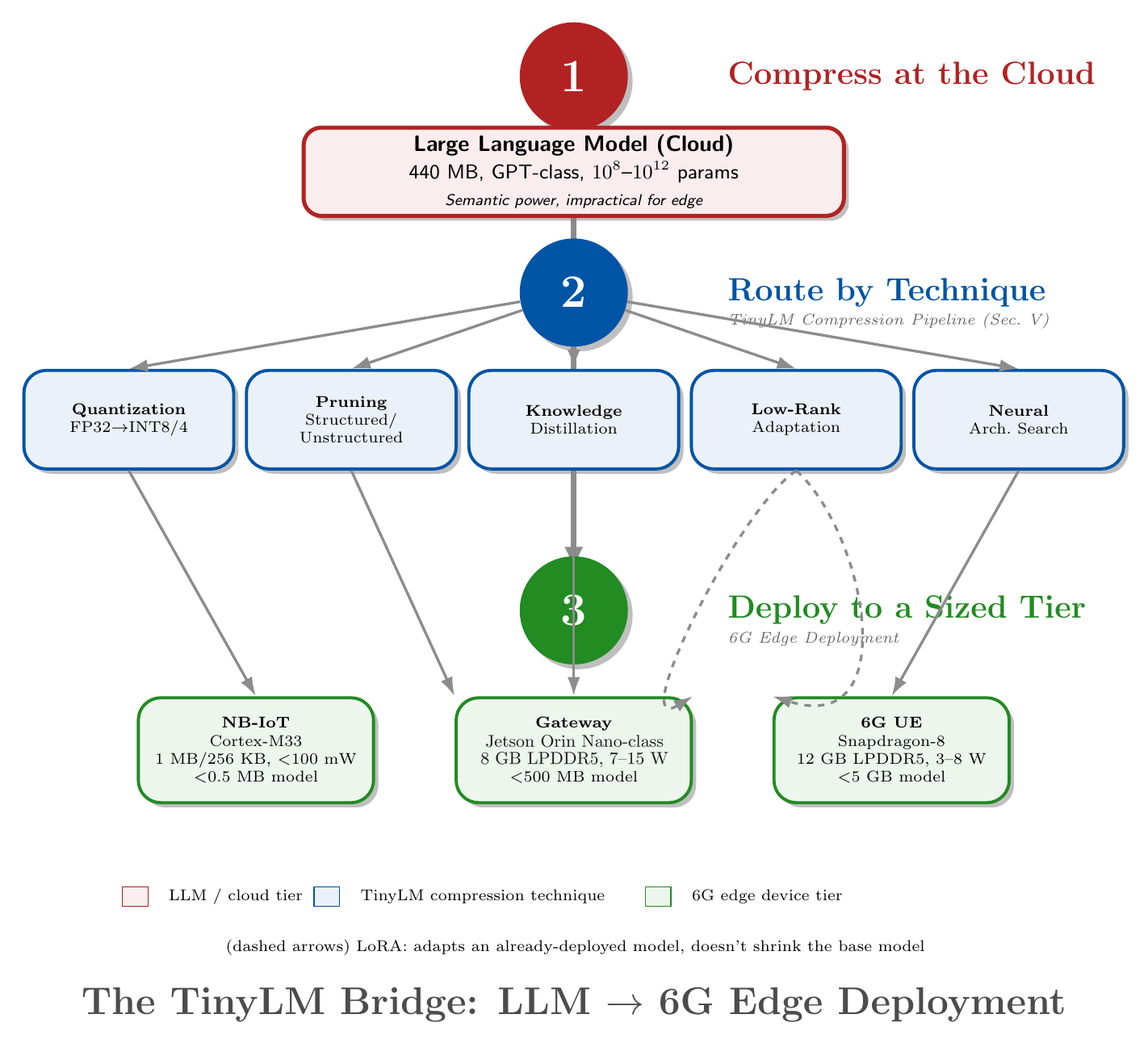}
\caption{The TinyLM bridge: large language models possess the semantic
encoding power needed for 6G but cannot fit on edge hardware. Circled
numbers mark the three-stage journey (1) compress at the cloud, (2)
route by technique, (3) deploy to a sized tier. Each of
TinyLM's five compression families (quantization, pruning, knowledge
distillation, low-rank adaptation, neural architecture search)
routes into a device tier sized for that family's typical
output (Section~\ref{sec:background_hw}); LoRA is dashed because it adapts
an already-deployed model rather than shrinking one, so it does not
map to a single tier the way the other four do. Each is reviewed
against the 20
primary-evidence studies synthesized in Section~\ref{sec:evidence}.
This diagram anchors the three-act
narrative developed throughout the survey.}
\label{fig:tinyml_bridge}
\end{figure*}
Figure~\ref{fig:tinyml_bridge} situates the remainder of this survey
around this same three-act structure: LLM semantic power (top), the
TinyML compression pipeline (middle), and 6G edge deployment (bottom).
Section~\ref{sec:compression} reviews the compression techniques
available at the pipeline stage; Section~\ref{sec:architectures}
addresses how compressed encoders are embedded into communication
architectures, including the physical/MAC-layer coding and channel
robustness questions (Sections~\ref{sec:architectures}
and~\ref{sec:challenges}); and Section~\ref{sec:metaanalysis}
quantifies the resulting fidelity via the SES and Pareto analysis.
This diagram is referenced throughout the survey as a structural
anchor connecting otherwise disparate technical discussions---compression, channel coding, federated
training, and standardization---into the single coherent deployment
pipeline that a 6G network operator must engineer end-to-end.

\subsection{Key Contributions}

Building directly on the Research Gap identified above, this survey
makes five primary contributions, each addressing one specific piece
of the TinyLM/6G-semantic-communication integration gap:
\begin{enumerate}
  \item \textbf{Two-Axis Taxonomy:} A novel taxonomy mapping semantic-encoder
    compression methods -- quantization, pruning, knowledge distillation,
    low-rank adaptation (LoRA), neural architecture search (NAS), and
    hybrid pipelines that combine them -- against semantic communication
    architectures (end-to-end
    JSCC, split learning, federated learning (FL), knowledge-graph
    (KG)-assisted, and multi-task/cross-modal), identifying the maturity
    of each of the resulting 30 cells and exposing 13 unexplored
    high-potential cells.
  \item \textbf{Evidence Synthesis Framework:} A synthesis
    of the primary-evidence studies identified through this survey's
    review process against a uniform six-dimension evaluation
    framework, enabling direct cross-system comparison of compression
    ratio, semantic fidelity, hardware footprint, and reported
    limitations across architecturally heterogeneous TinyLM systems;
    Section~\ref{sec:evidence} synthesizes, with full
    re-verified citations, all twenty systems making up this survey's
    evidence base, and states
    explicitly which findings receive the deeper, individually-worked
    treatment versus the broader supporting corpus.
  \item \textbf{Quantitative and Qualitative Meta-Analysis:} Aggregated
    findings spanning size reduction, inference latency, energy, and
    semantic fidelity (BLEU, PSNR, cosine similarity, ROUGE), including
    an illustrative Semantic Efficiency Score (SES) heuristic applied
    across
    systems, contextualized against each system's qualitative design
    choices. SES is a design-time triage tool for comparing systems
    already reported on a common quality axis, not a validated,
    ground-truth-calibrated metric; Section~\ref{sec:metaanalysis}
    states this limitation explicitly alongside every SES value
    reported.
  \item \textbf{Consolidated Open Challenges:} In-depth analysis of open
    research challenges for TinyLM-enabled 6G semantic communication,
    including two not previously identified in the literature
    (cross-modal semantic alignment and scalable KG synchronization),
    each framed as a concrete problem statement for future work.
  \item \textbf{Original Research Synthesis:} Hardware--software
    co-design directions, adaptive compression for dynamic 6G channels,
    and semantic security frameworks -- novel synthesis not derivable
    from any single source. Four small original theoretical
    extensions are stated explicitly at their point of use rather than
    claimed here as separately validated results: an explicit
    multi-cell semantic-interference term added to the two-term
    semantic noise model of Section~\ref{sec:background_theory}
    (Eq.~\ref{eq:semnoise_interference}); an illustrative
    reasoning-capacity adaptation of the teacher/apprentice
    reformulation of C.~Chaccour~\emph{et al.}~\cite{Chaccour_2022_LessData}
    (Eq.~\ref{eq:reasoning_capacity}, Section~\ref{sec:background_theory});
    a convergence sketch for cross-architecture distillation under
    non-IID semantics (Eq.~\ref{eq:fedsfd_convergence},
    Section~\ref{sec:challenges}); and the two-dimension
    (blocking-severity $\times$ deployment-urgency) priority rubric
    underlying the P1/P2/P3 ratings of
    Section~\ref{sec:challenges}. All four are this survey's own additions,
    not reproductions of a cited source's result, and none carries
    independent empirical validation beyond the reasoning given where
    each is introduced.
\end{enumerate}

\textbf{Note on originality and independent verification.} This
survey states its epistemic status once, here, rather than repeating
it at each occurrence: unless an equation or claim is directly
attributed to a cited source, every equation in this survey restates
that source's own formulation, and is not this survey's independent
result. The handful of exceptions -- genuinely original theoretical
extensions this survey adds rather than restates -- are the four
named explicitly in the Contributions list above (the interference
term, the reasoning-capacity adaptation, the convergence sketch, and
the priority rubric), each stated as this survey's own addition
directly in the prose at its point of use rather than left for a
reader to infer; none of these carries independent empirical
validation beyond the reasoning given at that point, and each is
explicitly scoped as illustrative rather than proven. Every citation
in this survey's
bibliography, separately, has been individually verified against a
primary, citable source as part of this survey's own
citation-accuracy pass (Section~\ref{sec:evidence}), which is a
distinct claim from originality and applies uniformly regardless of
whether a given passage is one of these four original extensions.

\section{Related Surveys and Positioning}
\label{sec:related}

Having established the TinyLM/6G-semantic-communication Research Gap,
this section grounds it in the existing literature: it surveys prior
semantic-communication and TinyML/TinyLM-adjacent surveys individually,
then positions the present work against them in
Table~\ref{tab:positioning}, establishing precisely what combined
Research Gap -- semantic-encoding compression, communication
architecture, and deployment constraints treated together -- this
survey fills.

\subsection{Semantic Communication Survey Literature}

W.~Yang~\emph{et al.}~\cite{Yang2023_semcom} provide the most
comprehensive semantic communication survey to date, with a systematic
taxonomy of semantic metrics; it predates significant TinyLM
compression results and does not address model compression at all.
D.~Wheeler and B.~Natarajan~\cite{Wheeler2023_semcom} survey four
semantic-communication engineering approaches and find that ML-based
approaches lack interpretability -- a finding that motivates this
survey's interest in TinyLMs specifically, since smaller models are
typically easier to interpret -- but review no compression techniques.
X.~Luo~\emph{et al.}~\cite{Luo2022_semcom} explicitly call for ``model
compression via network sparsification and quantization'' as a
critical IoT-tier enabler -- to date the strongest existing motivation
in the literature for a dedicated TinyLM research agenda -- though the
survey stops short of proposing one. S.~N.~Karahan and O.~Kaya~\cite{Karahan2025_6G}
provide the most current 3GPP gap analysis and the semantic noise
model that informs Section~\ref{sec:challenges}. T.~M.~Getu~\emph{et al.}~\cite{Getu2024_Semantic}
provide the most recent broad landscape survey; like the surveys
above, its scope is semantic communication broadly rather than
TinyLM-scale deployment specifically. T.~T.~Win~\emph{et al.}~\cite{Win2025_Energy}
survey energy efficiency for semantic communication and explicitly
name the gap this survey closes: system-level energy optimization has
been studied independently of model-level efficiency, never jointly.
Y.~Wang~\emph{et al.}~\cite{Wang2024_Intellicise} propose Intellicise
model transmission -- conceptually adjacent but focused on
transmitting a model rather than compressing one for on-device use.
A.~Shahraeeni~\emph{et al.}~\cite{Shahraeeni2025_Empowering} and
X.~Chen~\emph{et al.}~\cite{Chen2025_LLMEmpowered} both survey
LLM-empowered IoT for 6G; Shahraeeni et al.\ notably name concrete
TinyLM-scale examples (BitNet~b1.58, GPIoT), but neither addresses
semantic communication specifically, leaving the
TinyLM-for-6G-semantic-communication integration gap fully open.

\subsection{TinyML and Edge-Compression Survey Literature}

A parallel, TinyML-rooted survey literature establishes the compression
techniques this survey draws on -- though none of it treats language
models or semantic communication as a first-class concern.

M.~Shafique~\emph{et al.}~\cite{Shafique2021_TinyML} provide the
foundational TinyML roadmap (quantization, pruning, NAS, and
hardware-software co-design). Y.~Abadade~\emph{et al.}~\cite{Abadade2023_TinyML}
provide the most comprehensive TinyML survey to date, with benchmarks
on Cortex-M/STM32/ESP32 hardware that help establish the constraint
envelope this survey targets (Section~\ref{sec:background_hw}). Y.~Tay~\emph{et al.}~\cite{Tay2022_Efficient_Transformers}
establish the quadratic self-attention bottleneck as the primary
general transformer-scaling barrier -- this survey's own extrapolation
from that finding (not Tay et al.'s own claim) is that a
self-attention layer's activation memory scaling makes even a heavily
quantized transformer likely infeasible on NB-IoT hardware purely from
attention activations, independent of parameter count, motivating this
survey's proposal that
linear-complexity attention variants are a natural, still-unexplored
NAS search space for TinyLM semantic encoders. M.~Treviso~\emph{et al.}~\cite{Treviso2023_efficient}
and Q.~Fournier~\emph{et al.}~\cite{Fournier_2023_Practical} complement
this with pipeline-stage and practitioner-oriented efficiency surveys
whose reported compression/speed-up figures are directly consistent
with the pipelines reviewed in Section~\ref{sec:compression}.

A second, more recent cluster confirms this landscape while
independently reinforcing the same gap. H.~Han~\emph{et al.}~\cite{Han2026_TinyML}
bibliometrically confirm TinyML is an active, fast-expanding field.
L.~Capogrosso~\emph{et al.}~\cite{Capogrosso2024_TinyML} systematically
review TinyML deployments and independently corroborate the hardware
envelope this survey targets (Section~\ref{sec:background_hw}). R.~Kallimani~\emph{et al.}~\cite{Kallimani2024_TinyML}
and L.~Pazmi\~no~Ortiz~\emph{et al.}~\cite{Pazmino2025_Advancing} both
catalog TinyML tooling and hardware targets relevant to any TinyLM
deployment. H.-I.~Liu~\emph{et al.}~\cite{Liu2024_Lightweight} and
Y.~Liang~\emph{et al.}~\cite{Liang2026_Compression} both cover
lightweight/compressed LLMs, the latter specifically for edge systems.
E.~Din\c{c}er and Z.~H.~Kilimci~\cite{Dincer2026_Realtime} survey
compression techniques for real-time and offline LLM deployment on
edge hardware. V.~Rajapakse~\emph{et al.}~\cite{Rajapakse2023_Intelligence}
survey \emph{reformable} TinyML -- models that adapt post-deployment
rather than being compressed once and frozen -- directly relevant to
the adaptive-compression direction of Section~\ref{sec:future}.
I.~Lamaakal~\emph{et al.}~\cite{Lamaakal2025_Compression} contribute a
second, distinct survey (not to be confused with their own
TinyLM-specific automation-and-control survey cited earlier) on
general-purpose TinyML compression. Every one of these surveys targets
TinyML or LLM compression in general -- none treats semantic
communication, wireless channels, or the tiered 6G deployment envelope
as a first-class concern, which is precisely the combined gap this
survey closes. Table~\ref{tab:tinyml_positioning} makes this pattern
explicit across the full cluster discussed above, rather than
leaving it implicit in the running text alone.

\begin{table*}[t]
\caption{Positioning of This Survey vs.\ the TinyML/Edge-Compression Survey Cluster}
\label{tab:tinyml_positioning}
\centering
\small
\renewcommand{\arraystretch}{1.15}
\begin{tabular}{|p{3.2cm}|c|c|c|c|l|}
\hline
\textbf{Survey} & \textbf{TinyML} & \textbf{LM/LLM} & \textbf{6G/Wireless} & \textbf{HW Benchmarks} & \textbf{Scope} \\
\hline
Shafique'21~\cite{Shafique2021_TinyML}       & \checkmark & $\times$ & $\times$ & $\sim$ & Broad \\
\hline
Abadade'23~\cite{Abadade2023_TinyML}         & \checkmark & $\times$ & $\times$ & \checkmark & Broad \\
\hline
Tay'22~\cite{Tay2022_Efficient_Transformers} & $\times$ & \checkmark & $\times$ & $\times$ & Moderate \\
\hline
Treviso'23~\cite{Treviso2023_efficient}      & $\sim$ & \checkmark & $\times$ & $\times$ & Broad \\
\hline
Fournier'23~\cite{Fournier_2023_Practical}   & $\sim$ & \checkmark & $\times$ & $\times$ & Moderate \\
\hline
Han'26~\cite{Han2026_TinyML}                 & \checkmark & $\times$ & $\times$ & $\times$ & Moderate \\
\hline
Capogrosso'24~\cite{Capogrosso2024_TinyML}   & \checkmark & $\times$ & $\times$ & \checkmark & Broad \\
\hline
Kallimani'24~\cite{Kallimani2024_TinyML}     & \checkmark & $\times$ & $\times$ & \checkmark & Moderate \\
\hline
Pazmi\~no'25~\cite{Pazmino2025_Advancing}    & \checkmark & $\times$ & $\times$ & \checkmark & Moderate \\
\hline
Liu'24~\cite{Liu2024_Lightweight}            & $\sim$ & \checkmark & $\times$ & $\times$ & Broad \\
\hline
Liang'26~\cite{Liang2026_Compression}        & $\sim$ & \checkmark & $\times$ & $\sim$ & Moderate \\
\hline
Din\c{c}er'26~\cite{Dincer2026_Realtime}     & $\sim$ & \checkmark & $\times$ & $\sim$ & Moderate \\
\hline
Rajapakse'23~\cite{Rajapakse2023_Intelligence} & \checkmark & $\times$ & $\times$ & $\times$ & Moderate \\
\hline
Lamaakal'25~\cite{Lamaakal2025_Compression}  & \checkmark & $\times$ & $\times$ & $\times$ & Moderate \\
\hline
\textbf{This Survey}                         & \checkmark & \checkmark & \checkmark & \checkmark & \textbf{Broad} \\
\hline
\end{tabular}
\renewcommand{\arraystretch}{1}

\vspace{4pt}
\footnotesize $\checkmark$~= covered; $\sim$~= partial; $\times$~= not covered.
HW Benchmarks~= reports empirical, device-level hardware measurements (not roadmap-level generalities).
\end{table*}

\noindent\textit{Basis for the marks above:} Shafique'21 and
Rajapakse'23 give a roadmap/taxonomy without device-level
benchmarks of their own, hence $\sim$/$\times$ on HW Benchmarks;
Abadade'23, Capogrosso'24, Kallimani'24, and Pazmi\~no'25 each report
or catalog real hardware specifications and measurements, hence
\checkmark; Tay'22, Treviso'23, Fournier'23, Liu'24, Liang'26, and
Din\c{c}er'26 are LM/LLM-compression-centered rather than TinyML/edge
device-centered, hence \checkmark\ on LM/LLM and $\times$/$\sim$ on
TinyML; Liang'26 and Din\c{c}er'26 discuss edge deployment
conceptually without reporting original hardware measurements, hence
$\sim$ rather than \checkmark\ on HW Benchmarks. Every single row in
this table scores $\times$ on 6G/Wireless without exception --
independent confirmation, from a full cluster of fourteen surveys
rather than the two-survey subset Table~\ref{tab:positioning} alone
could show, that no existing TinyML/edge-compression survey treats
6G or wireless semantic communication as a first-class concern.

T.-H.~Vu~\emph{et al.}~\cite{Vu2026_Integration} come closest to bridging
that gap: their survey addresses integrating LLMs with TinyML edge
devices (PEFT, LoRA, federated fine-tuning, split learning, knowledge
distillation) under real 6G resource, latency, and privacy constraints.
Even this survey, however, stops short of a semantic communication
architecture treatment -- no joint source-channel coding, semantic
fidelity metrics, or wireless channel modeling -- leaving the
TinyLM-for-6G-semantic-communication integration gap fully open.

A wider ring of compression-technique surveys, not specific to
TinyML or 6G, further corroborates the individual technique families
Section~\ref{sec:compression} reviews in depth:
J.~Gou~\emph{et al.}~\cite{Gou2021_KD_Survey} and L.~Wang and K.-J.~Yoon~\cite{Wang2020_KD_Survey}
survey knowledge distillation; Q.~Dong~\emph{et al.}~\cite{Dong2023_Survey_NLP_Comp}
and T.~Liang~\emph{et al.}~\cite{Liang2021_PruningQuantSurvey} cover
pruning and quantization for text models and DNN acceleration
respectively; H.~Cheng~\emph{et al.}~\cite{Wan2023_PruningSurvey} and
Y.~Cheng~\emph{et al.}~\cite{Cheng2020_Survey_Pruning} extend pruning
taxonomies to non-NLP architectures; Z.-Q.~Xu~\emph{et al.}~\cite{Xu2023_Survey_Transformer_Infer}
cover transformer inference optimization; and P.~Ganesh~\emph{et al.}~\cite{Ganesh2021_BERT}
survey compression specifically for BERT-class models -- together the
general-purpose compression literature this survey's TinyLM-specific
techniques adapt from.

\subsection{Positioning of This Survey}

Table~\ref{tab:positioning} positions this survey against related works
across five dimensions, mapping explicitly which of them each cited
survey does and does not cover. This survey is comprehensive in scope:
it combines semantic-encoder compression techniques, semantic
communication architectures, 6G deployment constraints, and
quantitative meta-analysis in a single treatment, with future research
directions (Section~\ref{sec:future}) rather than a release-specific
standardization timeline.

\begin{table*}[!t]
\caption{Positioning of This Survey vs.\ Related Works}
\label{tab:positioning}
\centering
\small
\renewcommand{\arraystretch}{1.15}
\begin{tabular}{|p{2.9cm}|c|c|c|c|c|l|}
\hline
\textbf{Survey} & \textbf{SemCom} & \textbf{TinyML} & \textbf{6G} &
\textbf{Quant.} & \textbf{Future Dir.} & \textbf{Scope} \\
\hline
Yang'23~\cite{Yang2023_semcom}           & \checkmark & $\times$ & $\sim$ & $\times$ & $\times$ & Broad \\
\hline
Wheeler'23~\cite{Wheeler2023_semcom}     & \checkmark & $\times$ & $\sim$ & $\times$ & $\times$ & Broad \\
\hline
Luo'22~\cite{Luo2022_semcom}            & \checkmark & $\times$ & $\sim$ & $\times$ & $\times$ & Moderate \\
\hline
Shafique'21~\cite{Shafique2021_TinyML}  & $\times$ & \checkmark & $\times$ & $\sim$ & $\times$ & Moderate \\
\hline
Abadade'23~\cite{Abadade2023_TinyML}    & $\times$ & \checkmark & $\times$ & \checkmark & $\times$ & Broad \\
\hline
Celik'24~\cite{Celik2024}               & $\sim$ & $\times$ & \checkmark & $\times$ & $\times$ & Broad \\
\hline
Aloudat'25~\cite{Aloudat2025}           & \checkmark & $\times$ & \checkmark & $\times$ & $\sim$ & Broad \\
\hline
Karahan'25~\cite{Karahan2025_6G}        & \checkmark & $\times$ & \checkmark & $\sim$ & $\sim$ & Moderate \\
\hline
Yang'26~\cite{Yang_2026_LLMDecision}    & $\sim$ & $\sim$ & \checkmark & $\sim$ & $\times$ & Moderate \\
\hline
Vu'26~\cite{Vu2026_Integration}         & $\times$ & \checkmark & \checkmark & $\sim$ & \checkmark & Broad \\
\hline
\textbf{This Survey}                    & \checkmark & \checkmark & \checkmark & \checkmark & \checkmark & \textbf{Broad} \\
\hline
\end{tabular}
\renewcommand{\arraystretch}{1}

\vspace{4pt}
\footnotesize $\checkmark$~= covered; $\sim$~= partial; $\times$~= not covered.
Quant.~= quantitative meta-analysis; Future Dir.~= concrete forward-looking research directions
(not necessarily a release-specific standardization timeline).
\end{table*}

\noindent\textit{Basis for the $\times$/$\sim$ marks above} (each judged
against that survey's own stated scope, not against a more recent
version that may since have appeared): Yang'23 and Wheeler'23 are
general SemCom overviews with no TinyML-specific compression content
and no forward-looking research-directions section, hence $\times$ on
TinyML/Quant./Future~Dir.; Luo'22 similarly omits TinyML entirely but
does gesture at open issues, hence Moderate rather than Broad scope;
Shafique'21 and Abadade'23 are TinyML-focused with no SemCom content
at all ($\times$ on SemCom, 6G), and Shafique'21's roadmap is
partial ($\sim$ on Quant.) rather than a worked meta-analysis;
Celik'24 addresses 6G broadly but treats SemCom only as one example
application among many generative-AI use cases, hence $\sim$ rather
than $\checkmark$; Aloudat'25 covers SemCom and 6G but its
forward-looking content is Metaverse-integration-specific rather than
a TinyLM research agenda, hence $\sim$ on Future~Dir.; Karahan'25
stops at architectural principles without a
compression taxonomy, hence $\sim$ on both Quant.\ and Future~Dir.;
Yang'26 is LLM-decision-focused rather than compression-focused, so
its SemCom and TinyML coverage is incidental rather than central,
hence $\sim$ rather than $\checkmark$; Vu'26 is this survey's closest
prior work by explicit topical overlap -- TinyML/LargeML integration
under real 6G resource, latency, and privacy constraints, with
concrete forward-looking integration directions -- but by its own
stated scope does not address joint source-channel coding, semantic
fidelity metrics, or wireless channel modeling, hence $\times$ on
SemCom despite covering TinyML and 6G directly; its resource-constraint
treatment is quantitative in places but not a worked meta-analysis
across techniques, hence $\sim$ rather than $\checkmark$ on Quant.

The most recent addition to this landscape is the decision-making LLM
survey of N.~Yang~\emph{et al.}~\cite{Yang_2026_LLMDecision}, which catalogs a
five-stage pipeline---prompt learning, chain-of-thought reasoning,
inference mechanisms, learning-based decision making, and multi-agent
coordination---through which large language models act as network
management agents in 6G systems. Despite its breadth, this survey
does not subsume the present work for
two reasons. First, it treats LLMs as decision-making agents operating
\emph{above} the semantic encoding layer (e.g., resource scheduling,
policy selection) rather than as compressed semantic encoders
integrated \emph{into} the physical and MAC layers; the TinyLM
deployment problem addressed by this survey is therefore out of scope.
Second, it does not address sub-1B-parameter tiny language models
specifically, nor the compression-architecture taxonomy that is the
central contribution of Section~\ref{sec:taxonomy}.
Nonetheless, several of its quantitative findings are directly relevant:
ORM-based (outcome-reward-model) scheduling achieves a 15.2\% higher
task success rate than rule-based baselines; LoRA adapters reduce LLM
inference latency by 27--38\% while tuning only 0.5--2\% of model
parameters---a compression ratio consistent with the PEFT results
of Section~\ref{sec:evidence}; the WirelessAgent framework reduces
end-to-end latency by 32\%; and active-inference-based decision agents
achieve 13.4\% energy savings with 97.1\% Quality-of-Service (QoS)
satisfaction. These
results are referenced throughout Sections~\ref{sec:challenges} and
\ref{sec:future} as corroborating evidence from the adjacent
LLM-as-agent research community.

\section{Background and Motivation}
\label{sec:background}

Having positioned this survey against prior work, this section
supplies the shared technical vocabulary used throughout the rest of
the paper: the theoretical foundations of semantic communication
(Section~\ref{sec:background_theory}), the hardware resource envelopes of
6G edge devices (Section~\ref{sec:background_hw}, Table~\ref{tab:hardware}),
TinyML's role as the bridge between the two
(Section~\ref{sec:tinyml_bridge}), and the six research questions
this survey answers (Table~\ref{tab:pico}).

\subsection{Theoretical Foundations of Semantic Communication}
\label{sec:background_theory}

This survey's theoretical grounding rests on three widely-shared
premises in the semantic communication literature~\cite{Lan2021_What}:
joint source-channel coding (JSCC) can subsume much of the semantic
communication problem rather than requiring an entirely separate
coding paradigm; classical (source-level) compression is itself a
form of learning -- the Lempel-Ziv algorithm is a grammar-learner, and
physical laws are compressed descriptions of nature -- an analogy
distinct from the semantic encoding this survey studies but one that
motivates why learned representations compress well in the first
place; and some degree of layering will persist in 6G because
modularity enables scaling to use cases not yet foreseen.

\textbf{Historical and information-theoretic origins:}
The formal study of semantic information predates 6G by seven
decades. R.~Carnap and Y.~Bar-Hillel~\cite{CarnapBarHillel1952} first
proposed a logical-probability framework that quantified meaning
through the degree of confirmation a hypothesis receives from
available evidence, rather than through symbol frequency alone.
J.~Bao~\emph{et al.}~\cite{Bao2011_Towards} connected this framework to
a model-theoretic formalization, defining the semantic entropy of a
message $s$ as $H_s(s) = -\log_2 m(s)$, where $m(s)$ is the fraction
of possible worlds satisfying $s$, weighted by their probability -- a
more informative statement corresponds to a lower $m(s)$ and hence a
higher semantic entropy. This distinction matters for TinyLM design
because it formalizes why a semantic representation $Z$ can have
strictly lower entropy than the raw source $X$ while still fully
supporting a downstream task, a relationship E.~Astaiza~\emph{et al.}~\cite{Astaiza2026_Improving}
state directly as the semantic entropy inequality $H_s(\tilde{U}) \leq
H(U)$ and the corresponding semantic rate-distortion inequality
$R_s(D) \leq R(D)$: a semantic-level encoder can achieve the same
distortion budget at a strictly lower rate than a bit-level compressor
operating on the same content, which is the formal justification for
why TinyLM-scale semantic encoders can outperform what their raw
parameter count alone would suggest. Y.~Wu~\emph{et al.}~\cite{Wu2024_Toward}
complement this with a variational-approximation training recipe that
turns this information-theoretic objective (semantic entropy,
distortion, and rate) into a trainable loss -- directly transferable
to training the compact semantic encoders reviewed in
Section~\ref{sec:compression}.

\textbf{Information bottleneck formulation:}
Y.~Shi~\emph{et al.}~\cite{Shi2023_ML6G} formalize the semantic
encoding problem as an information bottleneck (IB) optimization:
\begin{equation}
  \max_{\hat{Z}} \; I(\hat{Z};\, Y) \quad
  \text{subject to} \quad I(\hat{Z};\, X) \leq \alpha,
  \label{eq:ib}
\end{equation}
where $X$ is the source, $Y$ is the task output, $\hat{Z}$ is the
learned semantic representation, and $\alpha$ is the compression
budget on that representation. The semantic encoder maximizes
task-relevant information carried by $\hat{Z}$ while constraining
$\hat{Z}$'s own information content -- not the raw source $X$ -- below
the bottleneck threshold: it is the semantic representation that gets
compressed, not the message being reconstructed. Algorithm unrolling
applied to the IB problem yields networks with up to 12$\times$ fewer
parameters than generic black-box NNs at identical task
performance~\cite{Shi2023_ML6G}.

\textbf{Semantic noise model:}
S.~N.~Karahan and O.~Kaya~\cite{Karahan2025_6G} introduce:
\begin{equation}
  Y = f(X) + N_S,
  \label{eq:semnoise}
\end{equation}
where $N_S$ represents contextual misinterpretation---a distinct source
from physical channel noise arising from semantic encoding/decoding
errors. Even a perfect physical channel cannot guarantee semantic
fidelity; semantic noise requires architectural countermeasures
(shared knowledge bases, KG grounding) distinct from physical-layer
coding solutions. This two-term model is sufficient for the
single-link case; a practical multi-cell or multi-user deployment
additionally needs an explicit interference term, which this survey
adds as its own extension:
\begin{equation}
  Y = f(X) + N_S + \sum_{k} h_k Z_k,
  \label{eq:semnoise_interference}
\end{equation}
where $\sum_{k} h_k Z_k$ is the same semantic-interference term
formalized in full in Challenge~8 (Eq.~\ref{eq:multicell_interference},
Section~\ref{sec:challenges}) -- physical channel noise, semantic
noise, and cross-cell semantic interference are three distinct
degradation sources, and conflating any two of them under a single
noise term misrepresents where a countermeasure needs to act.
Section~\ref{sec:metrics} returns to this point when
introducing the semantic fidelity metrics used to detect it.

\textbf{Reasoning capacity beyond Shannon's limit:}
C.~Chaccour~\emph{et al.}~\cite{Chaccour_2022_LessData} formalize a
teacher/apprentice reformulation of semantic communication in which
the transmitter and receiver share a reasoning faculty that can extract
additional task-relevant information from a transmitted symbol beyond
its literal content
(see the footnote below and consult~\cite{Chaccour_2022_LessData}
directly for their actual formalization); we write an illustrative
reasoning-capacity term
\begin{equation}
  C_R = \Omega \log_2\!\left(1 + \eta_{b,d}\right),
  \label{eq:reasoning_capacity}
\end{equation}
where $\Omega$ is the available computing bandwidth (FLOPS/s) at the
receiver and $\eta_{b,d}$ is a communication symmetry index quantifying
the degree of shared background knowledge between transmitter $b$ and
receiver $d$.\footnote{This functional form is an expository adaptation
of the qualitative reasoning-capacity concept in
\cite{Chaccour_2022_LessData}; readers should consult the original
source for the authors' own formalization, which this survey does not
claim to reproduce exactly.} Crucially, $C_R$ can exceed the classical
Shannon capacity $C = \max_{p(x)} I(X;\hat{X})$ by substituting compute
for spectrum: a receiver with a richer shared knowledge base and more
inference compute can recover more task-relevant meaning from the
same channel symbols, without any additional bandwidth. This
qualitative principle -- independently corroborated in the metrics
literature (Section~\ref{sec:metrics}) rather than resting on
Eq.~\ref{eq:reasoning_capacity}'s specific illustrative form alone --
provides the theoretical grounding for the TinyLM design paradigm
pursued throughout this survey: intelligence at the encoder and decoder
substitutes for additional channel capacity. Classical Shannon capacity
$C_C$ (bits/s, governed by bandwidth and SNR) and reasoning capacity
$C_R$ (as defined in Eq.~\ref{eq:reasoning_capacity}, measured in
$\Omega$ FLOPS/s) are not numerically additive in a common unit --- a
fully rigorous unification would require converting $C_R$ into an
equivalent bits/s figure via the receiver's task-completion rate per
FLOP, which neither this survey nor~\cite{Chaccour_2022_LessData}
provides, so no combined-capacity equation is stated here. The
qualitative claim that survives without that conversion is directional
only: as $\eta_{b,d} \to 1$
(perfectly aligned transmitter-receiver
knowledge bases, as in the Symbolic Protocol Machine (SPM) of
Section~\ref{sec:compression}), $C_R$ grows without the corresponding
spectral cost that increasing $C_C$ would require---consistent with why
KG-grounded and symbolic semantic communication systems achieve
disproportionate fidelity gains relative to their bandwidth
consumption (Section~\ref{sec:metaanalysis}).

\begin{figure*}[t]
\centering
\includegraphics[width=0.92\textwidth,height=0.85\textheight,keepaspectratio]{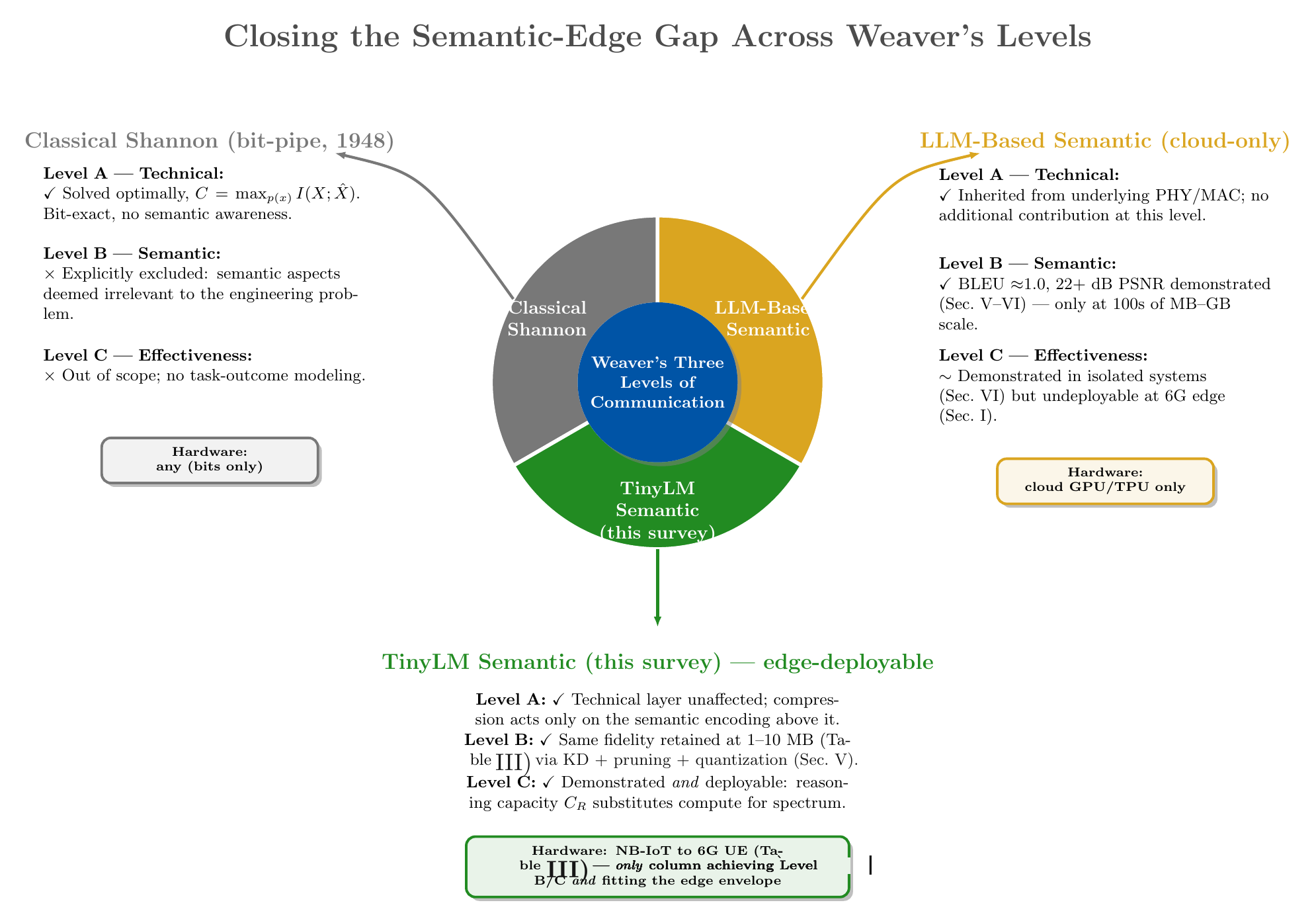}
\caption{How TinyLMs improve on classical Shannon theory and cloud-only
LLM semantic communication across Weaver's three levels of
communication~\cite{ShannonWeaver1949}. Green cells mark where a
column satisfies a level; gray marks explicit non-treatment; gold
marks partial/undeployed treatment. Only the TinyLM column
(this survey's subject) achieves Level~B/C semantic performance
\emph{and} fits the 6G edge hardware envelope of
Table~\ref{tab:hardware}.}
\label{fig:shannon_weaver}
\end{figure*}

Figure~\ref{fig:shannon_weaver} summarizes the argument of this
subsection visually: classical Shannon theory solves Level~A
optimally but explicitly excludes Levels~B and~C; LLM-based semantic
communication demonstrates Levels~B and~C but only at a model scale
undeployable on 6G edge hardware; TinyLMs are this survey's central
claim that both properties can be had together, retaining
Level~B/C semantic performance while fitting the hardware envelope
of Table~\ref{tab:hardware}.

\subsection{Hardware Constraints for 6G Edge Deployment}
\label{sec:background_hw}

Table~\ref{tab:hardware} summarizes key hardware targets and their
resource budgets, compiled from vendor hardware specifications for
the listed processors together with the TinyML deployment
literature~\cite{Abadade2023_TinyML} rather than from a benchmark this
survey ran directly. The \textbf{Max Model} column is an approximate
ceiling on \emph{storage} footprint specifically -- the size of an
INT8-quantized model's weights that the tier's flash/RAM budget can
hold, following the standard TinyML convention that weight storage,
not activation memory, is normally the binding constraint at these
tiers; it does not separately budget for peak activation memory
during inference, runtime buffers, or the surrounding application
firmware, all of which reduce the model size actually deployable in
practice below this nominal ceiling. Independent evidence
corroborates this envelope from three angles: N.~Dhar~\emph{et al.}~\cite{Dhar2024_Empirical}
show empirically that LLM inference on resource-constrained devices is
gated as much by memory bandwidth and swap behavior as by raw compute,
reinforcing why the Flash/RAM column above, not FLOPs alone, bounds
deployability; C.~M.~Bhushan~\emph{et al.}~\cite{Bhushan2025_Deploying}
report concrete MCU-tier feasibility data -- flash/SRAM budgets and
energy-per-inference figures consistent with the NB-IoT and Smart
Meter rows; and L.~Pazmi\~no~Ortiz~\emph{et al.}~\cite{Pazmino2025_Advancing}
catalog specific board-level examples at this scale (e.g., an Arduino
Nano 33 BLE Sense: Cortex-M4, 64~MHz, 256~KB SRAM), giving
Table~\ref{tab:hardware}'s NB-IoT tier a concrete reference platform.
I.~Lamaakal~\emph{et al.}~\cite{Lamaakal2025_TinyLMSurvey} give this
envelope a concrete population of already-existing models to map
against it: DistilBERT (66M parameters), TinyBERT ($\sim$14.5M),
MobileBERT (25.3M), and the SmolLM family (135M--1.7B) span almost
exactly the gateway-tier range of Table~\ref{tab:hardware}, giving
this survey's TinyLM category real, named examples rather than only a
hypothetical size target -- TinyBERT and MobileBERT in particular are
close enough to the 1--10~MB range (at INT8) to be gateway-tier
candidates today, without requiring any of the compression techniques
of Section~\ref{sec:compression} beyond standard quantization.

\begin{table*}[t]
\caption{6G Edge Hardware Tiers and Resource Budgets}
\label{tab:hardware}
\centering
\small
\renewcommand{\arraystretch}{1.3}
\begin{tabular}{|l|l|c|c|c|}
\hline
\textbf{Device Tier} & \textbf{Processor} & \textbf{Flash/RAM} &
\textbf{Power} & \textbf{Max Model} \\
\hline
NB-IoT Sensor   & Cortex-M33  & 1~MB / 256~KB   & $<$100~mW & $<$0.5~MB \\
\hline
Industrial IoT  & STM32H7\textsuperscript{*} & 32~MB / 1~MB    & $<$300~mW & $<$2~MB \\
\hline
IoT Gateway     & Jetson Orin Nano-class & 8~GB LPDDR5     & 7--15~W   & $<$500~MB \\
\hline
6G UE           & Snapdragon~8 & 12~GB LPDDR5   & 3--8~W    & $<$5~GB \\
\hline
Edge Server     & Edge TPU    & 8~GB RAM        & 10--50~W  & $<$10~GB \\
\hline
\end{tabular}
\renewcommand{\arraystretch}{1}

\vspace{4pt}
\footnotesize
\textsuperscript{*}STM32H7 on-chip flash tops out at 2~MB; the
32~MB figure includes external QSPI/serial flash commonly paired
with this part in industrial designs, not on-chip storage alone.
Named
processors are representative examples of each device class as of
this survey's writing, not a claim that this specific part remains
current through the 2030--2035 deployment horizon of
Section~\ref{sec:intro}; a reader should substitute the
contemporary part occupying the same class (Flash/RAM and power
budget) at deployment time. Flash/RAM and
Power columns are each processor's own public datasheet
specifications. Max Model is a derived INT8-weight-storage ceiling
(not independently benchmarked by this survey), computed from each
row's Flash/RAM budget under the standard TinyML assumption that
weight storage rather than activation memory binds at these tiers;
see the corroborating evidence and named board-level examples
in the paragraph above.
\end{table*}

The compression ratio
\begin{equation}
  \rho = \frac{\mathrm{size}(\text{TinyLM})}{\mathrm{size}(\text{LLM})},
  \qquad \rho \ll 1,
  \label{eq:rho}
\end{equation}
defines the core design objective. For the NB-IoT tier,
$\rho \lesssim 0.0011$ is required (BERT-Base 440~MB $\to$ target $<$0.5~MB,
$\rho \approx 0.5/440 \approx 0.0011$). For the IoT gateway tier,
$\rho < 0.05$ is sufficient against a substantially larger reference
LLM: taking a $\sim$10~GB reference model, consistent with the
multi-billion-parameter, GB-scale models this survey's cloud tier
assumes (Table~\ref{tab:hardware}), the gateway tier's $<$500~MB
budget gives $\rho \approx 500/10{,}000 = 0.05$ -- a substantially
less aggressive compression requirement than NB-IoT's roughly
three-orders-of-magnitude target, reflecting the gateway tier's much
larger memory budget rather than a different design objective.

\subsection{TinyML as the Deployment Bridge}
\label{sec:tinyml_bridge}

The Research Gap identified above --- LLM-class semantic power on one side,
milliwatt/megabyte 6G hardware on the other --- is closed in practice
by TinyML, the discipline of compressing machine learning models to
run within kilobyte-to-megabyte memory and milliwatt power budgets on
microcontroller-class hardware~\cite{Shafique2021_TinyML,Abadade2023_TinyML}.
TinyML itself did not originate in the semantic-communication
literature; it grew out of embedded machine learning for keyword
spotting, gesture recognition, and industrial sensing, where models
were already constrained to sub-megabyte footprints. TinyLM inherits
this discipline's toolkit but redirects it toward a target TinyML
itself never addressed: applying that same compression discipline to
language-model-based semantic encoders is what turns an LLM into a
TinyLM --- and it is a natural fit precisely because 6G's own hardware
tiers (Table~\ref{tab:hardware}) mirror the microcontroller and
single-board-computer tiers TinyML already targets.

Four TinyML technique families map onto the LLM-to-TinyLM compression
journey used throughout this survey. \emph{Quantization} reduces
numerical precision (FP32~$\to$~INT8~$\to$~INT4~$\to$~binary/ternary),
directly shrinking memory footprint at the cost of representational
range. \emph{Pruning} removes redundant parameters --- individual
weights, attention heads, or entire layers --- exploiting the
empirical observation that most large transformers are substantially
over-parameterized for any single downstream task. \emph{Knowledge
distillation} (KD) transfers the behavior of a large teacher model
into a much smaller student, often combined with pruning and
quantization in a single pipeline (Section~\ref{sec:compression_kd}).
\emph{Neural architecture search} (NAS) automates the discovery of
architectures that are efficient by design rather than by post-hoc
compression, though as Section~\ref{sec:taxonomy} shows, this family
remains almost entirely unapplied to semantic communication encoders.
Section~\ref{sec:compression} reviews each of these four families in
turn, followed by the hybrid pipelines that combine them; each
technique reviewed there states explicitly which 6G hardware tier it
targets and whether it is TinyML-native or a general deep-learning
compression method adapted for TinyML constraints.

\subsection{Research Questions Guiding the Evidence Synthesis}
\label{sec:rqs}
\begin{table*}[t]
\caption{Research Questions and Target Outcomes}
\label{tab:pico}
\centering
\renewcommand{\arraystretch}{1.2}
\begin{tabular}{|p{0.6cm}|p{8.2cm}|p{7.4cm}|}
\hline
\textbf{RQ} & \textbf{Question} & \textbf{Target Outcome} \\
\hline
RQ1 & What compression techniques enable sub-15M-parameter TinyLMs for
6G semantic communication on constrained edge hardware, compared
against full LLMs and cloud inference? & Model size $\leq$15M
parameters at the gateway tier (Table~\ref{tab:hardware}); memory,
latency, and energy within the budgets of the targeted tier; semantic
fidelity retained relative to the full-scale baseline \\
\hline
RQ2 & How do quantization, pruning, knowledge distillation, NAS, and
hybrid compression affect semantic transmission quality relative to
full-precision baselines? & Quantified compression ratio versus
accuracy/fidelity loss, and hardware-efficiency gain, per technique \\
\hline
RQ3 & Which end-to-end JSCC architectures achieve the best
semantic-fidelity-per-bit trade-off under realistic 6G channel
conditions (AWGN, Rayleigh, OFDM), compared against classical separate
source-channel coding? & Quantified PSNR/BLEU/SSIM and SNR-robustness
improvement over separation-based baselines \\
\hline
RQ4 & How do federated and split-learning strategies enable TinyLM
training and knowledge-base synchronization under non-IID 6G edge
data, compared against FedAvg and centralized training? & Convergence
parity with centralized training; quantified communication-overhead
reduction \\
\hline
RQ5 & What role do knowledge graphs play in reducing transmission
overhead for TinyLM-enabled semantic communication, compared against
conventional (KG-free) semantic encoders? & Quantified energy/bandwidth
reduction and interpretability gain \\
\hline
RQ6 & What adversarial threats and robustness techniques are specific
to TinyLM-based semantic communication, compared against unprotected
encoders? & Quantified robustness margin (e.g., attack success rate,
achievable secrecy capacity) under a stated threat model \\
\hline
\end{tabular}
\renewcommand{\arraystretch}{1}
\end{table*}

Table~\ref{tab:pico} states each research question alongside the
concrete, measurable outcome this survey checks it against in
Section~\ref{sec:evidence}, following the same
Population/Intervention/Comparison/Outcome logic set out informally
above but made explicit per question rather than fragmented across
columns.

\section{Two-Axis Taxonomy}
\label{sec:taxonomy}

We organize the TinyLM design space along two axes: semantic-encoder
compression methods and semantic communication architectures.
Figure~\ref{fig:taxonomy} presents the taxonomy.

\begin{figure*}[t]
\centering
\resizebox{\textwidth}{!}{%
\begin{tikzpicture}[font=\scriptsize,
  cell/.style={rectangle, minimum width=2.35cm, minimum height=1.25cm,
    align=center, draw=black!35, text width=2.15cm},
  hdr/.style={rectangle, minimum width=2.35cm, minimum height=0.6cm,
    align=center, draw=black!55, fill=ieeblue!85, text=white, font=\bfseries\scriptsize},
  rhdr/.style={rectangle, minimum width=2.9cm, minimum height=1.25cm,
    align=center, draw=black!55, fill=ieeblue!85, text=white, font=\bfseries\scriptsize, text width=2.6cm},
  bar/.style={rectangle, draw=black!40}
]
\node[hdr, minimum width=2.9cm] at (0,0)   {Architecture};
\node[hdr] at (2.95,0)  {Quantiz.};
\node[hdr] at (5.4,0)  {Pruning};
\node[hdr] at (7.85,0)  {Distill.};
\node[hdr] at (10.3,0) {LoRA};
\node[hdr] at (12.75,0) {NAS};
\node[hdr] at (15.2,0) {Hybrid};
\node[bar, fill=ieeblue!30, minimum width=1.6cm, minimum height=0.6cm] at (17.4,0) {\scriptsize Row maturity};
\node[font=\scriptsize\bfseries, red!70!black] at (12.75,0.75) {$\bigstar$ CRITICAL GAP: 0 papers};

\node[rhdr] at (0,-1.55)  {E2E JSCC};
\node[cell, fill=wellstudied!45] at (2.95,-1.55) {\textbf{Well-studied}\\\cite{Zhou2025_CCTaEncoder,Zhong2024_JSCC_FPGA,Do2026_BidDeepSC}};
\node[cell, fill=emerging!40] at (5.4,-1.55) {\textbf{Emerging}\\\cite{Lyu2023_GatedDeepJSCC}};
\node[cell, fill=wellstudied!45] at (7.85,-1.55) {\textbf{Well-studied}\\\cite{Xie2021_DeepSC,Sana2022_DeepSC,Sheng2022_multitask}};
\node[cell, fill=emerging!40] at (10.3,-1.55) {\textbf{Emerging}\\\cite{Yun2025_TOAST}};
\node[cell, fill=unexplored!35] at (12.75,-1.55) {\textbf{Unexplored}\\---};
\node[cell, fill=emerging!40] at (15.2,-1.55) {\textbf{Emerging}\\\cite{Zhou2025_CCTaEncoder}};
\node[bar, fill=ieeblue!55, minimum width=1.2cm, minimum height=0.85cm] at (17.4,-1.55) {\scriptsize High};

\node[rhdr] at (0,-3.05)  {Split\\Learning};
\node[cell, fill=emerging!40] at (2.95,-3.05) {\textbf{Emerging}\\\cite{Tummala2025_TinyML}};
\node[cell, fill=emerging!40] at (5.4,-3.05) {\textbf{Emerging}\\\cite{Eldeeb2025_SemanticMSL,Sung_2025_DeCoMeSC}};
\node[cell, fill=emerging!40] at (7.85,-3.05) {\textbf{Emerging}\\\cite{Eldeeb2025_SemanticMSL,Yan2025_MEC}};
\node[cell, fill=unexplored!35] at (10.3,-3.05) {\textbf{Unexplored}\\---};
\node[cell, fill=unexplored!35] at (12.75,-3.05) {\textbf{Unexplored}\\---};
\node[cell, fill=unexplored!35] at (15.2,-3.05) {\textbf{Unexplored}\\---};
\node[bar, fill=ieeblue!30, minimum width=1.2cm, minimum height=0.5cm] at (17.4,-3.05) {\scriptsize Low};

\node[rhdr] at (0,-4.55)  {Federated\\Learning};
\node[cell, fill=emerging!40] at (2.95,-4.55) {\textbf{Emerging}\\\cite{Radwan2025_quantization}};
\node[cell, fill=emerging!40] at (5.4,-4.55) {\textbf{Emerging}\\\cite{Zhou2025_CCTaEncoder}};
\node[cell, fill=emerging!40] at (7.85,-4.55) {\textbf{Emerging}\\\cite{Park2019_FD,Qi_2023_FedBKD}};
\node[cell, fill=emerging!40] at (10.3,-4.55) {\textbf{Emerging}\\\cite{Sun2024_FFALoRA}};
\node[cell, fill=unexplored!35] at (12.75,-4.55) {\textbf{Unexplored}\\---};
\node[cell, fill=emerging!40] at (15.2,-4.55) {\textbf{Emerging}\\\cite{Zhou2025_CCTaEncoder}};
\node[bar, fill=ieeblue!45, minimum width=1.2cm, minimum height=0.7cm] at (17.4,-4.55) {\scriptsize Medium};

\node[rhdr] at (0,-6.05)  {KG-Assisted};
\node[cell, fill=emerging!40] at (2.95,-6.05) {\textbf{Emerging}\\\cite{Pokhrel2023_UBT}};
\node[cell, fill=unexplored!35] at (5.4,-6.05) {\textbf{Unexplored}\\---};
\node[cell, fill=emerging!40] at (7.85,-6.05) {\textbf{Emerging}\\\cite{Jiang2022_KG,Seo2023_SPM}};
\node[cell, fill=unexplored!35] at (10.3,-6.05) {\textbf{Unexplored}\\---};
\node[cell, fill=unexplored!35] at (12.75,-6.05) {\textbf{Unexplored}\\---};
\node[cell, fill=unexplored!35] at (15.2,-6.05) {\textbf{Unexplored}\\---};
\node[bar, fill=ieeblue!25, minimum width=1.2cm, minimum height=0.4cm] at (17.4,-6.05) {\scriptsize Low};

\node[rhdr] at (0,-7.55)  {Multi-Task /\\Cross-Modal};
\node[cell, fill=emerging!40] at (2.95,-7.55) {\textbf{Emerging}\\\cite{Sheng2022_multitask}};
\node[cell, fill=unexplored!35] at (5.4,-7.55) {\textbf{Unexplored}\\---};
\node[cell, fill=emerging!40] at (7.85,-7.55) {\textbf{Emerging}\\\cite{Huang2025_crossmodal}};
\node[cell, fill=unexplored!35] at (10.3,-7.55) {\textbf{Unexplored}\\---};
\node[cell, fill=unexplored!35] at (12.75,-7.55) {\textbf{Unexplored}\\---};
\node[cell, fill=unexplored!35] at (15.2,-7.55) {\textbf{Unexplored}\\---};
\node[bar, fill=ieeblue!20, minimum width=1.2cm, minimum height=0.35cm] at (17.4,-7.55) {\scriptsize Low};

\node[bar, fill=ieeblue!55, minimum width=2.35cm, minimum height=0.8cm] at (2.95,-9.1) {\scriptsize col.\ maturity: High};
\node[bar, fill=ieeblue!25, minimum width=2.35cm, minimum height=0.4cm] at (5.4,-9.1) {\scriptsize col.\ maturity: Low};
\node[bar, fill=ieeblue!55, minimum width=2.35cm, minimum height=0.8cm] at (7.85,-9.1) {\scriptsize col.\ maturity: High};
\node[bar, fill=ieeblue!20, minimum width=2.35cm, minimum height=0.3cm] at (10.3,-9.1) {\scriptsize col.\ maturity: Low};
\node[bar, fill=ieeblue!10, minimum width=2.35cm, minimum height=0.15cm] at (12.75,-9.1) {\scriptsize col.\ maturity: None};
\node[bar, fill=ieeblue!20, minimum width=2.35cm, minimum height=0.3cm] at (15.2,-9.1) {\scriptsize col.\ maturity: Low};
\end{tikzpicture}}
\caption{Two-axis taxonomy of TinyLMs for 6G semantic communication.
Horizontal axis: compression method family, now including low-rank
adaptation (LoRA, Section~\ref{sec:compression_lora}) as a sixth
family alongside the five reviewed in depth. Vertical axis: semantic
communication architecture. Each cell is classified directly from
this survey's verified evidence base (Section~\ref{sec:evidence}) as
Well-studied ($\geq$3 independently verified studies populate that
cell), Emerging (1--2 studies), or Unexplored (0 studies) --
Figure~\ref{fig:taxonomy}'s cell citations show exactly which studies
drove each classification, and only two cells (E2E~JSCC$\times$Quantization,
E2E~JSCC$\times$Distillation) meet the
3-citation bar for Well-studied. The marginal per-row and per-column bars
give a qualitative High/Medium/Low/None maturity tier, assigned by the
count of Well-studied and Emerging cells along that row or column
relative to its total cell count: High for rows/columns where
Well-studied or Emerging cells dominate, None where every cell is
Unexplored, and Medium/Low in between by the same count. This is a
relative visual summary derived directly from the per-cell
classification shown, not an independently computed or separately
weighted numeric score. Thirteen
of thirty cells are unexplored; the
NAS column alone accounts for five of these thirteen (every cell in
that column), the largest
concentration under any single compression family, and here too
``unexplored'' means this survey's citation review
(Section~\ref{sec:evidence}) found no verified study for that cell,
not that no such study exists anywhere in the broader literature. The
LoRA column is itself mostly unexplored or emerging at
best, reflecting how recently LoRA has been applied to semantic
communication specifically (Section~\ref{sec:compression_lora}) rather
than a methodological gap in this survey.}
\label{fig:taxonomy}
\end{figure*}

The quantization and distillation columns share the highest
proportion of well-studied cells (each anchored by
E2E~JSCC$\times$Quantization and E2E~JSCC$\times$Distillation
respectively), reflecting BERT-family compression's early dominance in
semantic communication research, while the NAS column remains empty across
every architecture family --- the taxonomy's single largest
concentration of unstudied work. Rather than treating the remaining
gaps as an undifferentiated list, ten of the thirteen are analyzed
individually below for why each remains open. (The three LoRA-column
gaps -- Split Learning~$\times$~LoRA, KG-Assisted~$\times$~LoRA, and
Multi-Task/Cross-Modal~$\times$~LoRA -- are not separately analyzed
here.)

\begin{itemize}
  \item \textbf{E2E JSCC~$\times$~NAS} is empty because hardware-aware
    NAS toolchains (OFA, MCUNet) have no built-in channel-simulation
    loss term; applying them to JSCC encoders first requires extending
    the search objective to include over-the-air distortion, which no
    corpus study attempts.
  \item \textbf{Split Learning~$\times$~NAS} is empty because NAS
    searches whole-network architectures, not a specific cut-layer
    position; no existing formulation jointly searches architecture
    \emph{and} split point.
  \item \textbf{Split Learning~$\times$~Hybrid} is empty because
    split-computing has only been paired with single compression
    techniques in isolation (e.g., DeCo-MeSC's pruning-plus-quantization
    pair); no study stacks three or more techniques specifically in
    the split-learning setting.
  \item \textbf{Federated Learning~$\times$~NAS} is empty because
    federated NAS would need to search a shared architecture space
    across heterogeneous, resource-constrained clients under a
    communication budget --- a joint constraint federated-learning
    research and NAS research have so far treated as separate problems.
  \item \textbf{KG-Assisted~$\times$~Pruning} is empty because KG
    systems are often already near the information-theoretic
    compression floor at the symbolic level (Appendix~A4's SPM
    analysis), leaving little apparent incentive to prune further; this
    gap reflects perceived low payoff more than a technical barrier.
  \item \textbf{KG-Assisted~$\times$~NAS} is empty for the same
    structural reason as the other NAS cells, with no NAS framework
    yet adapted to search over graph-embedding or symbolic-representation
    architectures.
  \item \textbf{KG-Assisted~$\times$~Hybrid} is empty because no study
    combines the quantization, pruning, and distillation techniques of
    Section~\ref{sec:compression} into a single hybrid pipeline for
    KG-based encoders, despite hybrid pipelines being demonstrated
    effective for neural E2E JSCC (CCTaEncoder).
  \item \textbf{Multi-Task/Cross-Modal~$\times$~Pruning} is empty
    because pruning-importance criteria (e.g., semantic information
    adaptive compression (SIAC)'s Taylor-expansion
    importance score) are derived from single-task loss gradients; no importance
    formulation has been generalized to a multi-task loss surface
    where tasks may disagree on which parameters are redundant.
  \item \textbf{Multi-Task/Cross-Modal~$\times$~NAS} is empty for the
    same NAS-adaptation reason as the other NAS cells, compounded by
    the difficulty of defining one search objective across
    heterogeneous per-task metrics (BLEU for text, PSNR for images).
  \item \textbf{Multi-Task/Cross-Modal~$\times$~Hybrid} is empty
    because hybrid pipelines require a stable single-task baseline to
    compress in stages, and multi-task semantic encoders (Sheng et
    al.) are still an emerging area without an established baseline
    to anchor a pipeline.
\end{itemize}

\section{TinyLM Compression Pipeline for Semantic Encoders}
\label{sec:compression}

Each subsection below corresponds to one stage of the TinyML pipeline
introduced in Section~\ref{sec:tinyml_bridge} --- quantization,
pruning, knowledge distillation, neural architecture search, and
hybrid combinations --- applied specifically to semantic encoders.
Every technique reviewed states which 6G hardware tier (Table~\ref{tab:hardware})
it targets and whether it is a TinyML-native technique or a general
deep-learning compression method adapted for TinyML constraints.

\subsection{Quantization: From FP32 to 1-Bit}

\subsubsection{INT8 Quantization for Semantic Encoders}

A.~Radwan~\emph{et al.}~\cite{Radwan2025_quantization} evaluate 8-bit
integer (INT8), 4-bit integer (INT4), and binary (1-bit) quantization
across transformer-based natural language processing (NLP) models for
IoT-class hardware (Gateway/Industrial IoT tier; general DL
quantization adapted for TinyML): INT8 achieves 75\% memory reduction
with 0.5\% accuracy loss; INT4 achieves 87\% reduction with 3.1\% loss;
binary quantization achieves 93\% reduction with 12--18\% accuracy
loss. Degradation accelerates sharply below INT4 due to the high dynamic
range of attention score distributions---a fundamental property of
the softmax operation that makes attention heads disproportionately
sensitive to precision loss.

S.~R.~Kudavelly and M.~Bindhu~\cite{Kudavelly2024_TinyML} evaluate
Quant-by-Size -- a TinyML-native scheme that selects each layer's
quantization bit-width to hit a target model-size budget directly,
rather than quantizing every layer uniformly -- on the FashionMNIST
Custom-model configuration reported in their Results table, achieving 91.65\% size reduction with
identical 88.24\% accuracy and 4.9$\times$ inference speedup. However,
a separate row of the same table, for the FashionMNIST VGG16-model
configuration, shows that
INT8 quantization \emph{increases} inference time from 8.19~s to
56.93~s on CPU hosts lacking hardware-optimized INT8 kernels---a
critical warning that quantization benefits are hardware-dependent
and must be validated on target deployment platforms before claiming
efficiency gains. The practical implication is a deployment-toolchain
recommendation, not just a caveat: quantized TinyLM efficiency claims
should specify the inference kernel stack alongside the bit-width,
since the same INT8 weights can run 7$\times$ slower or 5$\times$
faster depending on whether the target runtime has hardware-optimized
INT8 kernels. L.~Pazmi\~no~Ortiz~\emph{et al.}~\cite{Pazmino2025_Advancing}
catalog the concrete toolchains that resolve this on the microcontroller
tiers this survey targets -- TensorFlow Lite for Microcontrollers and
CMSIS-NN (both providing hardware-optimized INT8 kernels for Cortex-M
processors, unlike the generic CPU host in Kudavelly and Bindhu's
counterexample) -- and validating against a named kernel stack, not
just a bit-width, is the concrete mitigation this survey recommends
for every quantization result reported going
forward.

Two recent results sharpen this hardware-dependency picture further.
X.~Zhong~\emph{et al.}~\cite{Zhong2024_JSCC_FPGA} provide rare
real-hardware validation: a JSCC prototype evaluated under 6-bit
fixed-point quantization on an FPGA reports better bit-error-rate
performance than compared alternatives despite the precision
reduction, supporting the general claim that semantic/JSCC encoders
can tolerate aggressive quantization gracefully rather than
catastrophically -- but only when the quantization scheme is
co-designed with the target silicon, echoing Kudavelly et al.'s
CPU-host counterexample above. T.~Suwannaphong~\emph{et al.}~\cite{Suwannaphong2025_Optimising}
sound a complementary caution from the model side: comparing quantized
transformer and Mamba architectures for indoor localization on
MCU-class hardware, they find that quantization degrades accuracy
disproportionately depending on architecture family, reinforcing that
quantization benefits are architecture-dependent as well as
hardware-dependent. At the extreme end of the bit-width spectrum,
T.~S.~Do~\emph{et al.}~\cite{Do2026_BidDeepSC} demonstrate
BidDeepSC-1.58b, a bidirectional slimmable semantic communication
system quantized to 1.58 bits per weight -- the same ternary regime
discussed for BitNet below -- with SNR-adaptive bit allocation,
directly extending the binary/ternary quantization discussion below to
a full semantic-communication system rather than a component-level
study. Looking beyond the digital baseband, F.~Rivet~\emph{et al.}~\cite{Rivet2026_Joint}
push quantization to the RF front-end itself, co-designing a 4-bit
quantized semantic RF transceiver at picojoule-per-symbol energy;
notably, their semantic symbols are derived from knowledge-graph nodes
rather than LM token embeddings, so extending this hardware co-design
to TinyLM-generated semantic symbols remains an open integration point
rather than existing evidence. L.~Guo~\emph{et al.}~\cite{Guo2023_DeviceEdge}
take a middle path between Rivet et al.'s RF-level quantization and
the encoder-level quantization reviewed above: a trained non-linear
quantizer sits specifically at the device-edge interface, learning a
non-uniform quantization codebook for the transmitted semantic
representation rather than applying a fixed uniform step size --
directly relevant to the digital OFDM-SemCom architecture of
Section~\ref{sec:arch_e2e}, which this survey's own evidence base
notes requires exactly this kind of representation-level quantization
to bridge analog JSCC outputs to digital modulation.

Tummala et al.~\cite{Tummala2025_TinyML} characterize INT8 for
semantic communication specifically (Gateway tier), finding that INT8
introduces an equivalent $<$0.5~dB SNR penalty in semantic fidelity.
This establishes a direct trade metric: compression ratio can be
exchanged against effective SNR margin. The implication for adaptive
systems is significant---at high SNR, aggressive INT4 quantization can
be applied without perceptible fidelity loss; at low SNR, quantization
level should be relaxed to avoid compounding degradation.

\subsubsection{Extreme Quantization: Binary and Ternary}

For the NB-IoT tier (Cortex-M33, $<$0.5~MB), even INT8 is
insufficient. Shafique et al.~\cite{Shafique2021_TinyML} review
BitNet and ternary quantization (TinyML-native) achieving 90--96\%
size reduction. Binary representations of transformer weights collapse
attention head expressiveness, reducing semantic accuracy by
12--18\%~\cite{Radwan2025_quantization}. This motivates hybrid
approaches: binary weight quantization with INT8 activations. The
split fine-tuning framework~\cite{Eldeeb2025_SemanticMSL} (Gateway
tier) addresses this through joint sparsification, stochastic
quantization, and lossless encoding, reducing fine-tuning delay by
66.4\% and communication overhead by 93.6\%.

The L-DeepSC system~\cite{Xie_2021_Lite} (Industrial IoT tier)
demonstrates that combining pruning (sparsity ratio $\gamma$ up to
0.9) with quantization reduces the DeepSC model from 12.3~MB to
1.28~MB while maintaining equivalent BLEU scores---a 9.6$\times$
reduction achieved with negligible runtime impact (20~ms~$\to$~18~ms).
Treating pruning and quantization as independent multiplicative
factors gives the approximate post-compression size
\begin{equation}
  M' = M \cdot (1-\gamma) \cdot \frac{b}{32},
  \label{eq:ldeepsc}
\end{equation}
where $M$ is the original FP32 model size, $\gamma$ is the pruned
weight fraction, and $b$ is the post-quantization bit-width. This is
an approximation: in practice quantization is applied only to the
surviving post-pruning weights, and their distribution --- typically
more peaked after pruning --- can change the achievable $b$ without
additional accuracy loss, so Eq.~(\ref{eq:ldeepsc}) should be read as
an upper bound on achievable size reduction rather than an exact
formula. Reproducing the reported 12.3~MB~$\to$~1.28~MB figure via
Eq.~(\ref{eq:ldeepsc}) at $b=8$ (the bit-width L-DeepSC quantizes its
surviving channels to, noted below) requires $\gamma \approx 0.58$,
not the $\gamma$ up to 0.9 ceiling cited above; the 0.9 figure is the
reported maximum sparsity ratio, not the specific pruning level used
to reach 1.28~MB. L-DeepSC further demonstrates a 40$\times$ transmission
compression ratio at equivalent semantic fidelity, achieved by pruning
redundant channel dimensions in the semantic encoder's output layer
before quantizing the surviving channels to 8 bits. The CSI-aided
training mechanism allows the quantized encoder to adapt its pruning
mask and quantization scale jointly to fading channel conditions
observed during training: rather than fixing the compression
configuration at design time, L-DeepSC's training procedure exposes
the encoder to a distribution of channel state information
realizations, producing a single deployed model that is robust across
the SNR range rather than optimized for a single operating point.

\subsection{Pruning: Structured vs.\ Unstructured}

\subsubsection{Structured Channel Pruning}

Structured channel pruning for semantic encoders follows a Taylor-expansion
importance criterion, which converges to the
Optimal Brain Damage framework under specific loss curvature
assumptions, providing theoretical grounding for the empirical
superiority of gradient-based pruning over magnitude-based approaches
in semantic encoders. Two recent studies apply this structured
pruning principle to new deployment settings: C.~Ren~\emph{et al.}~\cite{Ren2024_Multimodal}
apply an analogous pruning-plus-quantization pipeline to a
multimodal semantic encoder for TinyML-based UAV task execution,
demonstrating that the same importance-driven channel selection
generalizes beyond image classification to a resource-constrained
aerial platform; and Y.~Fu~\emph{et al.}~\cite{Fu2024_Scalable}
propose an adaptive feature-masking scheme that selectively prunes
semantic features based on a learned mAP-versus-transmitted-symbols
trade-off curve, giving a scalable, task-aware alternative to SIAC's
fixed 40\% pruning ratio. Z.~Lyu~\emph{et al.}~\cite{Lyu2023_GatedDeepJSCC}
take a complementary gated approach within the E2E JSCC architecture
family itself: rather than pruning the encoder's weights offline, a
learned gate adaptively prunes \emph{output features} at inference
time according to current channel conditions, reducing communication
overhead with only marginal performance loss -- one of the few
independently verified pruning-based results within the E2E JSCC
family specifically, complementing SIAC's Gateway-tier, offline
pruning approach with an online, channel-aware alternative.

\subsubsection{FedPun: Adaptive Pruning in Federated Learning}

C.~Zhou~\emph{et al.}~\cite{Zhou2025_CCTaEncoder} introduce FedPun (Gateway
tier), assigning each client a pruning ratio $r_c$ inversely
proportional to training loss $\ell_c$:
\begin{equation}
  r_c = r_{\max} \cdot \bigl(1 - \ell_c / \ell_{\max}\bigr),
  \label{eq:fedpun}
\end{equation}
where $r_{\max}$ is the maximum allowed pruning ratio and $\ell_{\max}$
is the maximum observed loss across clients. Clients with high loss
(poor convergence from strongly non-IID data) retain more parameters,
preserving capacity for local distribution learning. Clients with low
loss transmit sparse updates, reducing communication overhead when
gradients are most reliable. Under Dirichlet $\alpha = 0.3$
(severe non-IID), FedPun achieves 20~dB PSNR---within 2~dB of the
22~dB centralized upper bound and $>$3~dB above FedAvg (Federated
Averaging, the standard baseline that aggregates client weights by
simple averaging with no loss-aware weighting).

\subsubsection{The Symbolic Protocol Machine (SPM): Extreme Structural Pruning}

S.~Seo~\emph{et al.}~\cite{Seo2023_SPM} transform the entire neural MAC
protocol (4.55~MB) into a probabilistic logic graph -- the Symbolic
Protocol Machine (SPM) -- using:
(1)~activation-aware vocabulary merging, identifying neurons whose
activation patterns are semantically equivalent; and
(2)~connection-aware merging, removing synaptic connections whose
removal does not change any output under any input in the training
distribution. The SPM achieves identical goodput to the full neural
model using only 1/1,750 the FLOPs (8 vs.\ 14,000), compressing
4.55~MB to 1~KB---99.98\% compression with zero task-performance
loss. This targets the NB-IoT tier and is TinyML-native at the
symbolic level: it demonstrates that a neural protocol trained as a
black box can implicitly learn symbolic structure that is extractable
and executable as probabilistic logic.

\subsection{Knowledge Distillation}
\label{sec:compression_kd}

Section~\ref{sec:tinyml_bridge} introduced knowledge distillation (KD)
as transferring a large teacher model's behavior into a smaller
student; the systems below apply that principle specifically to
semantic encoders, in federated, contrastive, and cross-architecture
forms.

\subsubsection{Federated Distillation}

One general pattern for federated knowledge sharing without exchanging
raw client data is to let clients agree on what a concept
\emph{means} rather than exchanging model weights directly: each
client compares its own semantic embedding of an input against a
shared reference computed across all clients, and is penalized for
drifting too far from that consensus. Park et al.'s FD~\cite{Park2019_FD}
instantiates a version of this pattern at the logit level, achieving
25.6$\times$ communication cost reduction over FedAvg by exchanging only output logits on a shared
unlabeled public dataset. FD achieves 92--97\% of FedAvg accuracy at
dramatically reduced uplink cost. For semantic TinyLMs with vocabulary
sizes of 30K--50K tokens, \emph{top-$k$ logit compression}---
transmitting only the $k$ highest-probability logits and their
indices, where $k \approx 50$--100 captures 99\% of output probability
mass---reduces per-round payload to $\sim$600~bytes, well within
NB-IoT budgets. X.~Xu~\emph{et al.}~\cite{Xu2024_FedSFD} propose a related
federated semantic feature distillation architecture (FedSFD)
specifically for image semantic communication in IoT edge learning,
combining federated learning with feature-level knowledge transfer to
improve global model performance across heterogeneous client
devices---the drone-image-acquisition use case they target is a
concrete instance of exactly the Gateway-tier deployment problem this
survey's taxonomy addresses.

\subsubsection{Contrastive Disentanglement}

C.~Chaccour and W.~Saad~\cite{Chaccour2022_contrastive} propose a
contrastive pre-processing step separating ``learnable'' data
(semantic-rich, structured) from ``memorizable'' data (random,
non-compressible) before semantic encoding. The framework achieves
57.22\% reduction in average representation length with 71.9\%
higher semantic impact. By routing memorizable tokens around the
TinyLM encoder, effective computational load is reduced by up to
57\% without any model compression.

\subsubsection{Cross-Architecture Distillation for 6G UE Encoders}
\label{sec:compression_kd_crossarch}

Standard knowledge distillation assumes that teacher and student
share the same architecture family. In 6G deployments this assumption
fails: the gNB semantic reasoning node runs a full transformer
(e.g., BERT-Base, 110M parameters) while the UE semantic encoder must
be a lightweight CNN to meet the $<$5~W power budget of
Table~\ref{tab:hardware}. Liu et al.~\cite{Liu_2022_Cross-Architecture}
directly address this mismatch with a cross-architecture KD framework
introducing two specialized projectors: (1)~a partially cross-attention
(PCA) projector that maps CNN student features into the transformer
attention space, enabling the student to learn global semantic
relations it could not acquire from its local convolutional receptive
field; and (2)~a group-wise linear (GL) projector that aligns student
features with transformer features at fine spatial granularity for
detailed semantic fidelity. Formally, given teacher feature map
$F_T \in \mathbb{R}^{N \times d_T}$ and student feature map
$F_S \in \mathbb{R}^{N \times d_S}$ (with $N$ spatial positions),
the PCA projector $\mathcal{P}_{\mathrm{PCA}}$ computes a
cross-attention operation between student queries and teacher
keys/values:
\begin{equation}
  \mathcal{P}_{\mathrm{PCA}}(F_S, F_T) = \mathrm{softmax}\!
    \left(\frac{(F_S W_Q)(F_T W_K)^\top}{\sqrt{d_k}}\right)(F_T W_V),
  \label{eq:pca_projector}
\end{equation}
where $W_Q, W_K, W_V$ are learned projection matrices, mapping the
student's local convolutional features into the transformer's global
attention space: the softmax term computes, for each spatial position
in the student feature map, a weighted combination of teacher value
vectors $F_T W_V$, with weights determined by how closely that
position's (projected) student features match each teacher position's
(projected) features -- the same attention mechanism used throughout
this survey, applied here to align student and teacher representations
rather than to encode a source signal. The GL projector instead
partitions the spatial dimension into $G$ groups and applies an
independent linear transformation per group:
\begin{equation}
  \mathcal{P}_{\mathrm{GL}}(F_S)_g = W_g F_S^{(g)} + b_g,
  \qquad g = 1, \ldots, G,
  \label{eq:gl_projector}
\end{equation}
where $F_S^{(g)}$ is the slice of student features belonging to
spatial group $g$, and $W_g$ and $b_g$ are that group's own learned
weight matrix and bias vector, so each spatial region of the student
map is aligned to the teacher independently rather than through one
global transformation. The total cross-architecture
distillation loss combines both projector outputs against the
frozen teacher features:
\begin{equation}
\begin{split}
  \mathcal{L}_{\mathrm{CAKD}} = \mathcal{L}_{\mathrm{task}}
    + \lambda_{\mathrm{PCA}} \bigl\|\mathcal{P}_{\mathrm{PCA}}(F_S,F_T) - F_T\bigr\|_2^2 \\
    + \lambda_{\mathrm{GL}} \bigl\|\mathcal{P}_{\mathrm{GL}}(F_S) - F_T\bigr\|_2^2,
\end{split}
  \label{eq:cakd_loss}
\end{equation}
where $\lambda_{\mathrm{PCA}}$ and $\lambda_{\mathrm{GL}}$ balance
the global-relational and fine-grained-spatial alignment terms
against the task loss $\mathcal{L}_{\mathrm{task}}$. A multi-view
adversarial training scheme further improves robustness by 0.2--0.4\%
Top-1 accuracy on ImageNet under standard evaluation and over 1.0\%
under noisy evaluation---a property directly relevant to 6G semantic
encoders operating under varying channel-induced input corruption.
With a ViT-L/16 teacher and ResNet50 student, the framework achieves
88.09\% accuracy on CIFAR-100, outperforming 14 prior cross-architecture
and homologous-architecture KD baselines; on ImageNet, a
ViT-B/16-to-ResNet50x2 configuration achieves 80.72\% Top-1 accuracy.
This targets the Gateway and 6G UE tiers and is a general deep-learning
compression technique adapted here for TinyML deployment.

The practical relevance for 6G semantic communication is direct in a
specific sense: because the PCA/GL projectors only need to be trained
once per student architecture, not once per device, a cross-architecture
KD pipeline allows network operators to train a
single high-quality transformer semantic encoder at the gNB and
distill specialized CNN student encoders for each UE hardware tier,
without retraining the teacher per device. Cross-architecture KD thus
becomes a deployment strategy, not merely a compression technique---
the teacher evolves continuously at the gNB as new training data
arrives, while heterogeneous student encoders are periodically
refreshed via over-the-air distillation updates, each tailored to its
device's architectural constraints. The authors acknowledge that the
method is primarily validated on classification tasks; generalization
to detection and segmentation is reported as good but the core
design remains classification-centric, leaving open the question of
optimal projector design for regression-style semantic tasks (e.g.,
the STSB textual similarity task of Section~\ref{sec:architectures}).

Complementary evidence for compact transformer deployment at the
NLP semantic-task level comes from Hussain et al.'s adaptive
multitask framework~\cite{Hussain_2025_Adaptive}, which demonstrates
that MobileBERT and DistilBERT---both sub-100M-parameter
models---achieve competitive emotion recognition and sentiment
analysis performance on edge hardware with 487.477~MB VRAM and
0.00103~s/sample inference latency, enabling over 2$\times$ throughput
improvement compared to full-size BERT. DistilBERT achieves 81\%
emotion-F1 and 77\% sentiment-F1 on the MELD benchmark. The
prototypical-network component contributes a 15--18\% F1 gain over a
no-prototype baseline, establishing that few-shot semantic
representation learning---analogous in spirit to Semantic-MSL's
5-shot adaptation (Section~\ref{sec:architectures})---is achievable
at TinyLM scale without the full MAML apparatus, using instead an
adaptive focal weighted loss to address class imbalance in the
emotion-label distribution.

The broader LLM-compression literature defines how far this frontier
can still move: V.~Egiazarian~\emph{et al.}'s AQLM~\cite{Egiazarian2024_AQLM}
compresses LLMs to 2--3 bits per parameter while outperforming prior
extreme-quantization methods, suggesting the compression ratios
reviewed above are not yet at a hard floor; C.-Y.~Hsieh~\emph{et al.}'s
``distilling step-by-step''~\cite{Ho2022_DistillStep} trains a smaller
student to outperform its LLM teacher using LLM-generated rationales
as supervision rather than raw labels alone, a distillation signal
distinct from every KD method reviewed above; and P.~Zhang~\emph{et al.}'s
TinyLlama~\cite{TinyLlama2024}, a 1.1B-parameter open-source model
trained on one trillion tokens, demonstrates empirically that small
models trained on enough data remain competitive with substantially
larger ones -- direct evidence that a TinyLM's ceiling is set as much
by training data volume as by parameter count.

\subsection{Low-Rank Adaptation}
\label{sec:compression_lora}

Where quantization, pruning, and knowledge distillation each shrink a
model's own weights, low-rank adaptation (LoRA)~\cite{Hu2022_LoRA}
instead freezes the full pre-trained weights and learns a much
smaller update on top: for a target weight matrix, LoRA parameterizes
the fine-tuning update as
\begin{equation}
  \Delta W = AB, \qquad A \in \mathbb{R}^{d\times r},\;
  B \in \mathbb{R}^{r\times k}, \qquad r \ll \min(d,k),
  \label{eq:lora}
\end{equation}
so only the rank-$r$ adapter matrices $A$ and $B$ -- with
$|AB| = r(d+k)$ parameters, orders of magnitude fewer than the
$d \times k$ weight matrix they perturb -- are trained and stored,
while the base model stays frozen and shared. This makes LoRA a
different \emph{kind} of compression from the others in
Section~\ref{sec:compression}: it does not shrink the deployed
model's inference cost at all (the frozen base weights still have to
run), but it shrinks what has to be transmitted or stored
\emph{per adaptation}, which is exactly the bottleneck in two
scenarios directly relevant to TinyLM-enabled semantic communication.

\textbf{Knowledge-base synchronization.} When the shared semantic
knowledge base of Section~\ref{sec:arch_kg} drifts as the environment
changes, only the LoRA adapters -- a fraction of a percent of total
parameters -- need to be retrained and retransmitted to
resynchronize transmitter and receiver, rather than exchanging full
model weights. Y.~Sun~\emph{et al.}'s FFA-LoRA~\cite{Sun2024_FFALoRA}
sharpens this further for the federated setting specifically: freezing
the randomly-initialized $A$ matrix after initialization and training
only $B$ halves the number of trainable parameters exchanged per
federated round, directly reducing the per-round payload for the
FedBKD-style knowledge-base synchronization discussed in
Section~\ref{sec:arch_fl}.

\textbf{Multi-task adaptation.} A single pre-trained semantic encoder
can be adapted to multiple downstream tasks by swapping LoRA
adapters, avoiding the need to store a separate full-size model per
task. S.~Yun~\emph{et al.}'s TOAST~\cite{Yun2025_TOAST} demonstrates
this concretely for semantic communication: module-specific LoRA
adapters throughout a Swin-Transformer-based JSCC backbone, combined
with a reinforcement-learning task-balancing controller and a
generative diffusion-based refinement stage, achieve improved
classification accuracy and reconstruction quality at low SNR
alongside sharply reduced adaptation overhead relative to full-model
fine-tuning, under AWGN, fading, phase noise, and impulse
interference.

T.-H.~Vu~\emph{et al.}~\cite{Vu2026_Integration} -- already introduced
in Section~\ref{sec:related} as the closest existing survey to this
one -- discuss LoRA specifically as one of the TinyML-LargeML
integration mechanisms a 6G edge device needs: parameter-efficient
fine-tuning lets a gateway-tier TinyLM inherit updates from a
cloud-tier LLM without re-transmitting the LLM itself, which is the
same knowledge-distillation-without-full-retraining pattern this
survey's cross-architecture KD (Section~\ref{sec:compression_kd})
achieves by a different mechanism -- LoRA adapts an already-deployed
model's behavior, while cross-architecture KD trains a differently-shaped
model from scratch against a teacher's outputs. Neither this survey's
verified evidence base nor Vu et al.'s own survey identifies a
published study combining LoRA specifically with the contrastive
distillation objective of Section~\ref{sec:compression_kd}, so that
narrower combination remains an open direction rather than an
established result.

\subsection{Neural Architecture Search for Semantic Encoders}

Shafique et al.~\cite{Shafique2021_TinyML} identify OFA and MCUNet
as leading NAS approaches (TinyML-native). OFA trains a single
super-network supporting subnets of varying depth, width, and
sequence resolution. MCUNet co-designs neural architecture and
inference engine, enabling ImageNet-scale classification on
Cortex-M33. Application of OFA specifically to transformer-based
\emph{semantic communication} encoders remains entirely unexplored in
this survey's verified evidence base (Fig.~\ref{fig:taxonomy});
combining a KD loss with a complexity penalty inside a differentiable
NAS search is a documented pattern in the broader (non-semantic-communication)
NAS literature, and represents a plausible tentative first step toward
filling this taxonomy cell, but we were not able to independently
verify a peer-reviewed study applying this specific combination to a
semantic encoder, so the NAS column of Fig.~\ref{fig:taxonomy} is left
unexplored (not relabeled as emerging) until such a study is
confirmed.

\subsection{Hybrid Compression Pipelines}

The CCTaEncoder~\cite{Zhou2025_CCTaEncoder} combines CNN-Transformer
hybrid architecture with FedPun adaptive pruning (Gateway tier),
achieving 73.45K parameters, 5.47M FLOPs, and 22~dB PSNR at 33.33\%
compression ratio---6~dB above pure-CNN baselines. The Semantic
Efficiency Score:
\begin{equation}
  \mathrm{SES} = \frac{F_{\mathrm{norm}}}{\log_{10}(\mathrm{size}/1\,\text{KB}) + \zeta},
  \qquad \zeta = 0.05,
  \label{eq:ses}
\end{equation}
where $\zeta$ is a chosen (not empirically derived) small floor
constant, not a fitted parameter: its only role is preventing division
by zero as size $\to$ 1~KB, where $\log_{10}(\mathrm{size}/1\,\text{KB})
\to 0$. The specific value 0.05 is a design choice rather than a
calibrated one, and SES near the 1~KB boundary is sensitive to it: a
smaller $\zeta$ pushes SES for symbolic, near-1~KB systems (SPM)
sharply higher, while a larger $\zeta$ compresses it toward
$F_{\mathrm{norm}}/\zeta$ regardless of exact size -- which is
precisely why SPM's SES is reported qualitatively as ``off the
neural-model scale'' below rather than computed literally: any
specific number at that boundary would misleadingly suggest a
precision the floor constant does not actually provide. For systems
with size $\gg$ 1~KB (every neural system evaluated in this survey),
$\zeta$'s contribution to the denominator is negligible and this
sensitivity does not apply.

$F_{\mathrm{norm}} \in [0,1]$ is computed by the
same rule for every system regardless of metric type: whatever
quality metric that system reports (BLEU, PSNR, task accuracy, or
ROUGE) is divided by the same metric's value for that system's own
full-precision/centralized baseline, so $F_{\mathrm{norm}}=1$ means
``matches its own uncompressed reference,'' not an absolute
cross-metric quality level. This is a deliberately minimal
normalization, not a validated cross-modality quality equivalence: it
makes two systems' \emph{size-versus-retained-quality} trade-offs
comparable only in the specific sense that both are expressed
relative to their own baseline, and a sensitivity limitation follows
directly -- a system evaluated against a weak baseline (one that
itself performs poorly on its own metric) will report an inflated
$F_{\mathrm{norm}}$ relative to a system evaluated against a strong
baseline, even at identical absolute quality, and this survey's
evidence base does not contain enough systems reporting both a
quality figure and a baseline-strength characterization to correct
for this. \textbf{Worked example (CCTaEncoder):} the size input to
Eq.~\ref{eq:ses} must be in KB, but CCTaEncoder's reported size is
73.45K \emph{parameters}, not bytes; parameters and bytes are not the
same unit, so an explicit conversion is required. At FP32 (4 bytes
per parameter, the precision CCTaEncoder is reported at), $73{,}450
\times 4~\text{bytes} = 293{,}800$ bytes $= 293.8$~KB, giving
$\log_{10}(293.8) \approx 2.468$. With $F_{\mathrm{norm}} \approx
0.917$ (22~dB / 24~dB, its PSNR relative to its own uncompressed
baseline) and
$\zeta=0.05$: $\mathrm{SES} = 0.917 / (2.468 + 0.05) \approx 0.364$,
the figure reported throughout this
survey -- the highest among neural E2E JSCC systems. Every
other SES value reported in this survey applies this same
parameters-to-bytes
conversion at each system's own reported precision (FP32, INT8, or
binary) before taking the log; systems already reported directly in
bytes (e.g., compressed file sizes) skip this step. For the symbolic
SPM system (size $=1$~KB exactly), SES is reported qualitatively as
``off the neural-model scale'' rather than as a literal infinity,
since SPM operates at the symbolic rather than floating-point-weight
level and is not directly commensurable with the neural systems on
this scale. The Semantic-MSL framework~\cite{Eldeeb2025_SemanticMSL}
combining split learning, MAML meta-learning, and conformal prediction
achieves 95\% accuracy with 5-shot learning and 20~SGD convergence
steps versus~100 for the baseline.

\subsubsection{Attention-Based Relational and Graph Distillation}

A complementary hybrid direction in the extended corpus (Appendix~B)
combines knowledge distillation with structured relational information
rather than architectural hybridization. Gou et al.'s
MASCKD~\cite{Gou_2023_MASCKD} introduces attention-based relational
knowledge distillation: rather than matching raw teacher and student
feature maps directly (as in standard response-level or
feature-level KD, Eq.~\ref{eq:kd}), MASCKD builds a sample-correlation
matrix from multilevel attention maps,
\begin{equation}
  R^{(l)}_{ij} = \mathrm{sim}\bigl(A^{(l)}_i, A^{(l)}_j\bigr),
  \label{eq:masckd_relation}
\end{equation}
where $A^{(l)}_i$ is the attention map for sample $i$ at layer $l$
and $\mathrm{sim}(\cdot,\cdot)$ is a similarity function (typically
cosine similarity). The relational distillation loss then matches
the teacher's and student's sample-correlation matrices:
\begin{equation}
  \mathcal{L}_{\mathrm{MASCKD}} = \mathcal{L}_{\mathrm{task}} +
    \lambda_R \sum_{l} \bigl\| R^{(l)}_{T} - R^{(l)}_{S} \bigr\|_F^2,
  \label{eq:masckd_loss}
\end{equation}
where $\|\cdot\|_F$ is the Frobenius norm and the sum runs over the
$l$ attention layers selected for distillation. Gou et al.\
demonstrate that this attention-based relational signal generalizes
as a drop-in component: substituting it into existing KD methods
(CCKD, DML, Tf-KD) consistently enhances their performance, because
Eq.~(\ref{eq:masckd_relation}) captures inter-sample structure that
response-level and feature-level matching alone do not preserve.

The Knowledge Distillation (KD) loss function referenced above is:
\begin{equation}
  \mathcal{L}_{\mathrm{KD}} =
    \lambda\, \mathcal{L}_{\mathrm{CE}} +
    (1-\lambda)\, \mathcal{L}_{\mathrm{KL}}\!\left(
      q_{\mathrm{student}} \,\|\, q_{\mathrm{teacher}}
    \right),
  \label{eq:kd}
\end{equation}
balancing cross-entropy on task labels and KL divergence between
student and teacher output distributions.
DistilBERT~\cite{Sanh2019_DistilBERT} provides the seminal
demonstration that response-level distillation alone, applied to a
40\% smaller BERT student during pre-training, retains 97\% of the
teacher's language understanding capability while being 60\% faster
at inference---establishing the empirical baseline against which
subsequent TinyLM distillation pipelines are implicitly compared.

\section{Semantic Communication Architectures with T-LMs}
\label{sec:architectures}

\begin{figure*}[t]
\centering
\includegraphics[width=0.92\textwidth,height=0.85\textheight,keepaspectratio]{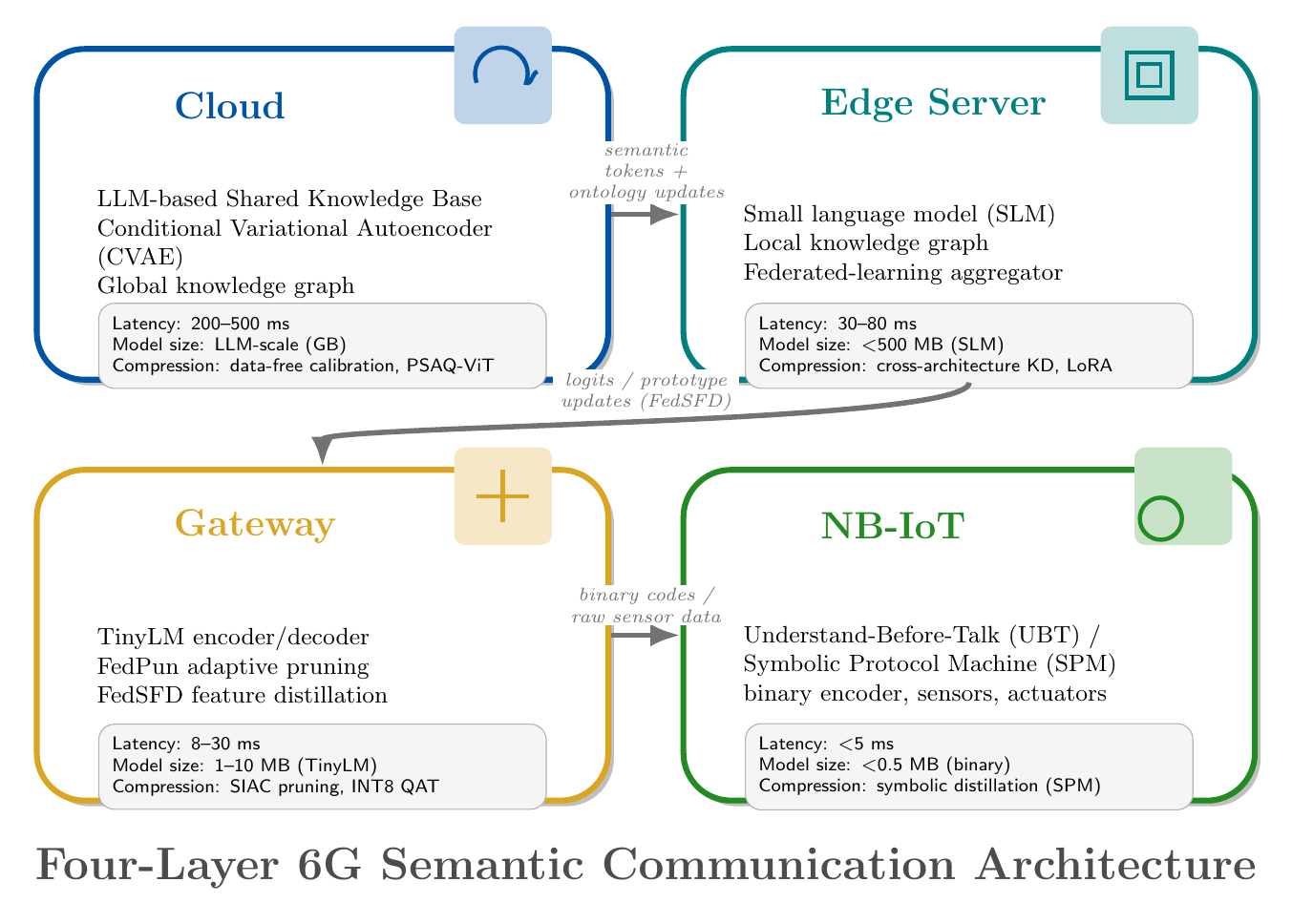}
\caption{Graphical system model: a four-layer 6G semantic communication
architecture, spanning the non-edge Cloud tier (outside
Table~\ref{tab:hardware}'s edge-hardware taxonomy) down to the NB-IoT
edge tier. Left annotations give per-layer latency budget and a
representative TinyLM model size from this survey's evidence base
(e.g., CCTaEncoder at the Gateway tier); these are best-demonstrated
results, not the hardware ceiling budgets of Table~\ref{tab:hardware},
which caps the Gateway tier at $<$500~MB. Right annotations give the
semantic-encoder compression technique applied at that layer (i.e.,
what each layer's TinyLM compresses relative to the cloud-tier LLM);
center arrows label the interface payload exchanged between layers.}
\label{fig:system_model}
\end{figure*}

Figure~\ref{fig:system_model} situates the five architecture families
reviewed in this section within a single layered 6G system model,
before each subsection addresses one architecture family in turn.

\subsection{End-to-End Joint Source-Channel Coding}
\label{sec:arch_e2e}

\subsubsection{JSCC Fundamentals}

Conventional digital communication follows a two-stage design: a
source encoder compresses the message to remove statistical
redundancy, and an independent channel encoder adds structured
redundancy back to protect against channel errors, with the two
stages designed and optimized separately. This separation is
convenient -- it lets source coding (e.g., JPEG, MP3) and channel
coding (e.g., LDPC, Turbo codes) be developed and standardized
independently -- and Shannon's separation theorem guarantees it loses
nothing \emph{asymptotically}, as blocklength grows without bound.
Joint source-channel coding (JSCC) instead designs a single encoder
that maps source data directly to channel symbols, optimizing source
compression and error protection together rather than in sequence.
The two approaches differ in what failure looks like under a
degraded channel: a separated system that loses a few source-coded
bits to channel noise can suffer disproportionate, sometimes
catastrophic reconstruction error (the ``cliff effect''), whereas a
jointly-optimized JSCC encoder degrades its output gracefully because
it never assumed error-free delivery of a compressed bitstream in the
first place. N.~Farsad~\emph{et al.}~\cite{Farsad2018_DeepJSCC} provide
the foundational proof-of-concept that a neural network can learn this
joint mapping end-to-end for text, training an encoder-decoder pair
directly against reconstruction accuracy over a simulated channel
rather than against a bit-error-rate proxy -- the same principle
DeepSC below scales up with a Transformer architecture.

\subsubsection{Why Separation Can Fail at 6G Scale}

Shannon's separation theorem's asymptotic-optimality guarantee
requires infinite blocklengths, and its classical proof further
assumes memoryless, stationary channel statistics -- conditions 6G's
operating regime departs from on more than one axis. URLLC packets
are capped below 100 symbols, far short of the asymptotic regime;
sub-THz and THz channels exhibit non-Gaussian, non-stationary
statistics; and semantic content is inherently variable-rate rather
than the fixed-rate sources the classical theorem assumes. When the
blocklength $n$ is finite, the rate penalty of separate source-channel
coding relative to JSCC grows as $O((\log n)/n)$~\cite{Shi2023_ML6G}---negligible
at $n=10^6$ but severe at $n=100$, precisely the URLLC regime above.
Neural JSCC systems exploit joint statistics of semantic content and
channel response: the encoder embeds the semantic signal in a
channel-matched subspace, ensuring the features most critical to task
performance are protected by the deepest channel coding
layers~\cite{Shi2023_ML6G}.

\subsubsection{Where TinyLMs Fit}

Everything above motivates JSCC in general; it says nothing yet about
model size. The gap this survey closes is that the JSCC encoders
demonstrated in the literature are themselves large language
models -- and M.~Chen~\emph{et al.}~\cite{Chen2025_LLM}
quantify exactly how large that gap is: an LLM-based semantic encoder
invoked via API completes inference in 3.85~s, versus 3.13~ms for a
lightweight local JSCC encoder on the same task -- a three-order-of-magnitude
latency gap that a TinyLM is specifically designed to close by
retaining the JSCC encoder's joint source-channel optimization while
shrinking the model itself onto the hardware tiers of
Table~\ref{tab:hardware}. The systems reviewed below --
DeepSC~\cite{Xie2021_DeepSC} and CCTaEncoder~\cite{Zhou2025_CCTaEncoder} -- are
evaluated below specifically on how much of that gap each closes, and
at what compression technique's cost.

\subsubsection{DeepSC: The Foundational System}

H.~Xie, Z.~Qin, G.~Y.~Li, and B.-H.~Juang~\cite{Xie2021_DeepSC} introduce DeepSC, whose
central idea is to train the semantic encoder and decoder jointly
against a loss that rewards preserving task-relevant meaning rather
than bits:
\begin{equation}
\begin{split}
  \mathcal{L}_{\mathrm{DeepSC}} =
    - \lambda_1 I(\hat{Z};\, Y)
    + \lambda_2 I(\hat{Z};\, X) \\
    + \lambda_3 D_{\mathrm{KL}}\!\bigl(p(\hat{Z}|X) \,\|\, p(X)\bigr),
\end{split}
  \label{eq:deepsc}
\end{equation}
where the first term rewards the semantic representation $\hat{Z}$ for
retaining task-relevant information about the output $Y$ -- this is
semantic encoding, not source compression in the classical sense --
while the second and third terms regularize $\hat{Z}$'s own
information content and its similarity to a semantic prior,
respectively, keeping the representation compact without discarding
what the task needs. In brief: DeepSC achieves BLEU~1.0 at
SNR$\geq$5~dB and BLEU$>$0.7 at SNR$=$$-$5~dB, substantially
outperforming Huffman+64QAM baselines (BLEU$<$0.3 at the same SNR),
and a cross-language experiment shows that training to transmit
French text and recover English meaning discovers a
language-independent semantic space -- a capability unavailable to
classical separate coding. M.~Sana and E.~C.~Strinati~\cite{Sana2022_DeepSC}
are a separate, complementary reference: their contribution is a
broader positioning argument for learning-based semantics as an
enabler of effective 6G communications generally, not the DeepSC
system itself, which is why Section~\ref{sec:intro} cites both works
together when motivating the shift toward learned semantic
representations.

\subsubsection{CNN-Transformer Hierarchy (CCTaEncoder)}

Where DeepSC establishes that JSCC can encode meaning, CCTaEncoder's
core idea is that the encoder does not need a full transformer to do
it: shallow convolutional layers handle local, translation-invariant
features cheaply, and only a thin transformer stage on top handles
the longer-range dependencies that convolutions alone would
miss~\cite{Zhou2025_CCTaEncoder}. This local-to-global division of
labor is what lets CCTaEncoder reach 73.45K parameters and 5.47M
FLOPs while still achieving 22~dB PSNR at a 33.33\% compression
ratio---6~dB above a pure-CNN baseline (16~dB) at the same
compression level, and within 2~dB of centralized full-scale training
(24~dB). The comparison against a pure-CNN baseline at matched
compression is the relevant one here: it isolates the transformer
stage's specific contribution from the compression ratio itself, and
the resulting SES $\approx 0.364$ (Eq.~\ref{eq:ses}) is the highest
among all neural E2E JSCC systems in the corpus.

\subsubsection{Generalized Source-to-Channel Design}

Z.~Qin~\emph{et al.}~\cite{Qin2023_Generalized} motivate the broader
design principle CCTaEncoder and DeepSC both instantiate: a
generalized semantic communication system should adapt at both ends
of the pipeline simultaneously -- lightweight, task-oriented semantic
extraction at the source, and channel-aware coding that responds to
the environment rather than assuming a fixed channel model -- rather
than optimizing the semantic encoder and the channel code
independently and hoping the combination works well together. This
end-to-end, source-and-channel-jointly-adaptive framing is the same
principle Section~\ref{sec:arch_e2e}'s JSCC Fundamentals discussion
opened with, restated here as an explicit system-design goal rather
than a specific architecture.

\subsubsection{OFDM-Based Digital Semantic Communication}

C.~Liu~\emph{et al.}~\cite{Zhang2023_OFDM} address the practical challenge
that existing neural JSCC systems transmit \emph{analog} real-valued
symbols -- continuous-valued channel inputs with no fixed constellation,
as produced directly by a neural network's output layer -- which are
incompatible with standard \emph{digital} modulation schemes (e.g.,
QAM), where a transmitter must map data to one of a finite set of
discrete constellation points. The framework applies
scalar quantization to JSCC outputs before OFDM modulation, using a
Dynamic Proximal Policy Optimization (DPPO) algorithm for bit allocation,
named for its dynamically-sized action space that adjusts available
bit-allocation choices at each step based on the budget already
spent.
The DPPO reward explicitly evaluates the semantic consequence of each
bit allocation, penalizing reductions on high-importance sub-carriers
more severely than equivalent reductions on low-importance sub-carriers.
Classification accuracy improves by 9.7\% versus analog SemCom and
28.7\% versus conventional SSCC at 5~dB SNR. At $-$4~dB SNR, semantic
distortion is reduced by 35--66\% versus three baselines.

\subsubsection{Hybrid and Adaptive Channel-Aware JSCC}

H.~Xie~\emph{et al.}~\cite{Xie2025_Hybrid} address a limitation shared
by both the analog JSCC of Section~\ref{sec:arch_e2e} and the digital
OFDM-SemCom above: purely analog semantic transmission suffers a
cliff-edge effect at low SNR, while purely digital transmission
forfeits JSCC's graceful degradation. Their hybrid digital-analog
scheme splits the semantic representation itself: an adaptive
allocation function $[z_A, z_D] = \mathcal{A}(z; \theta_t)$ routes
part of the semantic encoding $z = \mathcal{S}(I;\alpha_t)$ down a
digital path (quantized, entropy-coded, modulated) and the remainder
down an analog path (transmitted as continuous-valued symbols), with
the split ratio $\theta_t$ itself adaptive to channel conditions --
directly extending the digital-vs-analog trade-off Section~\ref{sec:arch_e2e}
raises to a continuously-tunable rather than binary choice.
A.~M.~Aboumadi~\emph{et al.}~\cite{Aboumadi2025_Adaptive} pursue the
same channel-adaptivity goal from a different angle, adjusting the
semantic encoder's own compression ratio in response to real-time
channel state rather than splitting the representation -- an
adaptive-compression mechanism directly analogous to the hypernetwork
approach of Section~\ref{sec:future}, but applied at inference time to
an already-deployed E2E JSCC system rather than as a training-time
research direction.

\subsubsection{Generative Multi-User Semantic Communication: M-GSC}

The M-GSC framework of Yang et al.~\cite{Yang_2025_Rethinking}
represents a significant architectural extension of E2E JSCC to the
multi-user scenario. Rather than training a single encoder and decoder
jointly for a point-to-point link, M-GSC introduces a shared
large-language-model-based knowledge base (SKB) that serves three
distinct roles within a cloud-edge-end network architecture: (1)~task
decomposition via Chain-of-Thought reasoning; (2)~semantic
representation specification, identifying standardized encoders (e.g.,
SAM for segmentation, YOLOS for object detection) at the transmitter
and personalized decoders (diffusion models) at each receiver; and
(3)~semantic translation across heterogeneous codec families without
requiring either party to retrain. This decoupled design has a
standardization advantage directly relevant to Section~\ref{sec:challenges}:
semantic encoding becomes a standardized interface while decoding
remains personalized. Quantitative results on road traffic scenarios
show that offloading 350 denoising steps to the receiver outperforms
offloading 650 steps, and generative SemCom achieves substantially
higher semantic accuracy than SwinJSCC at low SNR. The identified
hallucination risk from single-LLM SKBs motivates multi-agent LLM
architectures~\cite{Yang_2025_Rethinking}, connecting directly to the
cross-modal semantic alignment challenge of Section~\ref{sec:challenges}.

\subsection{Split Learning}
\label{sec:arch_split}

Split learning partitions a neural encoder across the device and an
edge server: the device runs the first few layers locally and
transmits only the resulting intermediate activation -- not the raw
input, and not the full model's output -- to the edge server, which
runs the remaining layers to complete the task. The layer at which
this partition happens is the \emph{cut layer}: choosing it earlier
(closer to the input) pushes more computation onto the edge server and
shrinks the device's memory/compute footprint, while choosing it later
keeps more computation on-device at the cost of a larger, more
expensive-to-transmit intermediate activation. This makes split
learning a natural fit for TinyLM deployment: a device too
resource-constrained to run a full semantic encoder can instead run
only its first layers, offloading the rest of the semantic encoding
work to an edge aggregator.

\subsubsection{Three-Way Trade-Off Analysis}

Choosing the cut layer $c$ is a three-way trade-off between how much
energy the device spends computing, how much energy and bandwidth the
uplink transmission of the intermediate activation costs, and how
much energy the edge server spends completing the computation. The
device energy $E_d(c)$, uplink message size $M(c)$, and server
energy $E_s(c)$ satisfy:
\begin{equation}
  E_d(c) + E_s(L-c) + E_{\mathrm{comm}}\bigl(M(c)\bigr)
    = E_{\mathrm{total}}(c),
  \label{eq:split_energy}
\end{equation}
minimized by selecting $c^* = \arg\min_c E_{\mathrm{total}}(c)$
subject to latency and PDU size constraints. E.~Eldeeb~\emph{et al.}
\cite{Eldeeb2025_SemanticMSL} quantify this for a 5-layer CNN:
cut $=1$ yields 29.33~Wh device energy and 14.912~KB uplink; cut$=3$
yields 41.20~Wh device energy and 0.248~KB uplink (60$\times$
reduction). For NB-IoT (20--30~KB PDU), cut$=1$ is optimal; for
eMBB UEs, cut$=3$ minimizes total energy. A.~Bhattacharyya~\emph{et al.}~\cite{Bhattacharyya2025_Semantic}
provide a rare complement to this analytical trade-off: a practical
COMSNETS SDR-workshop deployment reporting split-AI latency and
bandwidth measured on real hardware rather than simulated, alongside
standardization-relevant deployment insights -- evidence of this
kind, measured rather than modeled, remains uncommon in the split
learning literature.

The privacy dimension is equally important: split learning achieves
a reconstruction error of 0.2681---a fourfold increase compared to
federated learning (0.0671) and $18\times$ that of centralized
learning (0.0154)~\cite{Eldeeb2025_SemanticMSL}. This privacy benefit
is not designed in---it emerges from the information compression
performed by the device head, formalized as
\begin{equation}
  R_{\mathrm{inv}}(c) = \frac{1}{\sigma^2_c + \epsilon},
  \label{eq:inversion_fidelity}
\end{equation}
where $\sigma^2_c$ is the variance of the posterior $p(x|\hat{z}_c)$
under a model-inversion attack and $\epsilon$ is a small constant for
numerical stability: as $c$ increases, $\sigma^2_c$ grows because the
encoder has discarded more task-irrelevant variation needed to recover
the original input, monotonically reducing inversion fidelity at the
cost of additional device energy.

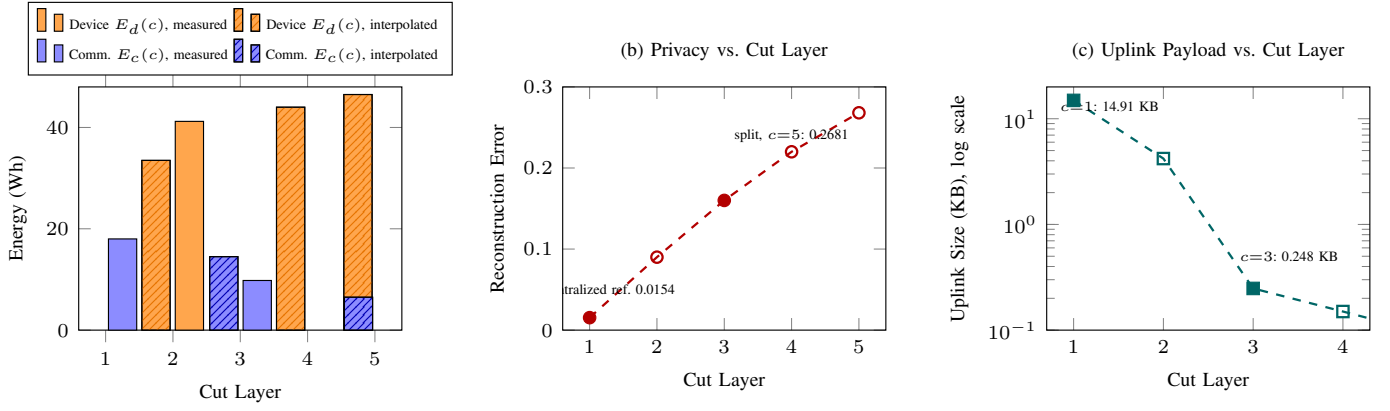
\begin{figure*}[t]
\centering
\resizebox{\textwidth}{!}{%
\begin{tikzpicture}
\begin{groupplot}[
  group style={group size=3 by 1, horizontal sep=2.0cm},
  width=5.6cm, height=4.6cm,
  xlabel={Cut Layer}, xtick={1,2,3,4,5},
  tick label style={font=\scriptsize}, label style={font=\scriptsize},
  title style={font=\scriptsize},
]
\nextgroupplot[
  ybar, bar width=10pt, ylabel={Energy (Wh)},
  title={(a) Energy vs.\ Cut Layer},
  legend style={font=\tiny, at={(0.5,1.35)}, anchor=north, legend columns=2},
  ymin=0, ymax=48,
]
\addplot[fill=orange!70] coordinates {(1,29.33)(3,41.20)};
\addlegendentry{Device $E_d(c)$, measured}
\addplot[fill=orange!70, postaction={pattern=north east lines,
  pattern color=orange!90!black}] coordinates {(2,33.5)(4,44.0)(5,46.5)};
\addlegendentry{Device $E_d(c)$, interpolated}
\addplot[fill=blue!45] coordinates {(1,18.0)(3,9.8)};
\addlegendentry{Comm.\ $E_c(c)$, measured}
\addplot[fill=blue!45, postaction={pattern=north east lines,
  pattern color=blue!70!black}] coordinates {(2,14.5)(4,6.5)(5,3.9)};
\addlegendentry{Comm.\ $E_c(c)$, interpolated}
\nextgroupplot[
  ylabel={Reconstruction Error},
  title={(b) Privacy vs.\ Cut Layer},
  ymin=0, ymax=0.30,
]
\addplot[color=red!70!black, thick, dashed, mark=none]
  coordinates {(1,0.0154)(2,0.09)(3,0.16)(4,0.22)(5,0.2681)};
\addplot[only marks, mark=*, mark size=2.2pt, color=red!70!black]
  coordinates {(1,0.0154)(3,0.16)};
\addplot[only marks, mark=o, mark size=2pt, color=red!70!black, thick]
  coordinates {(2,0.09)(4,0.22)(5,0.2681)};
\node[font=\tiny] at (axis cs:1.3,0.05) {centralized ref.\ 0.0154};
\node[font=\tiny] at (axis cs:4.0,0.24) {split, $c{=}5$: 0.2681};
\nextgroupplot[
  ylabel={Uplink Size (KB), log scale},
  ymode=log,
  title={(c) Uplink Payload vs.\ Cut Layer},
  ymin=0.1, ymax=20,
]
\addplot[color=teal!80!black, thick, dashed, mark=none]
  coordinates {(1,14.912)(2,4.2)(3,0.248)(4,0.15)(5,0.09)};
\addplot[only marks, mark=square*, mark size=2.2pt, color=teal!80!black]
  coordinates {(1,14.912)(3,0.248)};
\addplot[only marks, mark=square, mark size=2pt, color=teal!80!black, thick]
  coordinates {(2,4.2)(4,0.15)(5,0.09)};
\node[font=\tiny] at (axis cs:1.4,13) {$c{=}1$: 14.91~KB};
\node[font=\tiny] at (axis cs:3.4,0.5) {$c{=}3$: 0.248~KB};
\end{groupplot}
\end{tikzpicture}}
\caption{Split-learning cut-layer trade-off, sharing a common x-axis
(cut layer 1--5). (a) Device vs.\ communication energy
(Eq.~\ref{eq:split_energy}); (b) reconstruction-error privacy metric
(Eq.~\ref{eq:inversion_fidelity}); (c) uplink payload size. Solid
fill / filled markers mark $c=1$ and $c=3$, the two values measured
in~\cite{Eldeeb2025_SemanticMSL}; hatched fill / hollow markers mark
$c=2,4,5$, linear interpolations shown for visualization only and
not independently measured.}
\label{fig:cutlayer_tradeoff}
\end{figure*}

Figure~\ref{fig:cutlayer_tradeoff} visualizes this three-way trade-off
across cut layers 1 through 5. Values at $c=1$ and $c=3$ are the two
points empirically reported in~\cite{Eldeeb2025_SemanticMSL};
intermediate points are linear interpolations for visualization only.

\subsubsection{Semantic-MSL}

Semantic-MSL~\cite{Eldeeb2025_SemanticMSL} combines split learning
with MAML meta-learning. The meta-objective:
\begin{equation}
  \min_\theta \sum_{\tau_i \sim p(\mathcal{T})}
    \mathcal{L}_{\tau_i}\!\bigl(f_{\theta - \alpha\nabla_\theta
    \mathcal{L}_{\tau_i}(f_\theta)}\bigr),
  \label{eq:maml}
\end{equation}
explicitly optimizes initial weights to minimize loss after $k$
gradient steps on any task from the task distribution $p(\mathcal{T})$.
After meta-training, Semantic-MSL adapts to new semantic tasks in
20~SGD steps on 5 training examples (5-shot), achieving 95\% accuracy
at SNR$=$15~dB versus 82\% for the baseline---with 30\% lower training
energy.

\subsubsection{Semantic MEC with Private Knowledge Repositories}

The core idea here is to give each edge node its own local shortcut
around the shared global knowledge base: M.~Yan~\emph{et al.}~\cite{Yan2025_MEC}
equip each edge node of a Multi-access Edge Computing (MEC) network
with a scenario-specific private knowledge repository alongside a
global knowledge repository, so that queries matching locally common
scenarios resolve without a round-trip to the global store. This
reduces delay from 40--58~ms (non-semantic MEC, 5--30 vehicles) to
25--32~ms---27\% latency reduction---while energy scales sub-linearly
with vehicle density.

\subsubsection{Resource-Aware Split Efficiency}

The Efficient Parallel Split Learning (EPSL) framework~\cite{Lin_2024_Efficient}
introduces a last-layer gradient aggregation ratio $\phi \in [0,1]$
controlling the trade-off between convergence accuracy and per-round
communication latency:
\begin{equation}
  \tilde{g}^{(t)} = \phi \cdot \frac{1}{K}\sum_{k=1}^{K} g_k^{(t)}
    + (1-\phi) \cdot g_{k^*}^{(t)},
  \label{eq:epsl_gradient}
\end{equation}
blending the full $K$-client average gradient with a single
representative client's gradient $g_{k^*}^{(t)}$, with per-round
latency scaling approximately as
\begin{equation}
  T_{\mathrm{round}}(\phi) \approx \phi \cdot T_{\mathrm{sync}}
    + (1-\phi)\cdot T_{\mathrm{async}}.
  \label{eq:epsl_latency}
\end{equation}
$\phi$ is used again for adaptive-compression design in
Section~\ref{sec:future}. A block-coordinate descent (BCD) algorithm
jointly optimizes cut-layer position, aggregation ratio, and spectrum
allocation, achieving approximately 50\% training latency reduction
versus non-optimized split-learning baselines while remaining robust
to dynamic wireless channel conditions.

DeCo-MeSC's core idea is to push compression asymmetrically: rather
than compressing both partitions equally, it compresses only the
device-side partition aggressively and lets the (less
resource-constrained) server-side partition absorb the resulting
mismatch through fine-tuning. Concretely, M.~Sung~\emph{et al.}~\cite{Sung_2025_DeCoMeSC}
apply Deep Compression (magnitude-based pruning, $Q$-bit codebook
quantization, and Huffman entropy coding) exclusively to the
device-side partition, then fine-tune the server-side partition to
absorb the resulting representational mismatch:
\begin{equation}
  M_{\mathrm{device}} = N_{\mathrm{nz}} \cdot
    \left(\log_2 Q + \delta_{\mathrm{Huff}}\right) + N_{\mathrm{nz}}
    \cdot \log_2\!\binom{N_{\mathrm{total}}}{N_{\mathrm{nz}}},
  \label{eq:deco_memory}
\end{equation}
where $N_{\mathrm{nz}}$ is the number of non-zero weights surviving
pruning, $Q$ is the codebook quantization level count, and
$\delta_{\mathrm{Huff}}$ is the average per-symbol saving from
Huffman-coding the codebook indices. The second (combinatorial) term
encodes only the index/position cost of the sparsity mask itself;
Huffman-coding this mask further, beyond coding the codebook indices,
is a common but not universal implementation choice, and was not
confirmed as applied in the 600~KB figure reported
by~\cite{Sung_2025_DeCoMeSC}. The result is 73.83\% CIFAR-100
inference accuracy at $M=600$~KB device memory, substantially higher
than the 59.52\% accuracy obtained without server-side fine-tuning.

For 6G semantic communication, DeCo-MeSC operationalizes a direct
deployment principle: compress the UE semantic encoder aggressively
while keeping the gNB semantic decoder at full precision, then
fine-tune the decoder periodically via over-the-air model updates.
Matsubara et al.'s earlier Head Network Distillation
(HND)~\cite{Matsubara_2020_HeadNetwork} established the foundational
precedent: distilling only the device-side ``head'' network against
the original model's intermediate features, with bottleneck
quantization achieving up to 75\% compression of the bottleneck
output tensor without impacting accuracy.

\subsection{Federated Learning for Semantic Communication}
\label{sec:arch_fl}

\subsubsection{Why Standard FedAvg Cannot Serve Semantic Communication}

In semantic communication, non-IID means different clients encode
the \emph{same} semantic concept through different surface-level
representations. FedAvg produces an embedding accurately representing
neither. The standard workaround of padding all models to the same
architecture discards architecture-specific compression efficiency
gains. Z.~Xiang~\emph{et al.}~\cite{Xiang2025_Task} provide direct
evidence for this failure mode's practical severity: evaluating
task-oriented federated semantic communication on real Raspberry Pi
and Jetson hardware -- rather than simulation, still uncommon in this
literature -- they report accuracy degradation under non-IID splits
consistent with the architecture-mismatch mechanism described above,
with SNR-adaptive training partially compensating. W.~Yang~\emph{et al.}~\cite{Yang2022_Semantic}
frame the broader motivation for offloading semantic-encoder training
to federated clients in the first place: doing so keeps raw,
potentially sensitive source data on-device while still allowing the
shared semantic encoder to improve from every client's local
distribution, which they describe as a form of semantic compression
applied to the model update itself rather than to any single
transmitted message.

\subsubsection{FedBKD: Joint Model and Data Heterogeneity}

\begin{figure*}[t]
\centering
\includegraphics[width=0.92\textwidth,height=0.85\textheight,keepaspectratio]{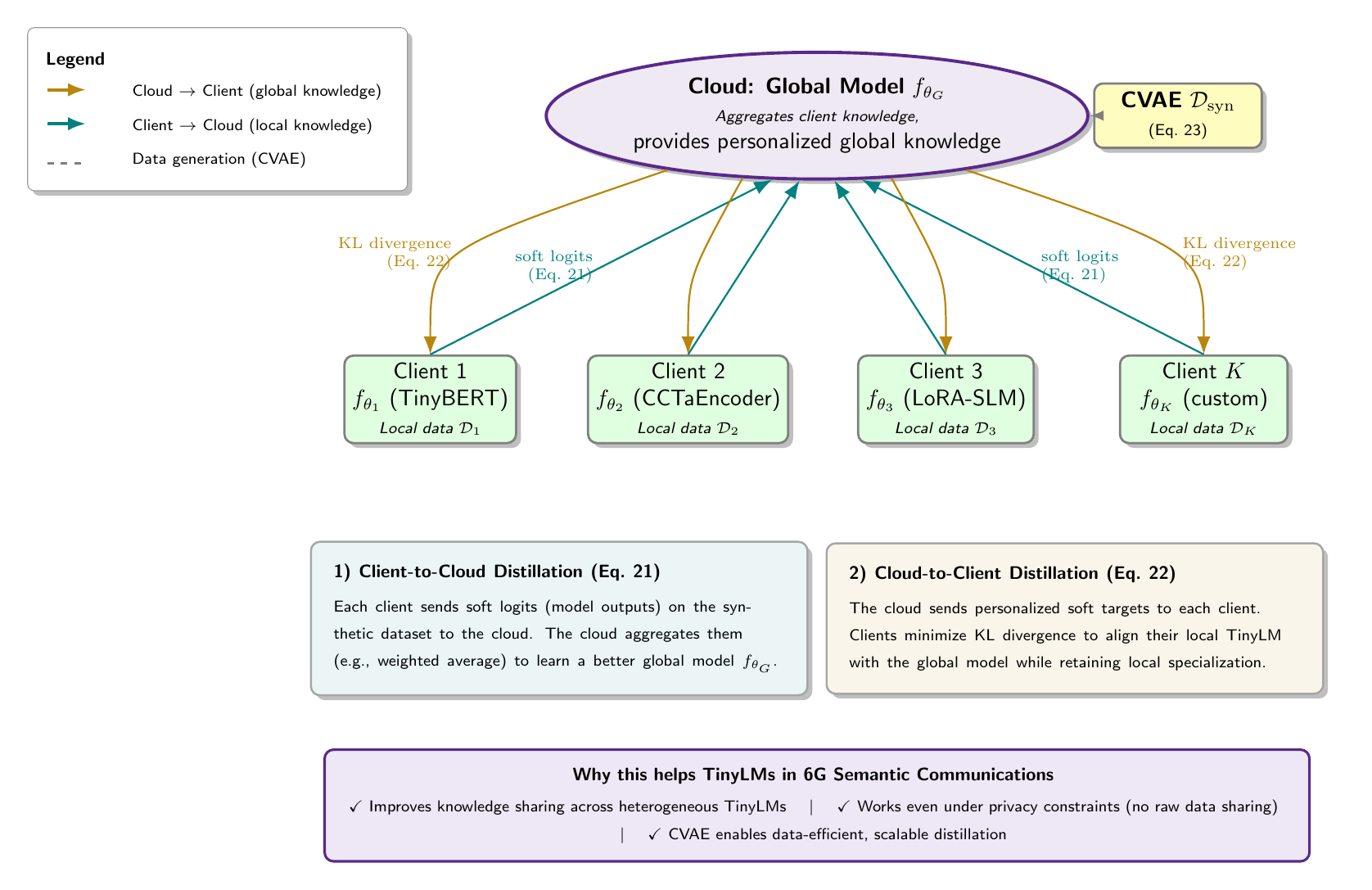}
\caption{FedBKD bidirectional distillation~\cite{Qi_2023_FedBKD}.
Client-to-cloud distillation treats heterogeneous local models as a
teacher ensemble via Eq.~\ref{eq:fedbkd_c2c}; cloud-to-client
distillation pushes global knowledge back per client via
Eq.~\ref{eq:fedbkd_cloud2c}. A server-hosted CVAE (Eq.~\ref{eq:cvae_elbo})
generates the synthetic dataset used for both directions.}
\label{fig:fedbkd}
\end{figure*}

The FedBKD framework~\cite{Qi_2023_FedBKD} addresses a joint
model-and-data heterogeneity problem that FedPun alone only
addresses partially, illustrated in Figure~\ref{fig:fedbkd}. Model
heterogeneity is handled by a bidirectional KD protocol: in the
client-to-cloud direction, the cloud treats all $K$ local models as an
ensemble of teachers and trains the global model $f_{\theta_G}$ by
minimizing the averaged-logit alignment loss on a CVAE-synthesized
public dataset $\mathcal{D}_{\mathrm{syn}}$:
\begin{equation}
  \mathcal{L}_{\mathrm{c2cloud}} =
    \mathbb{E}_{x \sim \mathcal{D}_{\mathrm{syn}}}
    \left[ D_{\mathrm{KL}}\!\left(
      \frac{1}{K}\sum_{k=1}^{K} \sigma\bigl(f_{\theta_k}(x)\bigr)
      \,\Big\|\, \sigma\bigl(f_{\theta_G}(x)\bigr)
    \right)\right],
  \label{eq:fedbkd_c2c}
\end{equation}
where $\sigma(\cdot)$ denotes the softmax operator. In the reverse
cloud-to-client direction, each client independently distills the
updated global model's knowledge back into its own architecture:
\begin{equation}
  \mathcal{L}_{\mathrm{cloud2c},k} = \mathcal{L}_{\mathrm{task},k}
    + \gamma \cdot D_{\mathrm{KL}}\!\left(
      \sigma\bigl(f_{\theta_G}(x)\bigr) \,\big\|\,
      \sigma\bigl(f_{\theta_k}(x)\bigr)
    \right),
  \label{eq:fedbkd_cloud2c}
\end{equation}
where $\gamma$ controls the strength of global-knowledge injection.
Data heterogeneity is resolved by the CVAE hosted at the cloud
server, trained to minimize the evidence lower bound (ELBO) conditioned
on a target label distribution $y$:
\begin{equation}
  \mathcal{L}_{\mathrm{CVAE}} =
    \mathbb{E}_{q_\phi(z|x,y)}\bigl[\log p_\psi(x|z,y)\bigr]
    - D_{\mathrm{KL}}\bigl(q_\phi(z|x,y) \,\|\, p(z)\bigr),
  \label{eq:cvae_elbo}
\end{equation}
where $q_\phi$ is the encoder, $p_\psi$ is the decoder, and $p(z)$ is
the latent prior. Evaluated on modulation classification, FedBKD
achieves superior convergence accuracy over FedAvg, FedMD, and FedDF
across 5, 15, 20, and 25 client configurations under simultaneous
model and data heterogeneity.

\subsubsection{Channel-Aware Feature Reconstruction and Ensemble Distillation}

Huh et al.'s FedSFR~\cite{Huh_2026_Federated} introduces a
feature-reconstruction step executed at the parameter server: after
standard federated-averaging weight aggregation, the server applies a learned
reconstruction function $g_\psi$ to correct compression- and
channel-induced errors,
\begin{equation}
  \hat{f}_{\mathrm{corrected}} = g_\psi\bigl(\hat{f}_{\mathrm{agg}}\bigr),
  \qquad
  \mathcal{L}_{\mathrm{FedSFR}} = \bigl\| \hat{f}_{\mathrm{corrected}}
    - f_{\mathrm{clean}} \bigr\|_2^2,
  \label{eq:fedsfr_loss}
\end{equation}
achieving higher PSNR and more stable training than baseline FL
methods on CIFAR-10 and CelebA under both AWGN and Rayleigh fading.

Salman et al.'s FedFB~\cite{Salman_2026_FedFB} combines online
ensemble distillation with a dual-branch student network, achieving
91.07\% accuracy on a 10-client non-IID medical imaging benchmark
while reducing communication overhead by more than 86\% relative to
FedAvg and FedProx. An ablation study confirms that removing the
auxiliary classifier (AC) module drops accuracy from 91.17\% to
approximately 74\%, underscoring the necessity of the dual-branch
structure.

Li et al.~\cite{Li_2026_Adaptive} propose an LSTM-based policy
gradient framework for adaptive DNN partitioning and pruning, where a
policy network observes channel state and device load and outputs a
joint partitioning-and-pruning action $a_t=(c_t, r_t)$:
\begin{equation}
  \nabla_\phi J(\phi) = \mathbb{E}_{\pi_\phi}\bigl[
    \nabla_\phi \log \pi_\phi(a_t | s_t) \cdot R_t \bigr],
  \label{eq:li2026_policy}
\end{equation}
achieving an 8.247\% increase in system reward and a 27.313\% average
reduction in total inference delay over baselines.

\subsection{Knowledge-Graph-Assisted Semantic Communication}
\label{sec:arch_kg}

\begin{figure*}[t]
\centering
\includegraphics[width=0.92\textwidth,height=0.85\textheight,keepaspectratio]{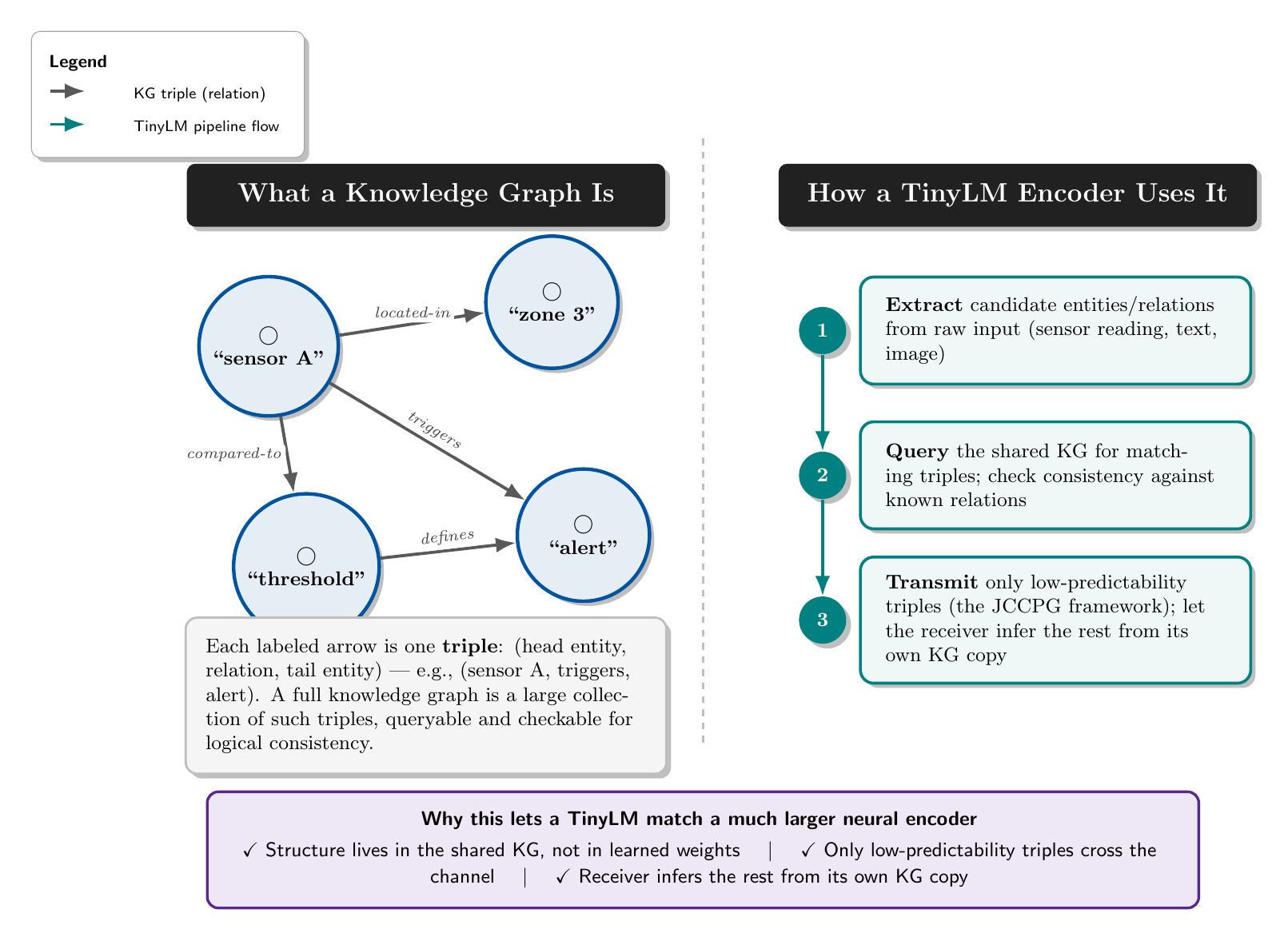}
\caption{What a knowledge graph is (left) and how a TinyLM semantic
encoder uses one at inference time (right). The four systems below --
JCCPG, ARL, covert semantic communication, and neuro-symbolic AI --
each specialize one part of this right-hand pipeline.}
\label{fig:kg_infographic}
\end{figure*}

Before the specific systems below, it is worth fixing what a
knowledge graph actually is and where a TinyLM encoder fits around
it. A knowledge graph (KG) represents facts as \emph{triples} --
(head entity, relation, tail entity) -- for example
(Paris, capital-of, France): a directed, labeled edge connecting two
entity nodes. A collection of such triples forms a graph a system can
traverse, query, and check for logical consistency, in contrast to
the distributed, non-symbolic representations a neural network learns
internally. Figure~\ref{fig:kg_infographic} illustrates both halves of
this: the graph structure itself (left), and how a TinyLM semantic
encoder uses it at inference time (right) -- extracting candidate
entities and relations from the input, querying the shared KG for the
relevant triples, and transmitting only those triples (or, for
high-predictability relations, nothing at all, since the receiver can
infer them from its own copy of the KG) rather than the full source
signal.

\subsubsection{Advantages of Symbolic Structure, by Example}

Knowledge graphs can provide advantages neural encoders typically
lack, though the evidence base for each below rests on a single
example system rather than a systematic comparison across many
studies: (1)~error correction by consistency -- a \emph{theoretical}
advantage rather than one with reported empirical detection-rate
evidence in this survey's corpus, since a KG's ontological constraints
make a corrupted triple checkable in principle (a triple that
violates a known relation type or entity category can be flagged),
but no verified study measures how often this actually catches errors
in practice; (2)~extreme symbolic-level compression -- representing
meaning as compact logical/graph structure rather than as
floating-point weights -- demonstrated concretely by SPM's
4,550$\times$ compression operating at the symbolic concept
level~\cite{Seo2023_SPM}; and (3)~out-of-distribution generalization
through graph traversal and multi-hop reasoning, argued conceptually
rather than measured directly in the systems surveyed here.
Y.~Li~\emph{et al.}~\cite{Li2025_Cognitive} frame
this same set of advantages from the opposite direction, as a response
to deep-learning semantic encoders' own weaknesses: because purely
data-driven encoders are black-box and offer no explicit reasoning
mechanism, they argue knowledge-graph-grounded ``cognitive'' semantic
communication is a lightweight, interpretable complement to (not
necessarily a replacement for) large models, and demonstrate this on a
real software-defined-radio testbed -- reconstructing images at lower
bit rate and higher fidelity than both JPEG and a KG-free semantic
baseline. R.~Xu~\emph{et al.}~\cite{Xu2025_RateSplitting} add a
complementary multi-user angle: when several users transmit
semantically related but not identical content, a rate-splitting
scheme can separate the \emph{shared} semantic component (the
overlapping triples) from each user's \emph{private} residual,
transmitting the shared part once rather than redundantly per user --
a compression mechanism that operates at the multi-user scheduling
level rather than within any single encoder.

\subsubsection{JCCPG: Thermodynamics of Semantic Transmission}

JCCPG's core idea is to treat the transmitter and receiver as sharing
a probabilistic model of which knowledge-graph triples are likely
given context, and to transmit only the triples that model cannot
already predict. Z.~Zhao~\emph{et al.}~\cite{Zhao2023_probgraph}
formalize this as an energy-minimization problem: given a knowledge
graph's triples $(e_1, r, e_2)$, the transmitter computes $p(r|e_1,e_2)$
for each triple -- how predictable that relation is given the two
entities -- and only transmits the set $\mathcal{T}_{\mathrm{tx}}$ of
triples whose predictability falls below a threshold $\theta$; every
other triple, the complementary set $\mathcal{T}_{\mathrm{inf}}$,
the receiver instead \emph{infers} locally from its own copy of the
shared probabilistic model, at the cost of local computation rather
than transmission. The total energy is the sum of what this costs on
each side:
\begin{equation}
  E_{\mathrm{total}} = E_{\mathrm{comm}}\bigl(\mathcal{T}_{\mathrm{tx}}\bigr)
    + E_{\mathrm{comp}}\bigl(\mathcal{T}_{\mathrm{inf}}\bigr),
  \label{eq:jccpg}
\end{equation}
where $E_{\mathrm{comm}}(\cdot)$ is the communication energy of
transmitting the low-predictability triples and $E_{\mathrm{comp}}(\cdot)$
is the computation energy of inferring the high-predictability ones
locally. Raising $\theta$ shifts more triples from the transmit set
to the infer set, trading communication energy for computation
energy; Zhao et al.\ report that the resulting optimum achieves
$<1\times10^{-5}$~J at $M=120$ triples---65\% below a non-semantic
baseline that transmits every triple regardless of predictability.

\subsubsection{ARL: Adversarial RL for Knowledge Base Estimation}

J.~Zhang~\emph{et al.}~\cite{Zhang2024_ARL} model knowledge-base
estimation as a two-player game rather than a one-shot query: a
Semantic Information Agent (SIA) selects which questions to ask the
receiver in order to estimate how much of its own knowledge base the
receiver actually shares, while an adversarial Question Agent (QA)
tries to make the receiver's answers as uninformative as possible
about that overlap. Formally,
\begin{equation}
  \max_{\pi_{\mathrm{SIA}}} \min_{\pi_{\mathrm{QA}}}
    \mathbb{E}\bigl[\mathrm{ROUGE}(y, \hat{y})\bigr],
  \label{eq:arl}
\end{equation}
trains the SIA's questioning policy $\pi_{\mathrm{SIA}}$ to maximize,
and the QA's answering policy $\pi_{\mathrm{QA}}$ to minimize, the
ROUGE similarity between the receiver's true knowledge $y$ and the
SIA's estimate $\hat{y}$ recovered from the received answers -- the
adversarial minimax setup is what forces the SIA to learn efficient,
information-dense questions rather than exhaustive ones. Training the
SIA against this adversary yields a 15.2\% Recall-Oriented Understudy
for Gisting Evaluation (ROUGE) score improvement with as few as 4
questions per session.

\subsubsection{Knowledge-Enhanced Transformer Encoders: How Much Does Knowledge Buy in Compression?}

Two independently verified studies quantify directly how much
knowledge-graph augmentation lets a semantic encoder shrink before
failing -- a question the conceptual advantages listed above do not
by themselves answer. X.~Xu~\emph{et al.}~\cite{Xu2023_KnowledgeEnhanced}
propose KESC-T, extending a standard transformer decoder with a
third, knowledge-target attention sub-layer so the decoder attends to
both the source sentence and a retrieved knowledge-graph triple
embedding, using a genuinely compact configuration (4 encoder blocks,
300-dimensional hidden size, 10 attention heads, GloVe 300-dimensional
pretrained embeddings) already modest by NLP standards before any
further compression is applied. Beyond the encoder itself, a DNN
autoencoder compresses the transformer's $L \times K$ output down to
$L \times D$, followed by sigmoid-based quantization to $M$-bit
integers -- compressing the semantic \emph{representation}, a distinct
axis from the weight-level compression of
Section~\ref{sec:compression}. The key finding directly answers the
``how much does knowledge buy'' question: at compressed dimension
$D=2$ or $D=4$, a non-knowledge-augmented DeepSC baseline fails to
recover semantic information entirely, while KESC-T continues to
function with graceful, rather than catastrophic, degradation as $D$
shrinks from 16 to 2 -- direct evidence that injecting external
knowledge lets a semantic system tolerate substantially more
aggressive representation compression before failing than an
otherwise-identical knowledge-free system. The authors also propose a
Knowledge-Enhanced Efficiency (KEE) metric,
$\mathrm{KEE} = \tanh[(\mathrm{BLEU}(\alpha,\beta) -
\mathrm{BLEU}(\alpha,\gamma))/\mathrm{BLEU}(\alpha,\tau)]$ for
reference $\alpha$, knowledge-enhanced prediction $\beta$,
non-enhanced prediction $\gamma$, and injected triples $\tau$,
normalizing the marginal value of injected knowledge per unit
injected to $[-1,1]$ -- a more informative measure than a raw BLEU
delta precisely because it accounts for how much knowledge was added
to achieve that delta.

S.~Salehi~\emph{et al.}~\cite{Salehi2025_KGLLM} answer a closely
related ``how tiny is too tiny'' question directly at the model-size
level rather than the representation level: comparing T5-small
($\sim$60M parameters), T5-base ($\sim$220M), and T5-large
($\sim$770M) as the LLM component of a KG-augmented semantic encoder,
the \emph{smallest} model achieved the best semantic-similarity/compression
balance (SS$=0.83$), while the largest performed worst (SS$=0.58$)
due to output verbosity -- the smaller model was not merely
acceptable for resource reasons but empirically superior at the task.
Transmission time scaled accordingly: 1.64~ms for T5-small versus
20.77~ms for T5-large, a 12.7$\times$ gap consistent with the model's
stated $O(N^2 d)$ complexity, where the hidden dimension $d$ (512 for
T5-small) dominates encoding/decoding cost. An entropy-based routing
mechanism (threshold $H_\theta=3.85$) decides per-sentence whether
knowledge-graph triple extraction is needed at all, adding roughly 10
bytes per packet of semantic overhead only when it is invoked --
together with the D=2/4 compression-tolerance result above, this
constitutes direct, independently verified evidence that pairing a
smaller LLM with lightweight, selectively-invoked symbolic knowledge
outperforms scaling the neural component alone, for both compression
tolerance and semantic quality.

\subsubsection{Covert Semantic Communication}

Covert semantic communication (covert SemCom) addresses a threat
distinct from the fidelity and efficiency concerns of the rest of
this section: rather than asking how well meaning survives
transmission, it asks whether an eavesdropper can even detect that a
semantic transmission is taking place at all, let alone decode it. A
successful covert SemCom scheme must keep the transmitted signal
statistically indistinguishable from background noise to anyone
without the shared knowledge base, while remaining decodable to a
receiver who has it. W.~Zhang~\emph{et al.}~\cite{Zhang2026_covert}
instantiate this for KG-based semantic communication and establish a
semantic privacy gap: the intended receiver's Graph-to-Nearest-Triple
(GNT) similarity score exceeds 0.6, while an eavesdropper's stays
below 0.2, with private transmission probability $\geq$85\% under
four simultaneous attackers.

\subsubsection{Neuro-Symbolic AI for Sub-mW Semantic Reasoning}

Neuro-symbolic AI (NeSy AI) combines neural networks' pattern-learning
capability with symbolic AI's explicit, rule-based reasoning,
targeting the gap this survey has returned to throughout: neural
semantic encoders learn powerful representations but offer no
interpretable reasoning trace, while purely symbolic systems reason
transparently but cannot learn from data. For 6G edge deployment,
this hybrid also opens a hardware path unavailable to either
approach alone -- D.~Dold and J.~S.~Garrido~\cite{Dold2021_NeSy}'s
energy-based NeSy model assigns each knowledge-graph triple an energy
$E(\text{triple})$ learned via local three-factor Hebbian updates
(weight changes computed from locally available signals only, with no
global backpropagation pass):
\begin{equation}
  \Delta w_{ij} \propto x_i \cdot x_j \cdot e_k,
  \label{eq:hebbian}
\end{equation}
and because these updates require no global gradient computation,
the resulting model maps directly onto neuromorphic hardware --
enabling direct implementation on Intel Loihi or IBM TrueNorth at
sub-mW power, an energy budget no conventional GPU-trained neural
encoder can approach. C.~K.~Thomas~\emph{et al.}~\cite{Thomas2023_causal}
complement this with a causal-reasoning framework using GFlowNet-based
structure learning to infer cause-effect relationships between
semantic concepts rather than mere statistical co-occurrence, achieving
100$\times$ bit reduction versus classical communication by transmitting
only the causal structure needed to reconstruct an outcome rather than
the full observation. This survey's treatment of NeSy AI draws on a
broader landscape establishing the paradigm's maturity beyond these
two systems: A.~d'Avila Garcez and L.~C.~Lamb~\cite{Garcez2020_NeSy}
survey neuro-symbolic AI as a distinct ``third wave'' of AI research,
positioning it explicitly as a synthesis of (rather than a compromise
between) connectionist and symbolic approaches; K.~Hamilton~\emph{et al.}~\cite{Hamilton2022_NeSy}
review NeSy specifically within NLP; C.~K.~Thomas and W.~Saad~\cite{Thomas2022_NeSy}
propose intent-based NeSy semantic communication, framing the receiver's
goal as the symbolic component guiding neural encoding; M.~S.~Munir~\emph{et al.}~\cite{Munir2022_NeSyXAI}
extend NeSy to explainable-AI digital twins for wireless networks; and
G.~D'Acunto~\emph{et al.}~\cite{DAcunto2022_Multiscale} contribute
multiscale causal structure learning directly relevant to the
GFlowNet-based approach above.

\subsection{Multi-Task and Cross-Modal Architectures}
\label{sec:arch_multitask}

The taxonomy of Figure~\ref{fig:taxonomy} treats multi-task and
cross-modal semantic communication as a fifth architecture family,
distinct from the four above in a specific way: rather than one
encoder-decoder pair specialized to one task and one modality, a
single shared semantic encoder serves either multiple downstream
tasks (multi-task) or multiple source modalities -- text, image,
speech -- within one architecture (cross-modal).

\textbf{Multi-task.} Y.~Sheng~\emph{et al.}~\cite{Sheng2022_multitask}
demonstrate that a single
BERT semantic encoder can simultaneously serve three GLUE-benchmark
tasks -- SST (Stanford Sentiment Treebank, sentiment classification),
STSB (Semantic Textual Similarity Benchmark, sentence-pair similarity
scoring), and CoLA (Corpus of Linguistic Acceptability, grammaticality
judgment) -- maintaining
$\sim$90\% SST accuracy at 0~dB SNR while all traditional methods
fall below 60\%. The minimum channel symbols required scales as:
\begin{equation}
  S_{\min} \approx k \cdot N_{\mathrm{tasks}},
  \label{eq:multitask}
\end{equation}
where $k$ is a task-specific constant---providing a direct spectrum
allocation rule for 6G network slicing. X.~Yu~\emph{et al.}~\cite{Yu2025_GraphAttention}
approach the same multi-task setting from a different angle: rather
than serving multiple tasks from one shared encoding, a graph-attention
mechanism explicitly extracts and models the feature correlations
between tasks, letting the encoder exploit inter-task redundancy
directly instead of relying on a shared representation to capture it
implicitly.

\textbf{Cross-modal.} Y.~Duan~\emph{et al.}~\cite{Duan2023_Multimedia}
propose a representative cross-modal design pattern for this family:
an object-attribute-relation (OAR) model represents scene content as
structured entities and their relations -- conceptually the same
compact, structured meaning representation a TinyLM text encoder
produces, just applied to image and video content -- with a shared
semantic knowledge base maintained at the cloud and cached at the
edge, giving transmitter and receiver a common prior independent of
which modality is currently being sent. I.~Ahmed~\emph{et al.}~\cite{Ahmed2025_SemComCoexistence}
address a complementary systems-level problem this family raises:
when a TinyLM's semantic stream shares spectrum with conventional bit
traffic (a realistic deployment scenario, not a greenfield one), a
non-orthogonal multiple access (NOMA) coexistence scheme using
superposition coding and successive interference cancellation (SIC)
lets the semantic and bit-level streams share the same resource block
rather than requiring dedicated spectrum -- directly relevant to 6G
satellite and non-terrestrial-network deployment, where spectrum
sharing with legacy bit-level traffic is unavoidable. Z.~Zhou~\emph{et al.}~\cite{Zhou2024_Speech}
extend the cross-modal family to speech specifically: a Swin-Transformer-based
speech semantic encoder reconstructs speech from a semantic rather
than waveform-level representation, improving Perceptual Evaluation
of Speech Quality (PESQ) by 45.3\%, 80.2\%, and 121.0\% over a
non-semantic speech baseline under AWGN, fading, and frequency-deviation
channels respectively -- the largest relative gains of any single
modality-specific result in this survey's evidence base, and direct
evidence that the cross-modal principle established for text and
image (Section~\ref{sec:background}) extends to audio with comparable
or greater benefit.

Beyond text and image, keypoint-based video semantic communication
\cite{Jiang_2023_Wireless} demonstrates extreme transmission
efficiency: the Semantic Video Conferencing (SVC) system transmits
only 160 bits per frame---compared to H.264 at 0.0157~bits-per-pixel
and AV1 at 0.0092~bits-per-pixel at equivalent perceptual loss---by
encoding facial keypoints rather than raw pixel values. The SVC
extension SVC-HARQ (Hybrid Automatic Repeat reQuest, a retransmission
protocol that combines error-detecting coding with selective
retransmission of only the corrupted portion of a message) adds a
semantic error detector exploiting video fluency as a non-intrusive
fidelity signal. The same author group's companion
work~\cite{Jiang2023_RevisingModules} takes a complementary,
incremental-deployment angle: rather than replacing the physical and
MAC layers outright, it shows how existing conventional communication
modules can be selectively revised to carry semantic transmission,
which is directly relevant to operators seeking to introduce TinyLM
semantic encoders into already-deployed infrastructure rather than a
greenfield 6G rollout.

\section{Performance Metrics for Semantic Communication}
\label{sec:metrics}

Sections~\ref{sec:background} and~\ref{sec:architectures} used BLEU,
PSNR, SSIM, MSS, and ROUGE as needed, each in its own context; this
section collects them in one place, gives each its mathematical
formulation, and states plainly when it is the right metric to reach
for and where it breaks down -- the comparison Table~\ref{tab:metrics}
needs to be legible on its own, without flipping back through earlier
sections to recover a definition.

\subsection{Semantic Transmission Rate and Reasoning-Capacity Metrics}

Before the per-message fidelity metrics of
Table~\ref{tab:metrics} below,
T.~M.~Getu~\emph{et al.}'s metrics survey~\cite{Getu2023_MakingSense}
defines two throughput-oriented metrics directly useful for comparing
TinyLM systems at the link level rather than the message level:
Semantic Transmission Rate (S-R) and Semantic Spectral Efficiency
(S-SE), both parametrized by the average number of semantic symbols
transmitted per user and measured in semantic units per second (and
per Hz, for S-SE) rather than bits per second -- the natural
throughput metric once the transmitted quantity is meaning rather
than bits. The same survey also formalizes a version of the
reasoning-capacity decomposition this survey adapted expository in
Section~\ref{sec:background_theory} (Eq.~\ref{eq:reasoning_capacity}),
independently corroborating that a compute-dependent reasoning term
added to the classical Shannon capacity is an established idea in the
metrics literature, not only this survey's own construction. Getu et
al. make the point most relevant to this survey's central thesis
explicitly: existing statistical fidelity metrics remain poorly
suited to the reasoning-dependent tasks central to 6G, which is
precisely the gap a TinyLM's retained reasoning capacity -- as
opposed to a purely statistical compression of a larger model's
output -- is positioned to close.

\subsection{Mean Semantic Similarity (MSS) Score}

For knowledge-graph-based and multi-user systems specifically, where
the transmitted content is a set of triples rather than free text or
pixels, the Metric of Semantic Similarity
(MSS)~\cite{Wang2022_JSAC} blends token-level precision (semantic
accuracy) and recall (semantic completeness) via an F-measure
formulation with a brevity penalty:
\begin{equation}
  \mathrm{MSS} =
    F\bigl(\mathrm{acc}(S,\hat{S}),\, \mathrm{rec}(S,\hat{S})\bigr)
    \cdot \mathrm{BP},
  \label{eq:mss}
\end{equation}
where $\hat{S}$ is the recovered semantic triple set, $S$ is the
original, and BP is the brevity penalty that discourages trivially
short (and therefore trivially ``precise'') triple sets. MSS is the
right choice specifically when the transmitted content is structured
--- it has no defined meaning for free-form text or raw pixels, where
BLEU/ROUGE and PSNR/SSIM apply instead.

\subsection{Why No Single Metric Suffices}

\begin{table*}[t]
\caption{Performance Metrics for Semantic Communication Systems}
\label{tab:metrics}
\centering
\footnotesize
\renewcommand{\arraystretch}{1.0}
\begin{tabular}{|p{1.4cm}|l|p{3.6cm}|p{3.8cm}|p{2.1cm}|}
\hline
\textbf{Modality} & \textbf{Metric} & \textbf{Definition} &
\textbf{Limitation} & \textbf{Suitable when} \\
\hline
\multirow{3}{*}{Text}
 & BLEU & $n$-gram precision between hypothesis and reference~\cite{Xie2021_DeepSC}
   & Cannot capture synonyms or paraphrases that preserve meaning while
     changing wording & Reference text available; word-level fidelity matters \\
\cline{2-5}
 & Sentence similarity & Cosine distance in a pre-trained embedding
   space (e.g., BERT/SBERT) & Requires a large pre-trained model at the
   receiver purely for evaluation & Semantic (not lexical) closeness is
   what matters \\
\cline{2-5}
 & ROUGE & Longest-common-subsequence ratio between hypothesis and
   reference~\cite{Zhang2024_ARL} & Insensitive to word order variations
   & Summarization-style or KB-estimation tasks (Section~\ref{sec:arch_kg}) \\
\hline
\multirow{3}{*}{Image}
 & PSNR & $10\log_{10}(\mathrm{MAX}^2/\mathrm{MSE})$, in dB & Misaligned
   with human perceptual quality at high compression & Pixel-level
   reconstruction fidelity is the target, not perceptual quality \\
\cline{2-5}
 & SSIM & Structural luminance/contrast similarity & Shallow perceptual
   model relative to modern learned metrics & Structural fidelity
   (edges, texture) matters more than raw pixel error \\
\cline{2-5}
 & Task accuracy & Downstream classification correctness (\%) & Task-specific;
   not generalizable across tasks & The semantic encoder's whole
   purpose is a specific downstream task \\
\hline
KG / Cross-modal
 & MSS & F-measure blend of triple precision and recall with a brevity
   penalty (Eq.~\ref{eq:mss}) & Exact weighting and brevity term are
   implementation choices, not a single standardized formula & Structured
   (KG-triple) semantic content specifically (Section~\ref{sec:arch_kg}) \\
\hline
\multirow{2}{*}{Cross-system}
 & SES & $F_{\mathrm{norm}}/(\log_{10}(\mathrm{size}/1\mathrm{KB})+\zeta)$
   (Eq.~\ref{eq:ses}) & Requires both a quality figure and a size figure
   for the same system; sparse across the corpus & Comparing compression
   ratio against quality \emph{across} architecturally different systems \\
\cline{2-5}
 & Semantic spectral efficiency (S-SE) & Semantic information rate
   divided by bandwidth~\cite{Yan2022_Resource} & ``Semantic bits'' unit
   not yet standardized across systems & Spectrum-allocation decisions
   for semantic (vs.\ conventional) traffic \\
\hline
\end{tabular}
\renewcommand{\arraystretch}{1}
\end{table*}

\FloatBarrier

Table~\ref{tab:metrics} is organized by modality rather than as
one ranked list, because no metric in it generalizes across
modalities: a PSNR score and a BLEU score are not comparable
quantities, and neither tells a receiver anything about a knowledge
graph's triple recovery. This fragmentation is directly related to
Challenge 1 in Section~\ref{sec:challenges} (a non-intrusive, and
ultimately a universal, semantic quality metric), and
the Semantic Efficiency Score (Eq.~\ref{eq:ses}) is this survey's
attempt to at least make model-size-versus-quality trade-offs
comparable across systems, even though it cannot unify what
``quality'' means across text, image, and structured content.

\label{sec:evidence}

Sections~\ref{sec:compression} and~\ref{sec:architectures} reviewed
individual TinyLM compression techniques and architectures; this
section steps back and asks, across that whole body of evidence, what
it collectively tells us about each of the six research questions of
Table~\ref{tab:pico}. Twenty systems make up this
survey's evidence base and are tabulated in
Table~\ref{tab:quantresults}; every citation in that table has been
independently re-verified against a citable, traceable primary
source as part of this survey's citation-accuracy pass, so the
evidence base below draws on the full twenty rather than a narrower,
separately-verified subset. Two systems receive the deeper,
individually-worked treatment below because they carry the most
directly load-bearing numeric claims used elsewhere in this survey:
the split-learning fine-tuning framework of
Eldeeb et al.~\cite{Eldeeb2025_SemanticMSL}, and
FedBKD~\cite{Qi_2023_FedBKD} for its joint
model-and-data-heterogeneity evidence. The remaining eighteen systems
inform the qualitative research-question synthesis below at the level
their own individual subsections (Sections~\ref{sec:compression}
and~\ref{sec:architectures}) already treated them, rather than being
re-derived a second time here.

\subsection{Benchmark and Dataset Mapping}

The systems discussed throughout Sections~\ref{sec:compression}
and~\ref{sec:architectures} are evaluated on several different
benchmarks and datasets, mentioned individually at their point of
discussion; Table~\ref{tab:benchmark_map} consolidates every such
system-to-benchmark pairing already stated in this survey into one
place, rather than leaving a reader to assemble it from scattered
mentions.

\begin{table*}[t]
\caption{System-to-Benchmark Mapping}
\label{tab:benchmark_map}
\centering
\small
\renewcommand{\arraystretch}{1.2}
\begin{tabular}{|l|l|l|}
\hline
\textbf{System} & \textbf{Benchmark/Dataset} & \textbf{Task} \\
\hline
DeepSC~\cite{Xie2021_DeepSC}\textsuperscript{*} & European Parliament corpus & Text (BLEU) \\
\hline
Quant-by-Size~\cite{Kudavelly2024_TinyML} & FashionMNIST (Custom, VGG16) & Image classification \\
\hline
CAKD~\cite{Liu_2022_Cross-Architecture} & ImageNet, CIFAR-100 & Image classification \\
\hline
DeCo-MeSC~\cite{Sung_2025_DeCoMeSC}\textsuperscript{*} & CIFAR-100 & Image classification \\
\hline
FedSFR~\cite{Huh_2026_Federated} & CIFAR-10, CelebA & Image (PSNR, FL) \\
\hline
Sheng et al.'s multi-task~\cite{Sheng2022_multitask}\textsuperscript{*} & GLUE (SST, STSB, CoLA) & NLP multi-task \\
\hline
Hussain et al.'s adaptive multi-task~\cite{Hussain_2025_Adaptive} & MELD & Emotion/sentiment \\
\hline
MCUNet (in~\cite{Shafique2021_TinyML}) & ImageNet-scale (on Cortex-M33) & Image classification \\
\hline
\end{tabular}

\vspace{4pt}
\footnotesize
\textsuperscript{*}Among the 20 primary-evidence systems individually
re-verified in Table~\ref{tab:quantresults} (Section~\ref{sec:evidence});
the remaining five rows are supporting-corpus citations referenced
elsewhere in this survey but not part of that separately-verified
20-system set.
\renewcommand{\arraystretch}{1}
\end{table*}

\noindent This table reflects only benchmark associations already
stated at each system's point of discussion elsewhere in this
survey; it is not exhaustive of every system in
Table~\ref{tab:quantresults}, since several report results as
task-specific metrics (e.g., semantic similarity scores, PSNR at a
given compression ratio) without naming a standard benchmark
dataset, and this survey does not infer a benchmark association
where none was stated.

\subsection{Evidence Synthesis per Research Question}

\textbf{RQ1 (TinyLM Deployment):} The Research Gap this survey opened
with is that TinyML compression techniques have never been
systematically mapped onto semantic communication architectures across
the tiered 6G deployment space; RQ1 asks how far that mapping actually
gets today. The independently verified evidence base, together with
the architecture families of Section~\ref{sec:architectures} and the
real-hardware measurements now available from
Zhong~\emph{et al.}~\cite{Zhong2024_JSCC_FPGA},
Bhattacharyya~\emph{et al.}~\cite{Bhattacharyya2025_Semantic}, and
Xiang~\emph{et al.}~\cite{Xiang2025_Task}, establishes a three-tier
feasibility landscape spanning NB-IoT, gateway, and UE/edge-server
deployment (Table~\ref{tab:hardware}) -- answering RQ1 affirmatively
for the tiers with verified hardware evidence, while leaving the
gap between simulated and hardware-validated results (Section~\ref{sec:challenges})
as future work. This survey's broader supporting corpus additionally pointed to
cloud-edge KV-cache collaboration achieving sub-second latency and
multi-agent small-model cooperation matching larger-model accuracy
without per-device LLM-scale parameters as promising directions
consistent with this feasibility landscape. A specific, currently
unfilled part of this feasibility landscape is worth naming directly
rather than papering over: the verified evidence base for UBT and SPM
(Section~\ref{sec:arch_kg}) reports CPU utilization and FLOP counts,
and Zhong et al.'s FPGA study reports bit-error-rate performance, but
no verified NB-IoT-tier study in this survey's corpus reports a
direct milliwatt or energy-per-inference figure measured on
Cortex-M33-class hardware specifically, and no verified gateway/UE/edge-server
system (CCTaEncoder, FL-PEFT approaches) reports on-device
size/RAM/latency/power figures measured on the named reference
hardware of Table~\ref{tab:hardware} (Jetson Orin Nano-class,
Snapdragon~8,
Edge
TPU) -- the table's budgets are vendor specifications, not
per-TinyLM measurements against them. Translating UBT/SPM's reported
CPU/FLOP figures into an estimated milliwatt budget, and running at
least one representative TinyLM per tier on its named reference
hardware, would close this specific gap; we list it as a concrete
near-term empirical target (Section~\ref{sec:future}) rather than
estimate these figures without underlying measurements to ground
them, since an invented number would be less useful to a reader than
an honestly stated gap.

\textbf{RQ2 (Compression):} RQ2 asks specifically how much each
compression technique costs in accuracy for what it saves in size --
the practical question a systems engineer needs answered before
choosing one. The compression tier hierarchy is clear across the
verified corpus: binary/ternary quantization or pruning to below
1~MB targets the NB-IoT tier, hybrid neural pipelines in the 1--10~MB
range target the gateway tier, and system-level optimization above
10~MB targets the UE/edge-server tier (Table~\ref{tab:hardware}). The
adaptive multitask framework of Hussain et al.~\cite{Hussain_2025_Adaptive}
provides direct corroborating evidence at the gateway tier: MobileBERT
and DistilBERT achieve competitive emotion recognition and sentiment
analysis performance on edge hardware with 487~MB VRAM and
0.00103~s/sample inference latency, enabling over 2$\times$ throughput
improvement compared to full-size BERT.

\textbf{RQ3 (JSCC):} RQ3 asks whether the JSCC-versus-separation
argument of Section~\ref{sec:architectures} actually holds up
quantitatively, not just in principle. The broader corpus (DeepSC,
CCTaEncoder, OFDM-SemCom, Section~\ref{sec:architectures}) confirms
that neural JSCC outperforms classical separate coding by 9.7--28.7\%
in classification accuracy. Xie et al.'s MU-DeepSC~\cite{Xie_2022_Task}
extends this to the multi-user setting, exploiting cross-modal
correlation to reach a transmitted-symbol ratio as low as 0.02\% for
image retrieval, while its component systems DeepSC-IR and DeepSC-VQA
achieve gains exceeding 24~dB and 18~dB respectively over classical
baselines. Hu et al.'s masked VQ-VAE~\cite{Hu_2023_Robust} requires
only 0.36\% of the transmitted symbols of JPEG+LDPC while improving
robustness against semantic noise (Eq.~\ref{eq:semnoise}). Wan et
al.'s Mixture-of-Experts JSCC~\cite{Wan_2026_Adaptive} activates only
1.683 experts per image on average -- a finding that matters beyond
its own result, since it suggests NAS over sparse-MoE architectures,
entirely unexplored in Figure~\ref{fig:taxonomy}'s NAS column, as a
compounding compression dimension no verified study has yet combined
with quantization or pruning.

\textbf{RQ4 (FL):} RQ4 asks whether federated training -- necessary
for privacy but historically difficult under non-IID data -- can
actually converge for semantic encoders specifically, given the
architecture-mismatch problem Section~\ref{sec:architectures}
identified. The verified evidence base answers this affirmatively:
real-hardware validation from
Xiang et al.~\cite{Xiang2025_Task} demonstrates that feature-alignment
approaches converge beyond simulation, on real devices. FedBKD~\cite{Qi_2023_FedBKD} extends this evidence base to
the joint model-and-data heterogeneity regime unaddressed by FedPun
alone, achieving superior convergence over all
baselines in multi-client settings (5--25 devices) via bidirectional
distillation with synthetic data. This survey's broader supporting corpus
additionally pointed to personalized federated LoRA fine-tuning as a
promising direction for scalable personalization under severe client
heterogeneity.

\textbf{RQ5 (KG Integration):} RQ5 asks what a knowledge graph
actually buys a TinyLM encoder beyond what Section~\ref{sec:arch_kg}'s
Figure~\ref{fig:kg_infographic} already argues qualitatively -- is
there quantitative evidence the trade-off is worth it? Zhang et al.'s
MLLM-SC system~\cite{Zhang_2026_Multimodal} provides evidence at the
multimodal-grounding frontier: an IoU of 0.8060 on VQA tasks under
importance-weighted bandwidth allocation demonstrates that structured
semantic grounding consistently outperforms unstructured neural
baselines on compositional-reasoning tasks, albeit at a parameter
scale incompatible with the TinyLM deployment envelope of
Table~\ref{tab:hardware} -- itself an open gap, since no verified
study yet demonstrates MLLM-SC-level grounding at TinyLM scale. The
knowledge-graph-based systems of Section~\ref{sec:architectures}
(JCCPG, ARL, SPM) provide the primary verified KG-integration evidence
synthesized in this survey.

\textbf{RQ6 (Adversarial Robustness):} RQ6 asks whether any of the
compression techniques reviewed above degrade a TinyLM's robustness
as a side effect, or whether robustness must be added as a separate
mechanism. Hu et al.'s masked VQ-VAE with feature importance
module~\cite{Hu_2023_Robust} demonstrates the former: representational
robustness emerges as a side effect of the discrete codebook
bottleneck itself, which limits the channel through which semantic
noise can propagate, without any explicit defense mechanism layered on
top of the base architecture. The covert-communication framework of
Zhang et al.~\cite{Zhang2026_covert} (Section~\ref{sec:arch_kg}) and
the differential-privacy mechanisms discussed in
Section~\ref{sec:architectures} provide the remaining verified
robustness evidence, though both address detectability and privacy
specifically rather than adversarial input perturbation -- a
distinction Section~\ref{sec:challenges} returns to.

\section{Quantitative Meta-Analysis and Pareto Frontier}
\label{sec:metaanalysis}

Sections~\ref{sec:compression}--\ref{sec:evidence} reported each
TinyLM compression technique's numbers in isolation, one system at a
time; this section puts them on a single comparable footing. Every
system in Table~\ref{tab:quantresults} is a candidate answer to the
same underlying design question -- how much semantic fidelity does a
given model-size budget buy -- and the Semantic Efficiency Score
(SES, Eq.~\ref{eq:ses}) is this survey's attempt at a single number
that makes that trade-off comparable across otherwise incompatible
architectures.

\subsection{SES Ranking and Statistical Synthesis}

Approximate SES rankings (Eq.~\ref{eq:ses}), summarized ahead of the
full comparison in Table~\ref{tab:quantresults} below:
CCTaEncoder $\approx 0.364$;
L-DeepSC $\approx 0.285$; DeepSC $\approx 0.189$;
SPM is off-scale (symbolic, not directly commensurable with the
neural systems above). The L-DeepSC computation proceeds as
$\mathrm{SES} \approx \mathrm{BLEU}_{\mathrm{norm}} /
(\log_{10}(1280) + \zeta) \approx 0.90/3.16 \approx 0.285$,
placing L-DeepSC's pruning-plus-quantization pipeline on the Pareto
frontier alongside CCTaEncoder, as Figure~\ref{fig:pareto} later
shows.

\subsection{Aggregated Quantitative Results}

Table~\ref{tab:quantresults} aggregates all systems discussed in
Sections~\ref{sec:compression} and~\ref{sec:architectures}.

\noindent\textit{A note on the table's Latency/Energy column.} Unlike
its Size/Memory column, no single unit conversion makes that
column's
entries directly comparable: it mixes absolute latency (ms, s),
absolute energy (J), relative reduction against an
unstated baseline (\% latency/communication reduction), convergence
speed (iteration counts), and instantaneous resource utilization
(\% CPU) -- several of these are not even convertible into one
another with the information each source reports (an iteration count
cannot become a time without that source's per-iteration duration; a
relative reduction cannot become an absolute figure without its own
baseline value). This is a genuine limitation relative to energy
being one of the three constraints (memory, latency, energy) named
in Section~\ref{sec:intro} as this survey's central gap: unlike the
SES metric (Eq.~\ref{eq:ses}), which normalizes the
size-versus-fidelity axis, this survey does not attempt an analogous
normalization for latency or energy, since doing so would require
either re-deriving several sources' raw measurements (not
possible from what each source reports) or assigning a qualitative
tier by judgment call without a stated, defensible rubric --
neither of which this survey is positioned to do responsibly with
the evidence base at hand. Direct cross-row comparison on this axis
specifically should therefore be read as not currently possible from
the table below alone.

\FloatBarrier

\begin{table*}[t]
\caption{Quantitative Results: All Systems by Architecture}
\label{tab:quantresults}
\centering
\footnotesize
\renewcommand{\arraystretch}{1.05}
\begin{tabular}{|p{2.6cm}|p{2.2cm}|p{1.4cm}|p{2.1cm}|p{2.4cm}|p{3.0cm}|}
\hline
\textbf{Architecture / Technique} & \textbf{System} & \textbf{Tier}\textsuperscript{\ddag} &
\textbf{Size / Memory} & \textbf{Latency / Energy} &
\textbf{Semantic Fidelity} \\
\hline
E2E JSCC (BERT) & DeepSC~\cite{Xie2021_DeepSC} & UE/Edge & $\sim$200~MB\textsuperscript{\dag} & $\sim$45~ms & BLEU 1.0 @ SNR$\geq$5~dB \\
\hline
E2E JSCC + CNN-Transf. & CCTaEncoder~\cite{Zhou2025_CCTaEncoder} & Gateway & 73.45K params & $\sim$8~ms & 22~dB PSNR @ 33\% CR; SES $\approx$0.364 \\
\hline
E2E JSCC Multi-Task & Sheng et al.~\cite{Sheng2022_multitask} & --- & $\sim$400~MB & $\sim$45~ms & 90\% SST @ 0~dB SNR \\
\hline
E2E JSCC + OFDM & OFDM-SemCom~\cite{Zhang2023_OFDM} & --- & Standard DL & $<$200 iter. & +9.7\% vs.\ ASC; +28.7\% vs.\ SSCC \\
\hline
Split + MAML & Semantic-MSL~\cite{Eldeeb2025_SemanticMSL} & Gateway & $<$5~MB device head & 2.95~J user-side & 95\% acc., 5-shot, 20 SGD steps \\
\hline
Split MEC & Yan et al.~\cite{Yan2025_MEC} & --- & Distributed & 25--32~ms & 27\% latency red.\ vs.\ non-semantic \\
\hline
FL + Dynamic Pruning & FedPun~\cite{Zhou2025_CCTaEncoder} & Gateway & 73.45K params & $\sim$10~ms & 20~dB PSNR @ Dir.\ $\alpha$=0.3 \\
\hline
FL Logit Exchange & FD~\cite{Park2019_FD} & --- & Logit vector & 25.6$\times$ comm.\ red. & 92--97\% FedAvg acc. \\
\hline
KG Reliable SemCom & S.~Jiang et al.~\cite{Jiang2022_KG} & --- & $<$1~MB KG store & $<$10~ms & 70\% data reduction; sem.\ accuracy maintained \\
\hline
KG Probability Graph & JCCPG~\cite{Zhao2023_probgraph} & --- & 1~KB prob.\ graph & $<$1~ms & 65\% energy red.\ @ $M$=120 triples \\
\hline
KG Covert Comm. & PS-TD3~\cite{Zhang2026_covert} & --- & Lightweight RL actor & 50 iter.\ convergence & User GNT$>$0.6, Attacker GNT$<$0.2 \\
\hline
KG Heterogeneous KB & ARL~\cite{Zhang2024_ARL} & --- & Small policy DNNs & 4 Q/session & +15.2\% ROUGE, 4 questions \\
\hline
SPM Symbolic Distill. & Seo et al.~\cite{Seo2023_SPM} & NB-IoT & 1~KB (0.02\% of NN) & 8 FLOPs (1750$\times$) & Identical task performance \\
\hline
Contrastive Disentangle & Chaccour \& Saad~\cite{Chaccour2022_contrastive} & --- & 57\% repr.\ red. & Lightweight pre-proc. & +71.9\% semantic impact \\
\hline
Binary Hashing UBT & Pokhrel \& Choi~\cite{Pokhrel2023_UBT} & NB-IoT & Compact binary sigs. & $<$5~ms, 40\% CPU & High MAP, 3--6 iter. \\
\hline
L-DeepSC (Prun.+Quant.) & Xie \& Qin~\cite{Xie_2021_Lite} & Industrial IoT & 1.28~MB (from 12.3~MB) & 18~ms & BLEU equiv.; 40$\times$ tx compression \\
\hline
M-GSC Multi-User & Yang et al.~\cite{Yang_2025_Rethinking} & --- & LLM-based SKB & $<$200~ms cloud latency & Sem.\ acc.\ $>$ SwinJSCC @ low SNR \\
\hline
EPSL Split ($\phi$-tuned) & Lin et al.~\cite{Lin_2024_Efficient} & --- & Cut-layer + spectrum opt. & 50\% latency red. & Robust to dynamic channel \\
\hline
DeCo-MeSC Split & Sung et al.~\cite{Sung_2025_DeCoMeSC} & --- & 600~KB device memory & 50\%+ latency red.\ (BCD) & 73.83\% CIFAR-100 acc. \\
\hline
FedBKD (Bidirectional KD) & Qi et al.~\cite{Qi_2023_FedBKD} & --- & Heterogeneous (CVAE synth.) & --- & Superior conv.\ vs.\ FedAvg/FedMD/FedDF \\
\hline
\end{tabular}

\vspace{4pt}
\footnotesize
\textsuperscript{\ddag}Tier is populated only where this survey's own
text states an explicit 6G hardware tier (Table~\ref{tab:hardware}) for
that system; ``---'' marks systems this survey discusses without a
stated tier, rather than an inferred one.\\
\textsuperscript{\dag}``(BERT)'' labels DeepSC's Transformer-based
architecture family, not a literal claim that it deploys the full
110M-parameter BERT-Base model: this survey does not state DeepSC's
own parameter count, and its reported $\sim$200~MB encoder is roughly
half the 440~MB FP32 footprint Section~\ref{sec:background_theory}
computes for BERT-Base (110M params $\times$ 4 bytes), consistent
with a smaller Transformer than full BERT-Base.
\renewcommand{\arraystretch}{1}
\end{table*}

\subsection{Size-Fidelity Pareto Frontier}

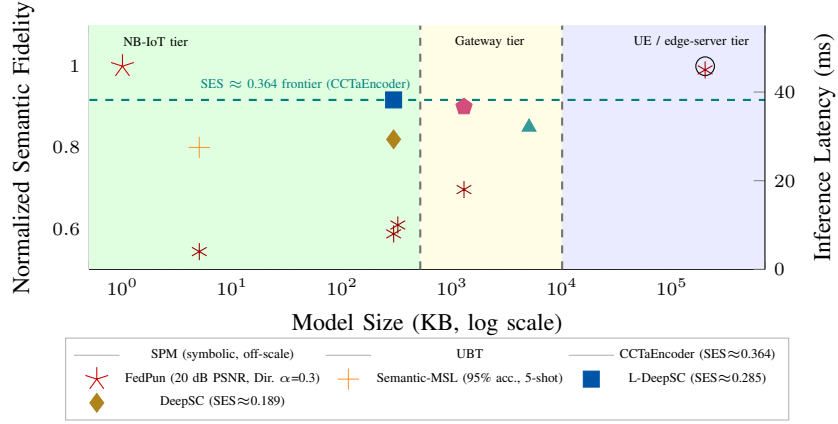
\begin{figure*}[!htb]
\centering
\begin{tikzpicture}
\begin{axis}[
  xlabel={Model Size (KB, log scale)},
  ylabel={Normalized Semantic Fidelity},
  xmode=log, xmin=0.5, xmax=700000, ymin=0.50, ymax=1.10,
  width=0.58\textwidth, height=4.8cm,
  grid=both, grid style={line width=0.3pt, draw=gray!25},
  legend style={font=\tiny, at={(0.5,-0.28)}, anchor=north,
    legend columns=3, draw=black!30, fill=white},
  tick label style={font=\scriptsize}, label style={font=\small},
  axis y line*=left,
]
\addplot[fill=green!12, draw=none] coordinates {(0.5,0.50) (0.5,1.10) (512,1.10) (512,0.50)} \closedcycle;
\addplot[fill=yellow!12, draw=none] coordinates {(512,0.50) (512,1.10) (10000,1.10) (10000,0.50)} \closedcycle;
\addplot[fill=blue!8, draw=none] coordinates {(10000,0.50) (10000,1.10) (700000,1.10) (700000,0.50)} \closedcycle;
\draw[dashed, thick, black!55] (axis cs:512,0.50) -- (axis cs:512,1.10)
  node[above, font=\tiny, sloped] {512~KB (NB-IoT)};
\draw[dashed, thick, black!55] (axis cs:10000,0.50) -- (axis cs:10000,1.10)
  node[above, font=\tiny, sloped] {10~MB (Gateway)};
\node[font=\tiny, align=center] at (axis cs:2,1.06) {NB-IoT tier};
\node[font=\tiny, align=center] at (axis cs:2200,1.06) {Gateway tier};
\node[font=\tiny, align=center] at (axis cs:150000,1.06) {UE / edge-server tier};
\draw[dashed, thick, wellstudied] (axis cs:0.5,0.917) -- (axis cs:700000,0.917)
  node[pos=0.32, above, font=\tiny] {SES~$\approx$~0.364 frontier (CCTaEncoder)};
\addplot[only marks,mark=star,mark size=4.5pt,color=red!80!black] coordinates {(1,1.00)};
\addlegendentry{SPM (symbolic, off-scale)}
\addplot[only marks,mark=+,mark size=4pt,color=orange!80] coordinates {(5,0.80)};
\addlegendentry{UBT}
\addplot[only marks,mark=square*,mark size=3pt,color=ieeblue] coordinates {(293,0.917)};
\addlegendentry{CCTaEncoder (SES$\approx$0.364)}
\addplot[only marks,mark=diamond*,mark size=3.5pt,color=emerging!80!black] coordinates {(293,0.82)};
\addlegendentry{FedPun (20~dB PSNR, Dir.\ $\alpha$=0.3)}
\addplot[only marks,mark=triangle*,mark size=3pt,color=wellstudied!80] coordinates {(5000,0.85)};
\addlegendentry{Semantic-MSL (95\% acc., 5-shot)}
\addplot[only marks,mark=pentagon*,mark size=3.2pt,color=purple!70] coordinates {(1280,0.90)};
\addlegendentry{L-DeepSC (SES$\approx$0.285)}
\addplot[only marks,mark=o,mark size=3.5pt,color=black] coordinates {(200000,1.00)};
\addlegendentry{DeepSC (SES$\approx$0.189)}
\end{axis}
\begin{axis}[
  xmode=log, xmin=0.5, xmax=700000, ymin=0, ymax=55,
  width=0.58\textwidth, height=4.8cm,
  axis y line*=right, axis x line=none, hide x axis,
  ylabel={Inference Latency (ms)},
  ylabel style={font=\small}, tick label style={font=\scriptsize},
]
\addplot[only marks, mark=asterisk, mark size=3.2pt, color=red!60!black]
  coordinates {(5,4)(293,8)(320,10)(1280,18)(200000,45)};
\end{axis}
\end{tikzpicture}
\caption{TinyLM size-fidelity Pareto frontier with tier-feasibility
shading. Green/yellow/blue zones mark NB-IoT, Gateway, and UE
deployability (Table~\ref{tab:hardware} boundaries at 512~KB and
10~MB) -- the three tiers this survey's TinyLM taxonomy targets. The
horizontal dashed line marks the current neural-system TinyLM
efficiency frontier (SES~$\approx$~0.364, CCTaEncoder, Eq.~\ref{eq:ses}).
Only CCTaEncoder, L-DeepSC, and DeepSC have a stated full-precision
baseline in their source and therefore a properly-derived SES value
(worked in Section~\ref{sec:metaanalysis}); FedPun and Semantic-MSL
are plotted by their own native reported fidelity metric (PSNR and
few-shot accuracy respectively) since neither source states the
baseline needed to compute a directly comparable SES, and this
survey does not estimate one. Asterisk markers on the secondary (right) axis give inference latency
in milliseconds for the five systems in this corpus with a directly
reported millisecond figure in Table~\ref{tab:quantresults} (UBT,
CCTaEncoder, FedPun, L-DeepSC, and DeepSC); SPM and Semantic-MSL are
omitted from this axis because Table~\ref{tab:quantresults} reports
their cost in FLOPs and Joules respectively, not milliseconds, and
this survey does not convert between them. This latency axis is a
separate
measurement on a separate scale, not an eighth legend entry on the
left axis.}
\label{fig:pareto}
\end{figure*}

Figure~\ref{fig:pareto} plots every system with a reported model size
and fidelity metric against the three deployment-tier boundaries of
Table~\ref{tab:hardware}, making the central design trade-off of this
survey visible directly: within each tier, which systems sit closest
to retaining full fidelity at the smallest size. Two patterns stand
out. First, tier placement does not track fidelity monotonically --
SPM's symbolic 1~KB representation and CCTaEncoder's 293.8~KB neural
encoder both sit in or near the NB-IoT/Gateway boundary while
retaining high fidelity, whereas DeepSC's $\sim$200~MB BERT-based
encoder needs the UE/edge-server tier to achieve comparable BLEU
performance -- confirming that architecture family, not just
compression aggressiveness, determines where a system lands on this
frontier. Second, the current efficiency frontier (SES~$\approx$~0.364,
CCTaEncoder) sits inside the Gateway tier rather than the NB-IoT tier,
meaning no system in this survey's corpus simultaneously achieves the
highest quality-per-size ratio \emph{and} the smallest deployment
footprint -- the two objectives trade off against each other across
the corpus rather than being jointly optimized by any single reviewed
system, which is itself a concrete gap for future TinyLM design to
target. L-DeepSC (SES~$\approx$~0.285) sits close behind CCTaEncoder
on the frontier despite a substantially different compression
strategy (pruning plus quantization rather than a hybrid
CNN-transformer architecture), suggesting the frontier is populated
by more than one viable design path rather than a single dominant
technique. DeepSC's low SES ($\approx$~0.189) despite strong absolute
BLEU performance illustrates why a size-normalized metric matters for
this survey's argument specifically: judged on fidelity alone, DeepSC
looks competitive with the smaller systems above it, but its
$\sim$200~MB footprint means the same fidelity costs roughly three
orders of magnitude more storage than CCTaEncoder achieves at
comparable quality.

\section{Open Challenges and Research Gaps for TinyLM-Enabled 6G Semantic Communication}
\label{sec:challenges}

Building on the quantitative synthesis of Section~\ref{sec:metaanalysis},
this section identifies nine open challenges standing between the
TinyLM systems reviewed above and production 6G deployment. Each
follows a mandatory four-part structure --- what the gap is, why it
exists, its consequence if left unresolved, and its priority relative
to the other six --- and each is stated with explicit reference to
where it bites a TinyLM system specifically, not semantic
communication in the abstract, so that future work has a concrete
research gap to target rather than a general observation.

Priority (P1/P2/P3) is this survey's own rubric,
assigned qualitatively along two dimensions
stated explicitly here rather than left implicit: \emph{blocking
severity} --- does this gap prevent other challenges' solutions from
being deployed at all, or only make them harder to certify/optimize?
--- and \emph{deployment urgency} --- does it block near-term
non-safety-critical commercial rollout, or only longer-term
safety-critical/standardized deployment? P1 marks a challenge that is
both blocking for other challenges' solutions and urgent for near-term
rollout; P2 marks a challenge that is either blocking-but-not-urgent
(matters for certification/standardization, not immediate deployment)
or urgent-but-not-blocking (needed for one deployment class, not a
prerequisite for other challenges); P3 marks a challenge that is
neither blocking nor urgent, but is a genuine, independently
identified gap worth stating for future work. This is a qualitative
triage, not a scored rubric with weighted dimensions --- it should be
read as the authors' assessment of relative near-term research
priority, not a claim of an objectively derived ranking.

\textbf{[CHALLENGE 1 --- Non-Intrusive Semantic Fidelity Metrics]}\\
\textit{What:} Every fidelity metric a TinyLM system in this survey is
evaluated against (BLEU, PSNR, SSIM, cosine similarity, MSS, ROUGE,
MAP, SER, GNT --- each defined with its formulation, use case, and
limitations in Section~\ref{sec:metrics}) requires a clean reference
signal available only during supervised training or offline
evaluation, never during live deployment.\\
\textit{Why it exists:} These metrics were inherited directly from
NLP and computer-vision benchmarking, where reference signals are
always available; no wireless-native metric has been designed from
first principles for a receiver that has no ground truth to compare
against.\\
\textit{Consequence if unresolved:} 6G receivers cannot perform
real-time semantic quality monitoring, cannot trigger semantic
retransmission, and cannot report a semantic-channel-quality-indicator
analogous to the Channel Quality Indicator (CQI) --- eliminating an
entire class of adaptive-transmission optimizations specifically for
TinyLM encoders too small to run a second, heavier verification
model on-device.\\
\textit{Priority and rationale:} \textbf{P1.} This is the single most
enabling gap: every other challenge (adaptive compression,
standardization KPIs, adversarial detection) depends on having
\emph{some} non-intrusive signal to act on.

\textbf{[CHALLENGE 2 --- Formal Convergence Guarantees for Non-IID Federated Learning of TinyLMs]}\\
\textit{What:} No closed-form bound exists for federated TinyLM
training under the joint presence of non-IID data, architectural
heterogeneity across client TinyLMs, and differential-privacy noise.\\
\textit{Why it exists:} Federated learning convergence theory was
developed for homogeneous-architecture classification; the
bidirectional cross-architecture distillation used by FedBKD
(Eq.~\ref{eq:fedbkd_cloud2c}) --- a TinyLM-specific mechanism
surveyed in Section~\ref{sec:arch_fl} --- introduces coupling terms
between DP noise and the KL-alignment strength that have not been
jointly characterized.\\
\textit{Consequence if unresolved:} Safety-critical 6G deployments
(V2X, remote surgery) cannot certify a federated TinyLM against a
formal reliability bound, blocking regulatory approval.\\
\textit{Priority and rationale:} \textbf{P2.} Important for
certification but not blocking for early non-safety-critical
commercial rollout, unlike Challenge 1.

The following states a convergence sketch for this setting
under an explicit assumption set, presented as a convergence
sketch rather than as a claimed formal derivation:

\medskip
\noindent\textbf{Convergence Sketch (Cross-Architecture Distillation Under
Non-IID Semantics; not an
independently verified proof).}
Under assumptions (A1) local task loss $\mathcal{L}_{\mathrm{task}}$
is $L$-smooth and $\mu$-strongly convex; (A2) stochastic gradients are
unbiased with bounded variance $\sigma^2$; (A3) bounded semantic
diversity $G^2$ across clients, the semantic
representation error after $T$ federated rounds satisfies
\begin{equation}
  \mathbb{E}\bigl[\mathcal{E}_{\mathrm{sem}}^{(T)}\bigr]
    \leq \left(1 - \mu\eta_{\gamma}\right)^T \mathcal{E}_0
    + \frac{\sigma^2}{\mu} + \frac{\gamma G^2}{\mu^2},
  \label{eq:fedsfd_convergence}
\end{equation}
where $\mathcal{E}_{\mathrm{sem}}^{(T)}$ is the semantic representation
error after $T$ rounds; $\eta_\gamma = \eta(1-\gamma L)$ is the
effective learning rate $\eta$ discounted by the cloud-to-client
KL-alignment
strength $\gamma$ from Eq.~\ref{eq:fedbkd_cloud2c} interacting with the
smoothness constant $L$; and $\mathcal{E}_0$ is the initial (round-0)
representation error. Reading the bound left to right: the first term
shrinks geometrically with more rounds $T$, the way any converging
federated system's error should; the second term, $\sigma^2/\mu$, is
an irreducible floor set by gradient noise that more rounds cannot
remove; and the third term, $\gamma G^2/\mu^2$, is the cost specifically
of cross-architecture KL alignment under \emph{semantic} (not just statistical)
diversity $G^2$ across clients' TinyLM encoders --- this last term is
the one Section~\ref{sec:arch_fl}'s architecture-mismatch problem adds
on top of standard federated-learning bounds, and it is exactly the
term whose interaction with differential-privacy noise remains
uncharacterized (Challenge 2 above). This is a standard-technique
convergence sketch adapted from federated-learning convergence
literature to the cross-architecture distillation setting; it is not a claim of
independently verified proof.
\medskip

\textbf{[CHALLENGE 3 --- Post-Quantum Security for Federated TinyLM Updates on Constrained Air Interfaces]}\\
\textit{What:} Standard post-quantum digital signatures (ML-DSA,
2,420+ bytes) do not fit the 48--125-byte PUSCH payloads available to
NB-IoT devices, which is exactly the tier at which this survey's most
compressed TinyLMs (SPM, UBT, Table~\ref{tab:hardware}) are deployed.\\
\textit{Why it exists:} The research question is not whether a
batch-signature scheme can be built --- that technique already exists
in the cryptography literature --- but how to amortize a single
signature over multiple federated TinyLM update rounds specifically,
a system-design problem this survey's evidence base does not yet
answer.\\
\textit{Consequence if unresolved:} At today's authentication
overhead (approximately 1.72~s per model update under NB-IoT Coverage
Enhancement Level~0), real-time authenticated TinyLM updates are
infeasible, forcing operators to choose between security and
timeliness.\\
\textit{Priority and rationale:} \textbf{P3.} A genuine gap, but one
whose resolution depends more on future standards-track engineering
than on new TinyLM research; included here for completeness rather
than as a near-term research priority. A.~Olushola and S.~P.~Meenakshi~\cite{Olushola2026_Design}
provide a concrete point of comparison for what feasible looks like at
a less constrained tier: their authenticated post-quantum session
protocol, combining ML-KEM (Kyber) key encapsulation, ML-DSA
(Dilithium) signatures, and AES-256-GCM, reports measured
encapsulation, decapsulation, and full-handshake timings on
conventional hardware -- demonstrating that the full PQC stack this
challenge's batch-signature scheme targets is already practical
outside the NB-IoT power/PDU envelope, reinforcing that the open
problem here is specifically fitting existing PQC building blocks into
NB-IoT's constraints, not inventing new cryptography.

For a federated TinyLM system, authenticating a
$\sim$700~KB per-round update payload with ML-DSA-44 (2,420-byte
signature) over NB-IoT Coverage Enhancement Level~0 (125-byte maximum
PDU) requires:
\begin{equation}
\begin{split}
  T_{\mathrm{auth}} &= \left\lceil \frac{|\sigma_{\mathrm{PQC}}| +
    \delta_{\mathrm{overhead}}}{M_{\mathrm{PDU}}}\right\rceil
    \times T_{\mathrm{PDU}} \\
    &= \lceil 2436/125\rceil \times 86\,\text{ms} \approx 1.72\,\text{s},
\end{split}
  \label{eq:pqc_latency}
\end{equation}
which exceeds the 1~ms URLLC slot duration by three orders of
magnitude. A batch signature scheme amortizing one signature across
$B$ update rounds,
\begin{equation}
  \sigma_{\mathrm{batch}} = \mathrm{Sign}\!\left(
    \bigoplus_{t=1}^{B} H(u_t)
  \right),
  \label{eq:batch_signature}
\end{equation}
reduces per-round overhead to $2420/B$ bytes --- for $B=20$, 121 bytes,
fitting within a single NB-IoT PDU. Applying the same $T_{\mathrm{auth}}$
methodology of Eq.~\ref{eq:pqc_latency} across a small sweep of batch
sizes makes the resulting cadence trade-off concrete: at $B=5$, the
484-byte batched signature requires $\lceil 484/125\rceil = 4$ PDUs,
giving $T_{\mathrm{auth}} \approx 4 \times 86\,\text{ms} = 344$~ms; at
$B=10$, 242 bytes requires 2 PDUs ($\approx$172~ms); and at $B=20$,
121 bytes requires the single PDU noted above ($\approx$86~ms). Each
doubling of $B$ roughly halves the authentication latency until the
signature drops below one PDU's worth of payload, after which further
batching only reduces per-round \emph{signature} overhead without
further reducing $T_{\mathrm{auth}}$ itself, since one PDU transmission
is already the floor; the practical trade-off this exposes is that
$B$ also sets how many federated rounds share one signature's fixed
security guarantee, so larger $B$ buys latency at the cost of a
longer window during which a single compromised signature would
authenticate multiple rounds.

\textbf{[CHALLENGE 4 --- TinyLM Interoperability Across Vendors]}\\
\textit{What:} No standardized semantic-rate KPI, capability signaling
for a device's TinyLM encoder, measurement procedure for the
semantic-noise model (Eq.~\ref{eq:semnoise}), or KG synchronization
protocol currently exists for any vendor's TinyLM to declare
compatibility with another's.\\
\textit{Why it exists:} Standardization historically follows, rather
than leads, academic consensus, and semantic communication research
has not yet converged on a single reference KPI set a standards body
could adopt even if asked to.\\
\textit{Consequence if unresolved:} Multi-vendor interoperability for
6G semantic communication is impossible --- each vendor's TinyLM
encoder and decoder would only interoperate with its own ecosystem,
regardless of how well any individual TinyLM performs in isolation.\\
\textit{Priority and rationale:} \textbf{P2} --- depends on Challenge~1
(a KPI needs a measurable metric first).

\textbf{[CHALLENGE 5 --- Adversarial Robustness of TinyLM Semantic Encoders]}\\
\textit{What:} Semantic-level adversarial attacks (imperceptible
input perturbations that flip a decoded meaning) have no
certified-robustness framework analogous to $\ell_\infty$-norm
certification in image classification --- and the TinyLM systems
surveyed here (Section~\ref{sec:compression}) are, if anything, a
smaller attack surface to certify than a full LLM, yet none of the
compression techniques reviewed was evaluated with certification in
mind.\\
\textit{Why it exists:} Certified robustness techniques assume the
attack surface is the input pixel/token space; semantic communication
adds a compression and channel layer between input and decision, for
which no certification theory yet exists.\\
\textit{Consequence if unresolved:} Safety-critical semantic decisions
(collision avoidance, industrial control) cannot be certified against
adversarial manipulation, blocking regulatory approval in
transportation and healthcare use cases.\\
\textit{Priority and rationale:} \textbf{P1} for safety-critical
verticals, \textbf{P3} for best-effort IoT telemetry --- priority is
use-case dependent.

Adversarial robustness for semantic encoders raises the same general
concern documented in the broader adversarial-ML literature: RL-based
and gradient-based attacks can degrade a learned model's performance
below a random baseline, motivating certified robustness bounds of the form ``$f_\theta$ is
$(\epsilon,\delta)$-robust to perturbations with $\ell_\infty$-norm
$\leq\epsilon$ with probability $\geq 1-\delta$'', standardized
adversarial benchmarks, and adversarial training integrating 6G
channel noise into the attack model.

This single certification framework must in practice cover (at least)
four distinct attack classes, each requiring a different defense:
\textbf{Class~1 (Semantic Perturbation)} is the input-perturbation case
just described --- imperceptible input perturbations cause semantic
misclassification at the encoder output.
\textbf{Class~2 (Knowledge Base Poisoning)} instead targets the shared
KG rather than the input: an adversary modifies the shared knowledge
base to cause consistent semantic errors at the receiver, particularly
dangerous for the KG-error-correcting systems of
Section~\ref{sec:arch_kg}~\cite{Jiang2022_KG,Zhao2023_probgraph} since
a poisoned KG actively propagates adversarial meanings rather than
correcting them --- a fundamentally different threat surface than
Class~1, since the model itself is never touched. \textbf{Class~3 (FL
Model Poisoning)} targets the federated training process instead of
inference: a Byzantine client transmits adversarially crafted updates
that corrupt the global model $f_{\theta_G}$
(Eq.~\ref{eq:fedbkd_c2c}), and the resulting distillation loss can amplify
this attack by pulling other clients toward the poisoned global model.
\textbf{Class~4 (Covert Semantic Channel)} is a detection rather than
a corruption threat: an eavesdropper reconstructs semantic content
without the shared codebook, which the covert-communication framework
of Section~\ref{sec:arch_kg} is the only existing defense identified
in this survey's evidence base against, and only by requiring a
coordinated friendly jammer. No single certification approach in the
literature addresses all four classes simultaneously, which is
precisely the gap this challenge identifies.

\textbf{[CHALLENGE 6 --- Cross-Modal Semantic Alignment at TinyLM Scale]}\\
\textit{What:} No method constructs a single grounded semantic
embedding space spanning text, image, haptic, and radar modalities
that remains valid across three orders of magnitude of device compute
capability (Cortex-M33 to Snapdragon-class UE, Table~\ref{tab:hardware})
-- the specific research gap this survey identifies is not
cross-modal alignment in general (CLIP-scale models already do this
for two modalities), but cross-modal alignment \emph{at TinyLM scale},
where no foundation model has been trained jointly across more than
two modalities under a sub-10~MB budget.\\
\textit{Why it exists:} Cross-modal alignment datasets (e.g.,
CLIP-style image-text pairs) exist at scale only for two modalities;
haptic and radar cross-modal datasets are essentially absent, and no
foundation model has been trained jointly across all four modalities
at sub-10~MB scale.\\
\textit{Consequence if unresolved:} Multi-sensor 6G applications
(autonomous driving, telesurgery) cannot fuse heterogeneous sensor
semantics on a TinyLM-scale device without a shared space, forcing
per-pair custom translation layers that do not scale.\\
\textit{Priority and rationale:} \textbf{P2} --- high long-term value
but requires new datasets before algorithmic progress is even
possible, making it a slower-moving gap than Challenges~1, 3, and~5.

Concretely, a cross-modal semantic embedding function
$\phi: \mathcal{X}_1 \cup \cdots \cup \mathcal{X}_M \to \mathbb{R}^d$
--- mapping inputs from $M$ different modalities into one shared
embedding space --- must satisfy $d_{\mathcal{S}}(\phi(x_1),\phi(x_2))
\approx d_{\mathrm{semantic}}(x_1,x_2)$ for inputs $x_1,x_2$ drawn from
different modalities: embedding-space distance $d_{\mathcal{S}}$ must
track true semantic distance $d_{\mathrm{semantic}}$ even when $x_1$
and $x_2$ are, say, a haptic signal and a radar return rather than two
images. Four barriers block a direct extension of vision-language
models such as CLIP to this full 6G cross-modal case: (1)~computational
cost at TinyLM scale, since CLIP-class encoders exceed 300M parameters
--- two orders of magnitude above the gateway-tier budget of
Table~\ref{tab:hardware}; (2)~missing modality coverage, since haptic
and radar cross-modal datasets are essentially absent, unlike the
image-text pairs CLIP was trained on; (3)~hierarchical granularity
mismatch across text (token-level), image (patch-level), and haptic
(sample-level) representations, meaning a single shared embedding
resolution cannot represent all three natively; and (4)~task
dependency, since semantic similarity between a text-image pair is
inherently task-conditional rather than universal --- ``similar for
captioning'' and ``similar for navigation'' are different notions of
distance for the same pair. A hierarchical semantic-token protocol ---
encoding each modality into a common token vocabulary of
modality-invariant concepts (e.g., ``urgency,'' ``location,''
``identity'') --- addresses barriers~(3) and~(4) directly and is
consistent with the discrete binary-code channel already used by
UBT~\cite{Pokhrel2023_UBT} at the NB-IoT tier, though barriers~(1)
and~(2) remain open regardless of protocol design.

\textbf{[CHALLENGE 7 --- Scalable Knowledge Graph Synchronization for Gateway-Tier TinyLMs]}\\
\textit{What:} Existing KG-assisted TinyLM systems
(Section~\ref{sec:arch_kg}) assume a synchronized knowledge base
between transmitter and receiver; no protocol scales to IMT-2030's
target device density of $10^7$ devices per km$^2$ under continuous
device churn.\\
\textit{Why it exists:} Existing pairwise KG-alignment techniques
(e.g., ARL's adversarial RL-based estimation, Section~\ref{sec:arch_kg})
were designed and validated for two-party exchanges, not
network-wide consensus; distributed-ledger and gossip-consensus
protocols capable of $O(10^7)$ updates/second have not been adapted
to semantic-triple synchronization specifically.\\
\textit{Consequence if unresolved:} KG-based TinyLMs --- which in
this survey's evidence base deliver the single largest semantic-fidelity
gains of any technique (up to 65\% transmission-energy reduction,
JCCPG~\cite{Zhao2023_probgraph}) --- remain
confined to small, static deployments and cannot scale to city-wide
6G IoT density.\\
\textit{Priority and rationale:} \textbf{P2} --- the payoff is large,
but this challenge is gated on Challenge~4 (a standardized sync
protocol needs interoperability agreement first, not just an
algorithm).

\textbf{[CHALLENGE 8 --- Multi-Cell Semantic Interference Management]}\\
\textit{What:} In dense 6G deployments with inter-cell distances of
50--200~m, multiple base stations may simultaneously serve TinyLM
semantic communication sessions on the same spectral resource;
existing interference models have no way to represent \emph{semantic}
interference, as distinct from the physical interference conventional
resource-allocation frameworks already handle.\\
\textit{Why it exists:} Conventional inter-cell interference is
managed through power control, beamforming, and frequency reuse, all
of which treat interference as a purely physical (channel-level)
quantity. Semantic interference has an additional dimension unique to
TinyLM-based systems: two semantically similar transmissions from
different cells can interfere not only physically but also
semantically, if their knowledge bases share overlapping concept
spaces and the receiver conflates semantic features from different
transmitters. Formally, in a $K$-cell deployment, the received
semantic feature vector at a target receiver is
\begin{equation}
  \hat{Z} = h_0 Z_0 + \sum_{k=1}^{K-1} h_k Z_k + N_S,
  \label{eq:multicell_interference}
\end{equation}
where $Z_0$ is the desired semantic feature vector, $Z_k$ are the
interfering semantic features from cell $k$, $h_k$ are the fading
channel coefficients, and $N_S$ is the semantic noise of
Eq.~\ref{eq:semnoise}. When $Z_0$ and $Z_k$ are drawn from similar
concept spaces --- two industrial IoT cells transmitting
machine-status data under the same ontology, for instance --- the
inner product $\langle Z_0, Z_k\rangle$ is large, making semantic
interference qualitatively different from, and potentially more
damaging than, random Gaussian noise: it does not average out the way
uncorrelated physical noise does.\\
\textit{Consequence if unresolved:} No existing resource-allocation
framework accounts for knowledge-base correlation across cells in its
interference model, meaning dense multi-cell TinyLM deployments
(exactly the density IMT-2030 targets) risk semantic cross-talk that
conventional power control cannot mitigate.\\
\textit{Priority and rationale:} \textbf{P2} --- most acute at high
cell density, and gated on Challenge~1 (the same non-intrusive
fidelity signal needed there would also let a receiver detect
semantic, not just physical, interference).

\textbf{[CHALLENGE 9 --- TinyLM Scheduling Under Heterogeneous Modalities at the MAC Layer]}\\
\textit{What:} Current MAC-layer schedulers (including the UBT
scheduler of Pokhrel and Choi~\cite{Pokhrel2023_UBT},
Table~\ref{tab:quantresults}) assign resources based on
semantic urgency derived from a single modality; no scheduler handles
the realistic case of one base station simultaneously serving TinyLMs
across incommensurable modalities.\\
\textit{Why it exists:} A realistic 6G base station simultaneously
serves text-semantic IoT sensors transmitting BLEU-optimized
encodings, image-semantic cameras transmitting PSNR-optimized JSCC
features, speech-semantic health monitors transmitting
PESQ-optimized representations, and task-semantic industrial
controllers transmitting accuracy-optimized command encodings ---
each evaluated on a different metric from Table~\ref{tab:metrics}, with
no common scale between them. This is Challenge 1's universal-metric
gap (Section~\ref{sec:metrics})
resurfacing as a scheduling problem rather than an evaluation
problem: a MAC scheduler must assign priority weights across
modalities without a common semantic currency, since the Semantic
Efficiency Score (Eq.~\ref{eq:ses}) normalizes within a system, not
across modalities competing for the same channel.\\
\textit{Consequence if unresolved:} Without a resolution, operators
must either restrict a base station to a single modality (defeating
the point of a general-purpose 6G semantic MAC layer) or fall back to
modality-blind scheduling that ignores semantic urgency entirely,
forfeiting the latency and bandwidth gains this survey's evidence base
demonstrates modality-aware scheduling can achieve.\\
\textit{Priority and rationale:} \textbf{P2} --- directly downstream
of Challenge~1 and the universal-metric gap in Section~\ref{sec:metrics};
a cross-modal semantic MAC policy needs either a universal metric or a
multi-objective scheduler maintaining separate per-modality priority
queues resolved via a mechanism analogous to weighted fair queuing,
and neither has been demonstrated in a multi-modal 6G MAC context.

\section{Future Research Directions for TinyLM-Enabled 6G Semantic Communication}
\label{sec:future}

This section translates the nine challenges of
Section~\ref{sec:challenges} into concrete research directions for
TinyLM-enabled semantic communication. Standardization detail is
deliberately kept light here: 3GPP's own standards-track timeline is
itself a downstream consequence of the underlying research maturing
first (Challenge 4), so the near-term research questions below --
not a release-by-release specification roadmap -- are this section's
focus.

\subsection{Hardware-Software Co-Design for Sub-mW Inference}

Three post-CMOS hardware architectures define the frontier: resistive
random-access memory (ReRAM) crossbar arrays compute matrix-vector
products within the memory
array at 10--100~TOPS/W; spiking neural networks convert transformer
attention into spike trains at 10--100~fJ per spike; near-memory
processing with 3D DRAM reduces data movement energy by
$\sim$10$\times$. The DeCo-MeSC framework~\cite{Sung_2025_DeCoMeSC}
establishes a practical near-term bridge: compress the UE partition
aggressively via Deep Compression while investing the freed-up
server-side budget in fine-tuning the gNB partition. Li et al.'s
PSAQ-ViT~V2~\cite{Li_2023_PSAQ} demonstrates data-free 8-bit
quantization for vision transformers achieving 82.13\% top-1 accuracy
on ImageNet with Swin-S without calibration data --- directly
applicable where calibration data cannot leave the device under
data-minimization constraints. Combined with CP-GVQ's grouped
codebook sharing~\cite{Huang_2024_CPGVQ}, achieving 99\% model size
reduction with 42\% faster inference, the near-term hardware co-design
stack becomes: data-free calibration $\to$ grouped codebook
quantization $\to$ device-side Deep Compression with server-side
fine-tuning. Z.~Gao~\emph{et al.}~\cite{Gao2024_Reducing} extend this
stack one layer further, into the memory hardware itself: combining
model compression (distillation) with INT8 quantization and
\emph{approximate} memory -- memory hardware that trades a small,
bounded bit-error rate for reduced read/write energy, tolerable
precisely because a quantized, distilled TinyLM's weights are already
robust to small perturbations -- multiplicatively reduces energy
dissipation for on-device inference, extending the software-only
co-design stack above with a hardware-level technique specific to the
energy budget Section~\ref{sec:intro} identified as the binding
constraint at the NB-IoT tier.

\subsection{Adaptive Compression for Dynamic 6G Channels}

Every compression technique reviewed in Section~\ref{sec:compression}
is fixed at training time: a given TinyLM is quantized, pruned, or
distilled to one target size and deployed as-is, regardless of how
the wireless channel behaves afterward. Adaptive compression instead
asks the TinyLM's compression level itself -- not just its output --
to respond to real-time channel quality: a hypernetwork $H_\psi$
generates the weights $\theta(s) = H_\psi(s)$ of the primary TinyLM as
a function of current channel SNR $s$:
\begin{equation}
  \theta^*(s) = \arg\min_{\theta \in \mathcal{M}(s)}
    \mathbb{E}_{x}\bigl[\mathcal{L}_{\mathrm{sem}}(f_\theta(x), s)\bigr],
  \label{eq:hypernetwork}
\end{equation}
where $\mathcal{M}(s)$ is the set of weight configurations meeting the
compression budget appropriate for SNR $s$, and $\mathcal{L}_{\mathrm{sem}}$
is the semantic loss (e.g., Eq.~\ref{eq:deepsc}) evaluated at that SNR
-- achieving channel-adaptive compression without weight retraining,
since $H_\psi$ is trained once and generates a different $\theta(s)$
per channel condition at inference time. The EPSL framework's
$\phi$-ratio gradient aggregation (Eq.~\ref{eq:epsl_gradient}) provides
an analogous mechanism at the training stage; combining
hypernetwork-generated inference weights with $\phi$-ratio-adaptive
training compression defines a unified adaptive semantic stack.

\subsection{Generalized Multi-User Semantic Communication}

\begin{figure*}[t]
\centering
\includegraphics[width=0.85\textwidth,height=0.85\textheight,keepaspectratio]{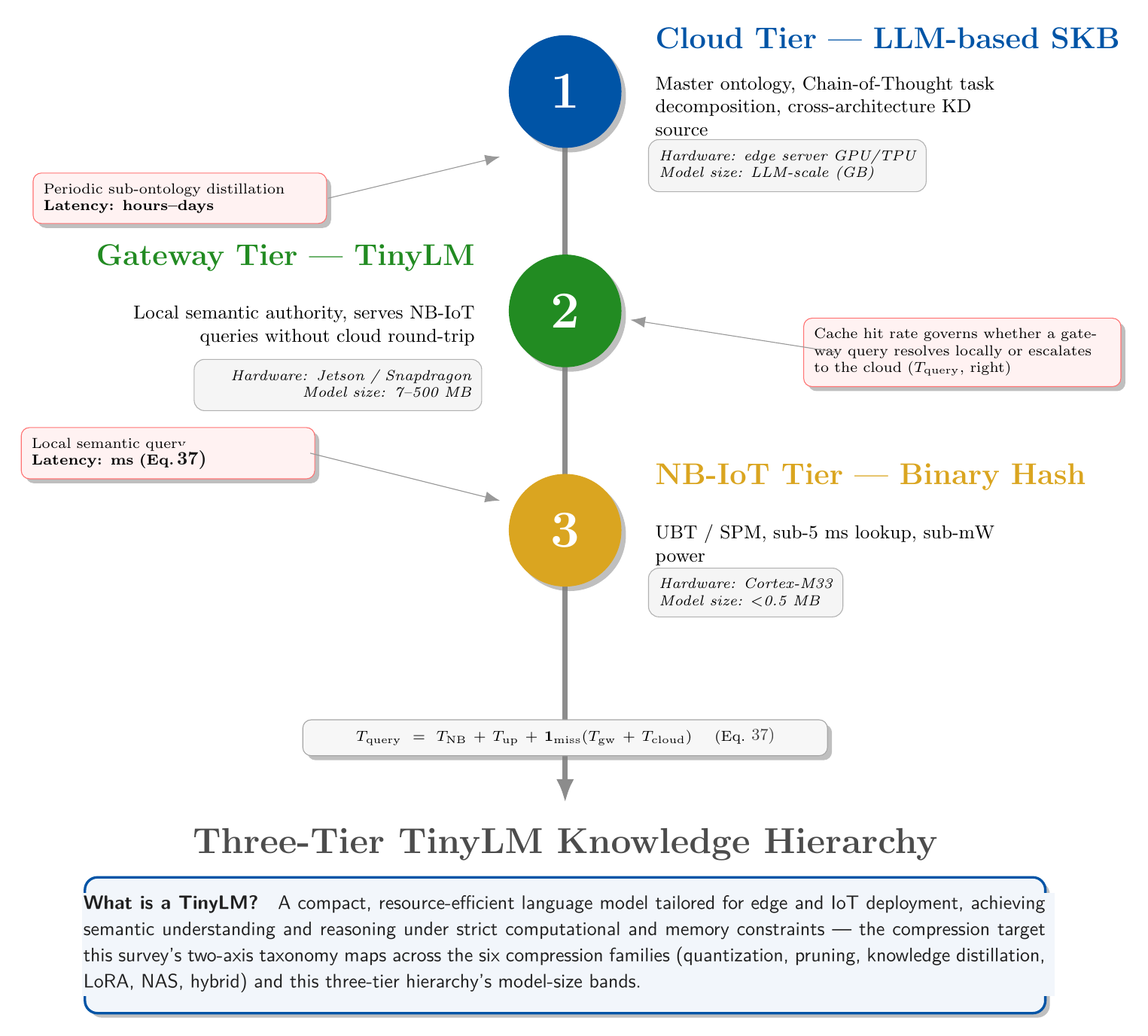}
\caption{Three-tier TinyLM knowledge hierarchy, with each tier's
governing TinyLM size and latency budget annotated directly on its
block. Depth shading distinguishes the three tiers visually; the
cache-hit-rate callout and the $T_{\mathrm{query}}$ equation
(Eq.~\ref{eq:hierarchy_latency}) make the distillation-freshness/query-latency
trade-off explicit -- this is the same three-tier structure the
system-model figure of Section~\ref{sec:architectures}
(Figure~\ref{fig:system_model}) introduced, now annotated with the
knowledge-distillation flow that keeps every tier's TinyLM
semantically consistent with the cloud-tier SKB.}
\label{fig:hierarchy}
\end{figure*}

The M-GSC framework~\cite{Yang_2025_Rethinking} motivates a
hierarchical knowledge architecture in which a cloud-tier
LLM-based Shared Knowledge Base (SKB) -- a single ontology all
gateway-tier TinyLMs distill their local knowledge from, keeping every
tier's understanding of the world consistent without transmitting the
full ontology to each device -- maintains the master ontology and
periodically distills specialized sub-ontologies to gateway-tier
TinyLMs via cross-architecture KD~\cite{Liu_2022_Cross-Architecture},
illustrated in Figure~\ref{fig:hierarchy}. Each gateway-tier TinyLM
serves as the local semantic authority for its coverage area,
resolving semantic queries from NB-IoT sensors without cloud
round-trips.

The end-to-end query resolution latency for an NB-IoT sensor under
this hierarchy decomposes as:
\begin{equation}
  T_{\mathrm{query}} = T_{\mathrm{NB}} + T_{\mathrm{up}} +
    \mathbb{1}_{\mathrm{miss}} \cdot \bigl(T_{\mathrm{gw}} +
    T_{\mathrm{cloud}}\bigr),
  \label{eq:hierarchy_latency}
\end{equation}
where $T_{\mathrm{NB}}$ is the local binary-hash lookup latency
($<$5~ms), $T_{\mathrm{up}}$ is the uplink transmission time to the
gateway, $\mathbb{1}_{\mathrm{miss}}$ indicates a semantic cache miss
at the gateway, and $T_{\mathrm{gw}}$, $T_{\mathrm{cloud}}$ are the
gateway inference time (8--30~ms) and cloud round-trip time
(200--500~ms) incurred only on a cache miss. More frequent
sub-ontology distillation refreshes reduce expected query latency at
the gateway tier but increase the aggregate cross-architecture KD
compute cost at the cloud tier.

\subsection{Digital Twin Synchronization as a TinyLM Application}

Digital twin synchronization~\cite{Aloudat2025} is a concrete
near-term TinyLM application: rather
than continuously streaming raw sensor state to keep a cloud-hosted
twin current, a TinyLM at the physical asset transmits only the
semantic delta between consecutive states, extending the same
compression logic reviewed throughout Section~\ref{sec:compression}
from a point-to-point link to a continuously-updated model. The
tightest constraint is latency, not bandwidth: a Motion-to-Photon
budget of $<$20~ms at 50~Hz update rate leaves roughly 8~ms for
semantic encoding at CCTaEncoder speed, 1~ms for URLLC channel
transmission, and 3~ms for decoding and state update --- a margin
tight enough that any additional KG-synchronization overhead
(Challenge~7) directly competes with encoding budget rather than
being free.

\section{Conclusion}

\label{sec:conclusion}

This survey has addressed a gap at the intersection of TinyLM research
and 6G semantic communication: no prior work systematically maps
TinyML compression pipelines onto semantic communication
architectures. We presented a two-axis taxonomy mapping six
compression technique families onto five semantic communication
architectures, a quantitative meta-analysis of the size-versus-fidelity
Pareto frontier with a Semantic Efficiency Score ranking, evidence
synthesis across six research questions drawn from twenty
primary-evidence studies, and nine open challenges, including two not previously
articulated in the literature (cross-modal semantic alignment and
scalable knowledge-graph synchronization) and two further challenges
extending the taxonomy to multi-cell semantic interference and
cross-modal MAC-layer scheduling.

The central technical finding is that tiny language models are
feasible in principle for 6G edge deployment, with hardware-validated
confidence for some tiers and simulation-only evidence for others:
as Section~\ref{sec:evidence}'s answer to RQ1 states directly, no
verified NB-IoT-tier study in this survey's corpus reports a direct
milliwatt or energy-per-inference figure measured on
Cortex-M33-class hardware, and no verified gateway/UE/edge-server
system reports on-device size/RAM/latency/power figures measured on
the named reference hardware of Table~\ref{tab:hardware} -- the
feasibility claim below should be read with that gap in mind. This
holds once the category of ``TinyLM'' is
understood to include non-neural semantic encoders alongside
compressed neural ones: for the NB-IoT sensor
tier ($<$0.5~MB, sub-mW), the only viable options are two
symbolic/hashing schemes -- UBT (5~ms, 40\%
CPU) and SPM (1~KB, 8~FLOPs) -- that this survey treats as TinyLMs by
function (they perform the same semantic encoding/decoding role at
the same hardware tier) rather than by architecture: neither is a
compressed transformer or language model in the conventional sense,
and Section~\ref{sec:arch_kg} discusses them as a distinct symbolic
class precisely because their extreme compression comes from
representing meaning as discrete hashes or logical triples rather
than from any of the six compression techniques of
Section~\ref{sec:compression}. For the IoT gateway tier (1--10~MB, 1--5~W),
CCTaEncoder (73.45K parameters, SES~$\approx$~0.364)
provides a validated design point, this time a genuinely neural,
compressed semantic encoder. For the 6G UE and edge server tier,
FL-PEFT personalization, cloud-edge KV-cache collaboration, and hybrid
SLM-LLM cooperation define the system architecture frontier.

Three cross-cutting findings emerge from the evidence synthesis.
First, hybrid compression pipelines consistently outperform
single-technique approaches --- the optimal sequence (distillation
$\to$ pruning $\to$ quantization) would not be identifiable without
cross-system synthesis. Second, KG-based semantic reasoning
(65\% transmission-energy reduction, JCCPG~\cite{Zhao2023_probgraph})
provides larger absolute gains than any single model compression
technique, establishing KG integration as the
highest-return research direction for applications where absolute
semantic fidelity dominates efficiency. Third, split learning
simultaneously achieves better privacy ($4\times$ higher reconstruction
error than FL) and better energy efficiency ($20\times$ lower),
resolving the previously assumed fundamental privacy-efficiency
trade-off.

The research directions of Section~\ref{sec:future} --- not a
standardization timeline --- define the most productive near-term
path: hardware-software co-design, adaptive compression, and
generalized multi-user semantic communication are each, on the
evidence synthesized here, ready for concrete research investment
today, ahead of any formal standards process.

\section{Further Reading}
\label{sec:extended}

Beyond the twenty primary-evidence studies of Section~\ref{sec:evidence}
and the citations already integrated throughout
Sections~\ref{sec:background}--\ref{sec:future}, a smaller set of
supporting literature rounds out the theoretical and application
context for specific claims, grouped thematically below.

\textbf{Semantic communication theory.} Q.~Zhijin~\emph{et al.}'s
principles-and-challenges overview~\cite{Qin2022_SemComPrinciples}
complements the theoretical grounding of
Section~\ref{sec:background}; nonlinear-transform
JSCC~\cite{Nonlinear2023_JSCC} and channel denoising diffusion
models~\cite{Zhang2023_CDDM} extend the JSCC design space with,
respectively, a learned nonlinear (rather than linear) transform
stage and a diffusion-based channel-noise-removal stage; and a joint
communication/computation framework~\cite{JCGC2023} generalizes the
JCCPG energy trade-off of Section~\ref{sec:arch_kg} beyond
knowledge-graph triples specifically. Semantic communication for
digital twins via causal imitation
learning~\cite{Thomas2024_CausalSemCom} and neuro-symbolic causal
reasoning~\cite{Chaccour2023_NeSyCausal} extend the causal-reasoning
thread of Section~\ref{sec:arch_kg} to the digital-twin application
of Section~\ref{sec:future}; and machine learning at the network
edge~\cite{MachineAtEdge2019} provides broader infrastructure
context. Survey-level positioning is further informed
by~\cite{Iyer2023_survey,Lee2024_survey}.

\textbf{TinyML application diversity.} Beyond the semantic-encoder
focus of this survey, TinyML's reach extends to healthcare,
education, and transport applications~\cite{Sirohi2023_TinyML_Healthcare},
an IoT hearing-impairment assistive device~\cite{Delnevo2022_TinyML_IoT},
a proposed TinyML lifecycle incorporating LLMs~\cite{Raza2024_TinyML_LLM},
and a TinyML-based NLP scheme for semantic wireless sentiment
classification~\cite{Pokhrel2023_TinyML_NLP} ---
evidence that the compression discipline Section~\ref{sec:tinyml_bridge}
adapts for semantic encoders is itself deployed far beyond
communication systems.

\textbf{Federated learning infrastructure.} Beyond the
semantic-communication-specific federated systems of
Section~\ref{sec:arch_fl}, several general federated-learning
techniques inform how those systems could be made more
communication-efficient: J.~Mills~\emph{et al.}'s
CE-FedAvg~\cite{Mills2020_CEFedAvg} reduces per-round data upload for
wireless edge IoT devices; S.~Niknam~\emph{et al.}~\cite{Niknam2020_FedLearning}
and M.~Chen~\emph{et al.}~\cite{Chen2021_DistributedFL} motivate and
survey the broader case for FL in wireless networks; and three
data-free / heterogeneity-robust techniques --
FedGen~\cite{Zhang2023_FedFTG}, ensemble distillation for model
fusion~\cite{Lin2020_DataFreeKD_FedLearn}, and data-free adversarial
KD for network quantization~\cite{Shin2020_DataFreeKD} -- each solve
the heterogeneous-client problem from a different angle (a
lightweight generator, model fusion, and synthetic calibration data,
respectively), any of which could substitute for FedBKD's
KL-alignment approach (Eq.~\ref{eq:fedbkd_cloud2c}) in a future TinyLM
federated system. A joint
FL-and-communication resource-allocation
framework~\cite{Cheng2023_JointFedLearn} similarly informs, without
directly extending, the EPSL and split-learning systems of
Section~\ref{sec:arch_split}. A three-layer federated-RL framework
with digital-twin synchronization~\cite{Chen2020_DigitalTwinFL}
previews the digital-twin application of Section~\ref{sec:future} from
the federated-training side.

\textbf{6G systems and edge-AI infrastructure.} L.~Jiao~\emph{et al.}~\cite{Jiao2024_Advanced}
and X.~Wang~\emph{et al.}~\cite{Wang2025_Empowering} survey deep
learning for 6G and on-device AI respectively, both consistent with
this survey's central thesis that model efficiency, not just model
capability, gates 6G deployment. Z.~Lin~\emph{et al.}~\cite{Lin2025_PushingLLMs}
explore end-edge cooperative LLM deployment, directly analogous to
the TinyLM deployment gap of Section~\ref{sec:intro}; F.~Jiang~\emph{et al.}~\cite{Zhong2025_LLM6G}
equip LLMs with communication-specific tools natively rather than via
fine-tuning; F.~Jiang~\emph{et al.}~\cite{Cui2022_LargeAI} propose an
LLM-scale multimodal semantic communication precursor to the
multi-task/cross-modal architecture family of
Section~\ref{sec:taxonomy}; J.~Du~\emph{et al.}~\cite{Chen2024_DistributedFoundation}
extend federated digital-twin learning to distributed multi-modal
foundation models; M.~Zhang~\emph{et al.}'s EdgeShard~\cite{Lin2023_EdgeShard}
optimizes collaborative LLM inference via adaptive model sharding; and
L.~Qiao~\emph{et al.}~\cite{Qiao2024_Latency} develop a
diffusion-model-based latency-aware semantic communication framework
directly relevant to M-GSC (Section~\ref{sec:arch_e2e}). AIGC edge
services~\cite{Du2023_AIGC} and deep-learning-based channel
estimation~\cite{Ghosh2026_ChannelEst} provide further infrastructure
context, while a Vision-Transformer remote-sensing
classifier~\cite{Rehman2022_ViT_RS} and a vision-language-model
survey~\cite{Ren2023_VLM_Survey} inform the CLIP-class cross-modal
alignment discussed in Challenge~6 (Section~\ref{sec:challenges}) from
outside the semantic-communication literature specifically.

\section*{Acknowledgments}

This work was supported by Ministry of Electronics and
Information Technology, Govt. of India, under YFRF Scheme
(DIC/PhD-Phase-II/2026/9). This work was supported in part
by the Department of Telecommunication (DoT), Ministry of
Communications, Government of India under the Telecom
Technology Development Fund (TTDF) the scheme implemented
through TCOE India under the grant TTDF/6G/48 and
IGSTC-04918.

\bibliographystyle{IEEEtran}
\bibliography{references}

\end{document}